\documentclass[aos,preprint]{imsart}

\RequirePackage[OT1]{fontenc}
\RequirePackage{amsthm,amsmath}
\RequirePackage[colorlinks,citecolor=blue,urlcolor=blue]{hyperref}

\usepackage{natbib} 
\usepackage{algorithm}
\usepackage{algpseudocode}

\usepackage{amsfonts}
\usepackage{amsgen,amsmath,amstext,amsbsy,amsopn,amssymb}
\usepackage[dvips]{graphicx}
\usepackage{comment}
\usepackage{enumitem}
\usepackage{natbib}
\usepackage{booktabs}
\usepackage{blkarray}
\usepackage{algorithm}
\usepackage{algpseudocode}
\usepackage{url}
\usepackage{longtable}

\usepackage[usenames]{color}
\usepackage{lipsum}
\usepackage{subcaption}

\newcommand{\red}{\color{red}}

\allowdisplaybreaks

\newtheorem{Theorem}{Theorem}

\newtheorem{Lemma}{Lemma}
\newtheorem{Remark}{Remark}

\newtheorem{Proposition}{Proposition}

\newcommand{\be}{\begin{equation}}
\newcommand{\ee}{\end{equation}}
\newcommand{\bea}{\begin{eqnarray}}
\newcommand{\eea}{\end{eqnarray}}
\newcommand{\beas}{\begin{eqnarray*}}
	\newcommand{\eeas}{\end{eqnarray*}}

\newcommand{\X}{{\mathbf{X}}}
\newcommand{\B}{{\mathbf{B}}}

\newcommand{\E}{{\mathbf{E}}}

\renewcommand{\S}{{\mathbf{S}}}

\newcommand{\W}{{\mathbf{W}}}
\newcommand{\A}{{\mathbf{A}}}
\newcommand{\Y}{{\mathbf{Y}}}
\newcommand{\Z}{{\mathbf{Z}}}

\newcommand{\bOmega}{\boldsymbol{\Omega}}

\newcommand{\rank}{{\rm rank}}

\newcommand{\diag}{{\rm diag}}

\newcommand{\argmin}{\mathop{\rm arg\min}}

\allowdisplaybreaks

\def\mybox#1{\vskip1mm \begin{center} \bf \red
		\hspace{.0\textwidth}\vbox{\hrule\hbox{\vrule\kern6pt
				\parbox{.95\textwidth}{\kern6pt#1\vskip6pt}\kern6pt\vrule}\hrule}
	\end{center} \vskip-5mm}

\arxiv{arXiv:1611.01129}

\startlocaldefs
\numberwithin{equation}{section}
\theoremstyle{plain}

\endlocaldefs

\begin{document}

\begin{frontmatter}
\title{Cross: Efficient Low-rank Tensor Completion}
\runtitle{Low-rank Tensor Recovery}

\thankstext{T1}{The research of Anru Zhang was supported in part by NSF Grant DMS-1811868 and NIH grant R01-GM131399-01.}
\begin{aug}
\author{\fnms{Anru} \snm{Zhang}\thanksref{T1}\ead[label=e1]{anruzhang@stat.wisc.edu}
\ead[label=u1,url]{www.stat.wisc.edu/{\raise.17ex\hbox{$\scriptstyle\sim$}}anruzhang/
}}

\runauthor{A. Zhang}

\affiliation{University of Wisconsin-Madison}

\address{Department of Statistics\\
University of Wisconsin-Madison\\
Madison, WI, 53706\\
\printead{e1}\\
\printead{u1}}

\end{aug}

\begin{abstract}
	The completion of tensors, or high-order arrays, attracts significant attention in recent research. Current literature on tensor completion primarily focuses on recovery from a set of uniformly randomly measured entries, and the required number of measurements to achieve recovery is not guaranteed to be optimal. In addition, the implementation of some previous methods is NP-hard. In this article, we propose a framework for low-rank tensor completion via a novel tensor measurement scheme we name Cross. The proposed procedure is efficient and easy to implement. In particular, we show that a third order tensor of Tucker rank-$(r_1, r_2, r_3)$ in $p_1$-by-$p_2$-by-$p_3$ dimensional space can be recovered from as few as $r_1r_2r_3 + r_1(p_1-r_1) + r_2(p_2-r_2) + r_3(p_3-r_3)$ noiseless measurements, which matches the sample complexity lower-bound. In the case of noisy measurements, we also develop a theoretical upper bound and the matching minimax lower bound for recovery error over certain classes of low-rank tensors for the proposed procedure. The results can be further extended to fourth or higher-order tensors. Simulation studies show that the method performs well under a variety of settings. Finally, the procedure is illustrated through a real dataset in neuroimaging.
\end{abstract}



\end{frontmatter}

\section{Introduction}\label{sec.intro}

Tensors, or high-order arrays, commonly arise in a wide range of applications, including neuroimaging \citep{zhou2013tensor,li2013tucker,guhaniyogi2015bayesian,li2016parsimonious,sun2016sparse}, recommender systems \citep{karatzoglou2010multiverse,rendle2010pairwise,sun2015provable}, hyperspectral image compression \citep{li2010tensor}, multi-energy computed tomography \citep{semerci2014tensor,li2014tensor}, computer vision \citep{liu2013tensor}, 3D light field displays \citep{wetzstein2012tensor} and scientific computation \citep{oseledets2009breaking}. With the revolutionary development of modern technologies, the rapid increase in data dimension, memory and time expenses outgrows the power of computing devices, which makes it difficult to work directly on the complete datasets and models.
For example, a tensor of dimension $10^4$-by-$10^4$-by-$10^4$ would be difficult to upload into the Random Access Memory (RAM) of a typical computer, making it hard to directly perform operations that involves all entries of the tensor. In order to conduct various statistical tensor data analyses, such as SVD or PCA \citep{richard2014statistical,zhang2017tensor} and Monte-Carlo algorithms for computations on large tensors \citep{guhaniyogi2015bayesian,johndrow2017tensor} when limited computation power is available, a fast and sufficient tensor compression is essential. To this end, a natural idea is to sample a small portion of entries from the original tensor dataset that preserves the important structural information and allows efficient recovery. By storing these entries to RAM, the follow-up tensor data analysis can be highly facilitated.

Tensor completion, whose central goal is to recover low-rank tensors based on limited numbers of measurable entries, is a plausible idea for compression and decompression of high-dimensional low-rank tensors. Such problems have been central and well-studied for order-2 tensors (i.e. matrices) in the fields of high-dimensional statistics and machine learning for the last decade. 
A large body of matrix completion literatures focused on the scenario of uniformly randomly sampled observations  \citep{keshavan2009matrix,candes2010power,koltchinskii2011nuclear,rohde2011estimation,negahban2011estimation,agarwal2012noisy}, but there exists another line of works where the observations are collected by other means, such as deterministically sampling patterns \citep{pimentel2016characterization},  column-subset-selection \citep{rudelson2007sampling,krishnamurthy2013low,wang2015provably,cai2016structured} and general sampling distributions \citep{klopp2014noisy}.
 There are efficient procedures for matrix completion with strong theoretical guarantees. For example, for a $p_1$-by-$p_2$ matrix of rank-$r$, whenever roughly $O(r(p_1+p_2)\text{polylog}(p_1+p_2))$ uniformly randomly selected entries are observed, one can achieve nice recovery with high probability using convex algorithms such as matrix nuclear norm minimization \citep{candes2010power,recht2011simpler} and max-norm minimization \citep{srebro2005rank,cai2016matrix}.
For matrix completion, the required number of measurements nearly matches the degrees of freedom, $O((p_1+p_2)r)$, for $p_1$-by-$p_2$ matrices of rank-$r$. 

Although significant progress has been made for matrix completion, similar problems for order-3 or higher tensors are far more difficult. There have been some recent literature, including \cite{gandy2011tensor,kressner2014low,yuan2014tensor,mu2014square,bhojanapalli2015new,shah2015optimal,barak2016noisy,yuan2016incoherent}, that studied tensor completion based on similar formulations. To be specific, let $\X \in \mathbb{R}^{p_1\times p_2 \times p_3}$ be an order-$3$ low-rank tensor, and $\boldsymbol{\Omega}$ be a subset of $[1:p_1]\times[1:p_2]\times[1:p_3]$. The goal of tensor completion is to recover $\X$ based on the observable entries indexed by $\bOmega$. Most of the previous literature focuses on the setting where the indices of the observable entries are uniformly randomly selected. For example, \cite{gandy2011tensor,liu2013tensor} proposed the matricization nuclear norm minimization, which requires $O(rp^2 \text{polylog}(p))$ observations to recover order-3 tensors of dimension $p$-by-$p$-by-$p$ and Tucker rank-$(r,r,r)$. 
Later, \cite{jain2014provable,bhojanapalli2015new} considered an alternative minimization method for completion of low-rank tensors with CP decomposition and orthogonal factors. \cite{yuan2014tensor,yuan2016incoherent} proposed the tensor nuclear norm minimization algorithm for tensor completion with noiseless observations and further proved that their proposed method has guaranteed performance for $p$-by-$p$-by-$p$ tensors of Tucker rank-$(r,r,r)$ with high probability when $|\boldsymbol{\Omega}| \geq O((r^{1/2}p^{3/2} + r^2p)\text{polylog}(p))$.
However, it is unclear whether the required number of measurements in this literature could be further improved or not. In addition, some of these proposed procedures, such as tensor matrix nuclear norm minimization, are proved to be computationally NP-hard, making them very difficult to apply in real problems. 
Recently, \cite{barak2016noisy} further showed that the completion of $p$-by-$p$-by-$p$ low-rank tensors is computationally infeasible when only $O(p^{3/2})$ uniform random entries are observable, unless a more efficient algorithm exists for boolean satisfiability problem.

The central goal of this paper is to address the following question: is it possible to perform efficient low-rank tensor completion with a small number of observable entries? If so, what is the sample complexity, i.e., the minimal number of entries one needs to observe, so that there exist fast algorithms for tensor completion with guaranteed performance? This problem is important to statistical learning theory and is inevitable in many high-dimensional tensor data analyses.
Given the previous discussions, to sample entries uniformly at random may not be an optimal strategy to achieve the central goal. Instead, we propose a novel tensor measurement scheme and the corresponding efficient low-rank tensor completion algorithm. 
We name our methods \emph{Cross Tensor Measurement Scheme} because the measurement set is in the shape of a high-dimensional cross contained in the tensor. We show that one can recover an unknown, Tucker rank-$(r_1, r_2, r_3)$, and $p_1$-by-$p_2$-by-$p_3$ tensor $\X$ with 
$$|\bOmega| = r_1r_2r_3 + r_1(p_1-r_1) + r_2(p_2-r_2) + r_3(p_3 - r_3)$$ 
noiseless Cross tensor measurements. This outperforms the previous methods in literature, and matches the degrees of freedom for all rank-$(r_1, r_2, r_3)$ tensors of dimensions $p_1$-by-$p_2$-by-$p_3$. To the best of our knowledge, we are among the first to achieve this optimal rate. We also develop the corresponding recovery method for more general cases where measurements are taken with noise. The central idea is to transform the observable matricizations by singular value decomposition and perform the adaptive trimming scheme to denoise each block.

To illustrate the properties of the proposed procedure, both theoretical analyses and simulation studies are provided. We derive upper and lower bound results to show that the proposed recovery procedure can accommodate different levels of noise and achieve the optimal rate of convergence for a large class of low-rank tensors. Although the exact low-rank assumption is used in the theoretical analysis, some simulation settings show that such an assumption is not really necessary in practice, as long as the singular values of each matricization of the original tensor decays sufficiently.

It is worth emphasizing that because the proposed algorithms only involve basic matrix operations such as matrix multiplication and singular value decomposition, it is tuning-free in many general situations and can be implemented efficiently to handle large scale problems. In fact, our simulation study shows that the recovery of a 500-by-500-by-500 tensor can be done stably within, on average, 10 seconds.  

We also apply the proposed procedure to a 3-d MRI imaging dataset that comes from a study on Attention-deficit/hyperactivity disorder (ADHD). We show that with a limited number of Cross tensor measurements and the corresponding tensor completion algorithm, one can estimate the underlying low-rank structure of 3-d images as well as if one observes all entries of the image.

This work also relates to some previous results other than tensor completion in the literature.  \cite{mahoney2008tensor} considered the tensor CUR decomposition, which aims to represent the tensor as the product of a sub-tensor and two matrices. However, simply applying their work cannot lead to optimal results in tensor completion since treating tensors as matrix slices would lose useful structures of tensors. \cite{krishnamurthy2013low} proposed a sequential tensor completion algorithm under adaptive samplings. Their result requires $O(pr^{2.5}\log(r))$ number of entries for $p$-by-$p$-by-$p$ order-3 tensors under the more restrictive CP rank-$r$ condition, which is much larger than that of our method. \cite{rauhut2016low} considered a tensor recovery setting where each observation is a general linear projections of the original tensor. However, their theoretical analysis heavily relies on a conjecture that is difficult to check. \cite{oseledets2008tucker} provided an existence proof for rank-$r$ Tucker-like approximations for $p$-by-$p$-by-$p$ tensors with $O(r^3+pr)$ parameters. \cite{caiafa2010generalizing} introduced representations for $p_1$-by-$p_2$-by-$p_3$ Tucker rank-$(r,r,r)$ tensors based on $r^3 + r(p_1+p_2+p_3)$ selected entries. In \cite{caiafa2015stable}, they further introduced a multi-way projection scheme for stable, robust, and fast low-rank tensor reconstruction, which requires $O(pr^2)$ measurements and some tuning parameters, such as the rank of the tensor, for implementation. To the extent of our knowledge, we are among the first to develop the tensor completion scheme that is efficient, easy to implement, tuning-free, and allows exact tensor completion in the noiseless setting and achieves optimal estimation error in the noisy setting under the minimal sample size.

The rest of the paper is organized as follows. After an introduction to the notations and preliminaries in Section \ref{sec.notation}, we present the Cross tensor measurement scheme in Section \ref{sec.Cross_scheme}. Based on the proposed measurement scheme, the tensor completion algorithms for both noiseless and noisy case are introduced in Sections \ref{sec.noiseless} and \ref{sec.noisy} respectively. We further analyze the theoretical performance of the proposed algorithms in Section \ref{sec.theory}. The numerical performance of algorithms are investigated in a variety of simulation studies in Section \ref{sec.simu}. We then apply the proposed procedure to a real dataset of brain MRI imaging in Section \ref{sec.real_data}. In Section \ref{sec.discussion}, we briefly discuss the extensions of main results. The proofs of the main results are finally collected in the supplement materials. 

\section{Cross Tensor Measurements \& Completion: Methodology}\label{sec.procedure}

\subsection{Basic Notations and Preliminaries}\label{sec.notation}

We start with basic notations and results that will be used throughout the paper. The upper case letters, e.g., $X, Y, Z$, are generally used to represent matrices. For $X \in \mathbb{R}^{p_1\times p_2}$, the singular value decomposition can be written as $X = U\Sigma V^\top$. Suppose $\diag(\Sigma) = (\sigma_1(X), \ldots, \sigma_{\min\{p_1, p_2\}}(X))$, then $\sigma_1(X)\geq \sigma_2(X)\geq\ldots \geq \sigma_{\min\{p_1, p_2\}}(X) \geq 0$ are the singular values of $X$. Especially, we note $\sigma_{\min}(X) = \sigma_{\min\{p_1, p_2\}}(X)$ and $\sigma_{\max}(X) = \sigma_{1}(X)$ as the smallest and largest singular value of $X$. Additionally, the matrix spectral norm and Frobenius norm are denoted as $\|X\| = \max_{u\in \mathbb{R}^{p_2}}\frac{\|Xu\|_2}{\|u\|_2}$ and $\|X\|_F = \sqrt{\sum_{i=1}^{p_1}\sum_{j=1}^{p_2}X_{ij}^2} = \sqrt{\sum_{i=1}^{\min\{p_1, p_2\}}\sigma_i^2(X)}$, respectively. We denote $\mathbb{P}_X \in \mathbb{R}^{p_1\times p_1}$ as the projection operator onto the column space of $X$. Specifically, $\mathbb{P}_X = X(X^\top X)^\dagger X^\top = X X^\dagger$. Here $(\cdot)^\dagger$ is the Moore-Penrose pseudo-inverse. Let $\mathbb{O}_{p, r}$ be the set of all $p$-by-$r$ orthogonal columns, i.e., $\mathbb{O}_{p, r} = \{V\in \mathbb{R}^{p\times r}: V^\top V = I_r\}$, where $I_r$ represents the identity matrix of dimension $r$. 

We use bold upper case letters, e.g., $\X, \Y, \Z$ to denote tensors. If $\X \in \mathbb{R}^{p_1\times p_2\times p_3}$, $E_t \in \mathbb{R}^{m_t\times p_t}$, $ t =1,2,3$. The \emph{mode products} (tensor-matrix product) is defined as
$$\X \times_1 E_1 \in \mathbb{R}^{m_1 \times p_2\times p_3},\quad \left(\X \times_1 E_1\right)_{ijk} = \sum_{s=1}^{p_1}E_{1, is} \X_{sjk},$$
where $i \in [1:m_1], j \in [1:p_2], k\in [1:p_3]$. The mode-2 product $\X \times_2 E_2$ and mode-3 product $\X \times_3 E_3$ can be defined similarly. Interestingly, the products along different modes satisfy the commutative law, e.g., $\X \times_t E_t \times_s E_s = \X \times_s E_s \times_t E_t$ if $s \neq t$.
The \emph{matricization} (or unfolding, flattening in literature), $\mathcal{M}_t(\X)$, maps a tensor $\X\in \mathbb{R}^{p_1\times p_2\times p_3}$ into a matrix $\mathcal{M}_t(\X)\in \mathbb{R}^{p_t \times \prod_{s\neq t}p_s}$, so that for any $i \in \{1,\cdots, p_1\}, j\in \{1,\cdots,p_2\}, k\in \{1,\cdots, p_3\}$,
\begin{equation*}
\begin{split}
\X_{ijk} = (\mathcal{M}_1(\X))_{[i, (j+p_2(k-1))]} = (\mathcal{M}_2(\X))_{[j, (k+p_3(i-1))]} = (\mathcal{M}_3(\X))_{[k, (i+p_1(j-1))]}.
\end{split}
\end{equation*}
The tensor Hilbert Schmidt norm and tensor spectral norm, which are defined as
\begin{equation*}
\|\X\|_{\rm HS} = \sqrt{\sum_{i=1}^{p_1}\sum_{j=1}^{p_2}\sum_{k=1}^{p_3} \X_{ijk}^2},\quad \left\|\X\right\|_{\rm op} = \max_{u\in \mathbb{R}^{p_1}, v\in \mathbb{R}^{p_2}, w\in \mathbb{R}^{p_3}}\frac{\X \times_1 u\times_2 v \times_3 w}{\|u\|_2\|v\|_2 \|w\|_2},
\end{equation*}
will be intensively used in this paper. It is also noteworthy that the general calculation of the tensor operator norm is NP-hard \citep{hillar2013most}. Unlike matrices, there is no universal definition of rank for third or higher order tensors. Standing out from various definitions, the \emph{Tucker rank} \citep{tucker1966some} has been widely utilized in literature, and its definition is closely associated with the following \emph{Tucker decomposition}: for $\X \in \mathbb{R}^{p_1\times p_2\times p_3}$,
\begin{equation}\label{eq:Tucker-decomposition}
\X =  \S \times_1 U_1 \times_2 U_2 \times_3 U_3,\quad \text{or equivalently}\quad \X_{i j k} = \sum_{i'j'k'} s_{i'j'k'} U_{1, ii'} U_{2, jj'} U_{3, kk'}.
\end{equation}
Here $\S\in \mathbb{R}^{r_1\times r_2\times r_3}$ is referred to as the \emph{core tensor}, $U_k \in \mathbb{O}_{p_k, r_k}$. The minimum number of triplets $(r_1, r_2, r_3)$ are defined as the \emph{Tucker rank} of $\X$ which we denote as $\rank(\X) = (r_1, r_2, r_3)$. The Tucker rank can be calculated easily by the rank of each matricization: $r_t = \rank(\mathcal{M}_t(\X))$. It is also easy to prove that the triplet $(r_1, r_2, r_3)$ satisfies $r_t\leq p_t, \max^2\{r_1, r_2, r_3\} \leq r_1r_2r_3$. For a more detailed survey of tensor decomposition, readers are referred to \cite{kolda2009tensor}. 

We also use the following symbols to represent sub-arrays. For any subsets $\Omega_1, \Omega_2$, etc., we use $X_{[\Omega_1, \Omega_2]}$ to represent the sub-matrix of $X$ with row indices $\Omega_1$ and column indices $\Omega_2$. The sub-tensors are denoted similarly: $\X_{[\Omega_1, \Omega_2, \Omega_3]}$ represents the tensors with mode-$t$ indices in $\Omega_t$ for $t = 1, 2, 3$. For better presentation, we use bracket to represent index sets. Particularly for any integers $a\leq b$, let $[a: b] = \{a, \ldots, b\}$ and let ``:" alone represent the whole index set. Thus, $U_{[:, 1:r]}$ represents the first $r$ columns of $U$; $\X_{[\Omega_1, \Omega_2, :]}$ represents the sub-tensor of $\X$ with mode-1 indices $\Omega_1$, mode-2 indices $\Omega_2$ and all mode-3 indices.

Now we establish the lower bound for the minimum number of measurements for Tucker low-rank tensor completion based on counting the degrees of freedom.
\begin{Proposition}[Degrees of freedom for rank-$(r_1, r_2, r_3)$ tensors in $\mathbb{R}^{p_1\times p_2\times p_3}$]\label{pr:degree_of_freedom}
	Assume that $r_1\leq p_1, r_2\leq p_2, r_3\leq p_3$, $\max^2\{r_1, r_2, r_3\} \leq r_1r_2r_3$, then the degrees of freedom of all rank-$(r_1, r_2, r_3)$ tensors in $\mathbb{R}^{p_1\times p_2\times p_3}$ is 
	$$r_1r_2r_3 + (p_1-r_1)r_1 + (p_2-r_2)r_2 + (p_3-r_3)r_3.$$ 
\end{Proposition}
\begin{Remark}\label{rm:dof}\rm 
	Beyond order-3 tensors, we can show the degrees of freedom for rank-$(r_1, \ldots, r_d)$ order-$d$ tensors in $\mathbb{R}^{p_1\times \cdots \times p_d}$ is $\prod_{t=1}^d r_t + \sum_{t=1}^d r_t(p_t-r_t)$ similarly.
\end{Remark}
Proposition \ref{pr:degree_of_freedom} provides a lower bound and the benchmark for the number of measurements to guarantee low-rank tensor completion, i.e., $r_1r_2r_3 + \sum_{t=1}^3 r_t(p_t - r_t)$. Since the previous methods are not guaranteed to achieve this lower bound, we focus on developing the first measurement scheme that can both work efficiently and reach this benchmark. 

\subsection{Cross Tensor Measurements}\label{sec.Cross_scheme}

\begin{figure}
	\makebox[\linewidth][c]{%
		\begin{subfigure}[b]{.42\textwidth}
			\centering
			\includegraphics[height=2in,width=2in]{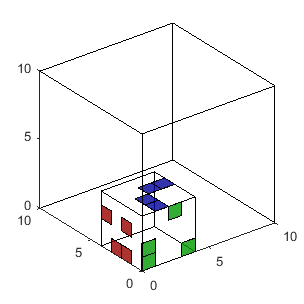}
			\caption{$\Omega_t, \Xi_t$ illustration}
		\end{subfigure}%
		\begin{subfigure}[b]{.42\textwidth}
			\centering
			\includegraphics[height=2in,width=2in]{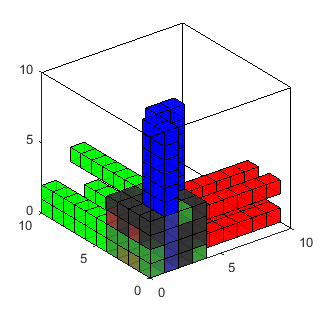}
			\caption{All measurements}
		\end{subfigure}%
		\begin{subfigure}[b]{.42\textwidth}
			\centering
			\includegraphics[height = 2in, width=2in]{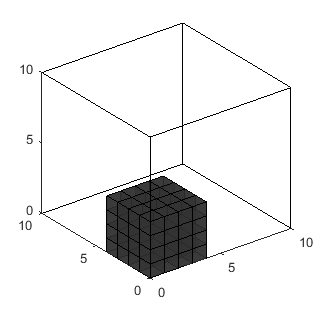}
			\caption{Body measurements $\Y_{[\Omega_1, \Omega_2 \Omega_3]}$}
		\end{subfigure}%
	}\\
	\makebox[\linewidth][c]{%
		\begin{subfigure}[b]{.42\textwidth}
			\centering
			\includegraphics[height = 2in, width=2in]{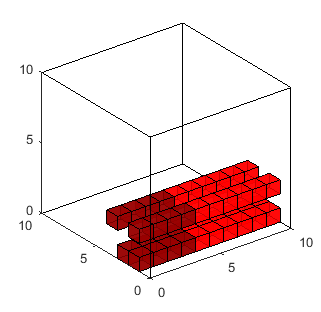}
			\caption{Arm measurements $\Y_{\Xi_1}$}
		\end{subfigure}%
		\begin{subfigure}[b]{.42\textwidth}
			\centering
			\includegraphics[height = 2in, width=2in]{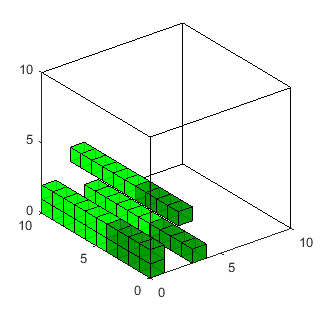}
			\caption{Arm measurements $\Y_{\Xi_2}$}
		\end{subfigure}%
		\begin{subfigure}[b]{.42\textwidth}
			\centering
			\includegraphics[height = 2in, width=2in]{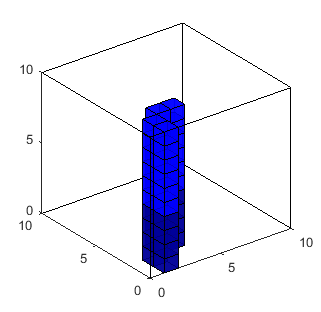}
			\caption{Arm measurements $\Y_{\Xi_3}$}
		\end{subfigure}%
	}\\	
	\makebox[\linewidth][c]{%
	\begin{subfigure}[b]{.42\textwidth}
		\centering
		\includegraphics[height = 2in, width=2in]{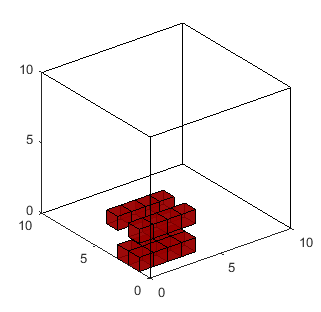}
		\caption{Joint measurements $\Y_{\Omega_1 \times \Xi_1}$}
	\end{subfigure}%
	\begin{subfigure}[b]{.42\textwidth}
		\centering
		\includegraphics[height = 2in, width=2in]{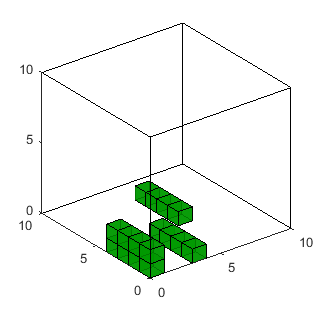}
		\caption{Joint measurements $\Y_{\Omega_2\times \Xi_2}$}
	\end{subfigure}%
	\begin{subfigure}[b]{.42\textwidth}
		\centering
		\includegraphics[height = 2in, width=2in]{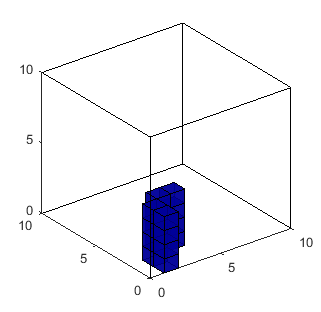}
		\caption{Joint measurements $\Y_{\Omega_3\times \Xi_3}$}
	\end{subfigure}%
}
	\caption{Illustrative example for Cross Tensor Measurements Scheme. For better illustration here we assume $\Omega_t = [1:m_t]$, $p_1 =p_2 = p_3 = 10, m_1 = m_2 = m_3 = g_1 = g_2 = g_3 = 4$.}\label{fig:measurements}\vspace{-.2in}
\end{figure}

In this section, we propose a novel Cross tensor measurement scheme. Suppose the targeting unknown tensor $\X$ is of $p_1$-by-$p_2$-by-$p_3$, we let
\begin{equation}\label{eq:index_Omega_Xi}
\begin{split}
& \Omega_1 \subseteq [1: p_1], \quad \Omega_2 \subseteq [1 : p_2], \quad \Omega_3 \subseteq [1:p_3],\quad |\Omega_t| = m_t,\quad t = 1, 2, 3;\\
& \Xi_1 \subseteq \Omega_2 \times \Omega _3, \quad \Xi_2 \subseteq \Omega_3 \times \Omega_1, \quad \Xi_3 \subseteq \Omega_1 \times \Omega_2, \quad |\Xi_t| = g_t,\quad t= 1, 2, 3.
\end{split}
\end{equation}
Then we measure the entries of $\X$ using the following indices set
\begin{equation}\label{eq:bOmega}
\bOmega = \left(\Omega_1\times \Omega_2\times \Omega_3\right)\cup \left([1:p_1]\times \Xi_1\right)\cup \left([1:p_2]\times \Xi_2\right) \cup \left([1:p_3]\times \Xi_3\right),
\end{equation}
where
\begin{equation}\label{eq:observations}
\begin{split}
\Omega_1\times \Omega_2\times \Omega_3 = \left\{(i, j, k): i\in \Omega_1, j\in \Omega_2, k\in \Omega_3\right\}\quad & \text{are \emph{body measurements;}} \\
\left.\begin{array}{ll}
~ [1:p_1]\times \Xi_1 =\left\{(i, j, k): i\in [1:p_1], (j, k)\in \Xi_1\right\}\\
~ [1:p_2]\times \Xi_2 =\left\{(i, j, k): j\in [1:p_2], (k, i)\in \Xi_2\right\}\\
~ [1:p_3]\times \Xi_3 =\left\{(i, j, k): k\in [1:p_3], (i, j)\in \Xi_3\right\}\\
\end{array}\right\} \quad & \text{are\emph{arm measurements}.}
\end{split}
\end{equation}
Meanwhile, the intersections among body and arm measurements, which we refer to as \emph{joint measurements}, also play important roles in our analysis:
\begin{equation}\label{eq:joint observations}
\begin{split}
&\Omega_1\times \Xi_1 = \left(\Omega_1\times \Omega_2 \times \Omega_3\right) \cap \left([1:p_1]\times \Xi_1\right) = \{(i, j, k): i\in \Omega_1, (j, k)\in \Xi_1\},\\
&\Omega_2\times \Xi_2 = \left(\Omega_1\times \Omega_2 \times \Omega_3\right) \cap \left([1:p_2]\times \Xi_2\right) = \{(i, j, k): j\in \Omega_2, (k, i)\in \Xi_2\},\\
&\Omega_3\times \Xi_3 = \left(\Omega_1\times \Omega_2 \times \Omega_3\right) \cap \left([1:p_3]\times \Xi_3\right) = \{(i, j, k): k\in \Omega_3, (i, j)\in \Xi_3\}.
\end{split}
\end{equation}
A pictorial illustration of the body, arm and joint measurements is provided in Figure \ref{fig:measurements}. Since the measurements are generally cross-shaped, we refer to $\bOmega$ as the \emph{Cross Tensor Measurement Scheme}. It is easy to see that the total number of measurements for the proposed scheme is $m_1m_2m_3 + g_1(p_1-m_1) + g_2(p_2-m_2) + g_3(p_3-m_3)$ and the sampling ratio is 
\begin{equation}\label{eq:sampling_ratio}
\frac{\# \text{Observable samples}}{\# \text{All parameters}} = \frac{m_1m_2m_3 + \sum_{t=1}^3 g_t(p_t-m_t)}{p_1p_2p_3}.
\end{equation}
Based on these measurements, we focus on the following model,
\begin{equation}\label{eq:Y_omega_observations}
\Y_{\bOmega} = \X_{\bOmega} + \Z_{\bOmega},\quad \text{i.e.}\quad Y_{ijk} = X_{ijk} + Z_{ijk},\quad (i, j, k)\in \bOmega,
\end{equation}
where $\X, \Y$ and $\Z$ correspond to the original tensor, observed values and unknown noise term, respectively.

\subsection{Recovery Algorithm -- Noiseless Case}\label{sec.noiseless}

When $\X$ is exactly low-rank and the observations are noiseless, i.e. $\Y_{ijk} = \X_{ijk}$, we can recover $\X$ with the following algorithm. We first construct the \emph{arm matricizations}, \emph{joint matricizations} and \emph{body matricizations} based on \eqref{eq:observations} and \eqref{eq:joint observations}, 
\begin{align}
& Y_{\Xi_t} = \mathcal{M}_t(\Y_{[1:p_t]\times\Xi_t}) \in \mathbb{R}^{p_t\times g_t}, \quad & \text{(Arm matricizations)} \label{eq:arm_measurements} \\
& Y_{t, \Omega} = \mathcal{M}_t(\Y_{[\Omega_1, \Omega_2, \Omega_3]}) \in \mathbb{R}^{m_t \times \prod_{s\neq t}m_s}, \quad  & \text{(Body matricizations)} \label{eq:body_matricization}\\
& Y_{\Omega_t\times \Xi_t} = \mathcal{M}_t(\Y_{\Omega_t\times \Xi_t})\in \mathbb{R}^{m_t\times g_t}.\quad & \text{(Joint matricizations)} \label{eq:joint_matricization}
\end{align}
In the noiseless setting, we propose the following formula to complete $\X$: 
\begin{equation}\label{eq:hat_X_noiseless}
\hat{\X} = \Y_{[\Omega_1, \Omega_2, \Omega_3]} \times_1 R_1 \times_2 R_2 \times_3 R_3,
\end{equation}
\begin{equation}\label{eq:R_noiseless}
\text{where}\quad R_t = Y_{\Xi_t}Y_{\Omega_t\times \Xi_t}^\dagger  \in \mathbb{R}^{p_t\times m_t}, \quad t = 1, 2, 3.
\end{equation}
The procedure is summarized in Algorithm \ref{al:noiseless}. The theoretical guarantee for this proposed algorithm is provided in Theorem \ref{th:noiseless}.
\begin{algorithm}
	\caption{Cross: Tensor Completion with Noiseless Observations}
	\begin{algorithmic}[1]
		\State Input: noiseless observations $\Y_{ijk}$, $(i, j ,k) \in \bOmega$ from \eqref{eq:bOmega}.
		\State 
		Construct $Y_{\Xi_1}, Y_{\Xi_2}, X_{\Xi_3}, Y_{\Omega_1\times\Xi_1}, Y_{\Omega_2\times\Xi_2}, Y_{\Omega_3\times\Xi_3}$ as \eqref{eq:arm_measurements}.
		\State Calculate
		\begin{equation*}
		R_1 = Y_{\Xi_1}Y_{\Omega_1\times \Xi_1}^\dagger  \in \mathbb{R}^{p_1\times m_1},~~ R_2 = Y_{\Xi_2}Y_{\Omega_2\times \Xi_2}^\dagger \in \mathbb{R}^{p_2\times m_2},~~ R_3 = Y_{\Xi_3}Y_{\Omega_3\times \Xi_3}^\dagger \in \mathbb{R}^{p_3\times m_3}. 
		\end{equation*}
		\State Calculate the final estimator
		$$\hat{\X} = \Y_{[\Omega_1, \Omega_2, \Omega_3]} \times_1 R_1 \times_2 R_2 \times_3 R_3. $$
	\end{algorithmic}\label{al:noiseless}
\end{algorithm}

\begin{Theorem}[Exact recovery in noiseless setting]\label{th:noiseless}
	Suppose $\X\in \mathbb{R}^{p_1\times p_2\times p_3}$, $\rank(\X) = (r_1, r_2, r_3)$. Assume all Cross tensor measurements are noiseless, i.e. $\Y_{\bOmega} = \X_{\bOmega}$. If $\rank(Y_{\Omega_t \times \Xi_t}) = r_t$ and $\min\{m_t, g_t\} \geq r_t$ for $t=1, 2, 3$ (so that $|\bOmega| \geq r_1r_2r_3 + r_1(p_1-r_1) + r_2(p_2 - r_2) + r_3(p_3-r_3)$), then
	\begin{equation}\label{eq:tensor-Schur-complement}
	\X =  \Y_{[\Omega_1, \Omega_2, \Omega_3]} \times_1 R_1 \times_2 R_2 \times_3 R_3,\quad R_t = Y_{\Xi_t}Y_{\Omega_t\times \Xi_t}^\dagger, \quad t = 1, 2, 3.
	\end{equation} 
	Moreover, if there are $\tilde{M}_t\in \mathbb{R}^{m_t\times r_t}$, $\tilde{N}_t \in \mathbb{R}^{g_t\times r_t}$ such that $\tilde{M}_t^\top X_{\Omega_t\times \Xi_t} \tilde{N}_t \in \mathbb{R}^{r_t\times r_t}$ is non-singular for $t = 1, 2, 3$, then we further have
	\begin{equation*}
	\X = \Y_{[\Omega_1, \Omega_2, \Omega_3]} \times_1 \tilde{R}_1 \times_2 \tilde{R}_2 \times_3 \tilde{R}_3,\quad
	\tilde{R}_t = Y_{\Xi_t}\tilde{N}_t^\top\left(\tilde{M}^\top Y_{\Omega_t\times \Xi_t} \tilde{N}\right)^{-1} \tilde{M}_t^\top.
	\end{equation*}
\end{Theorem}
Theorem \ref{th:noiseless} shows that, in the noiseless setting, as long as $\min\{m_t, g_t\} \geq r_t$, both $\mathcal{M}_t(\Y)$ and its $m_t$-by-$g_t$ submatrix $Y_{\Omega_t\times \Xi_t}$ are of rank $r_t$, exact recovery by Algorithm \ref{al:noiseless} can be guaranteed. Therefore, the minimum required number of measurements for the proposed Cross tensor measurement scheme is $r_1r_2r_3 + r_1(p_1 -r_1) + r_2(p_2 - r_2) + r_3(p_3-r_3)$ when we set $m_t = g_t = r_t$, which exactly matches the lower bound established in Proposition \ref{pr:degree_of_freedom} and outperforms the previous methods in the literature.

On the other hand, Algorithm \ref{al:noiseless} heavily relies on the noiseless assumption. In fact, calculating $Y_{\Omega_t\times \Xi_t}^\dagger = (X_{\Omega_t \times \Xi_t} + Z_{\Omega_t \times \Xi_t})^\dagger$ is unstable even with low levels of noise, which ruins the performance of Algorithm \ref{al:noiseless}. Since we rarely have noiseless observations in practice, we focus on the setting with non-zero noise for the rest of the paper.

\subsection{Recovery Algorithm -- Noisy Case}\label{sec.noisy}

\begin{figure}
	\makebox[\linewidth][c]{%
		\begin{subfigure}[b]{1.2\textwidth}
			\centering
			\includegraphics[height=3.4in]{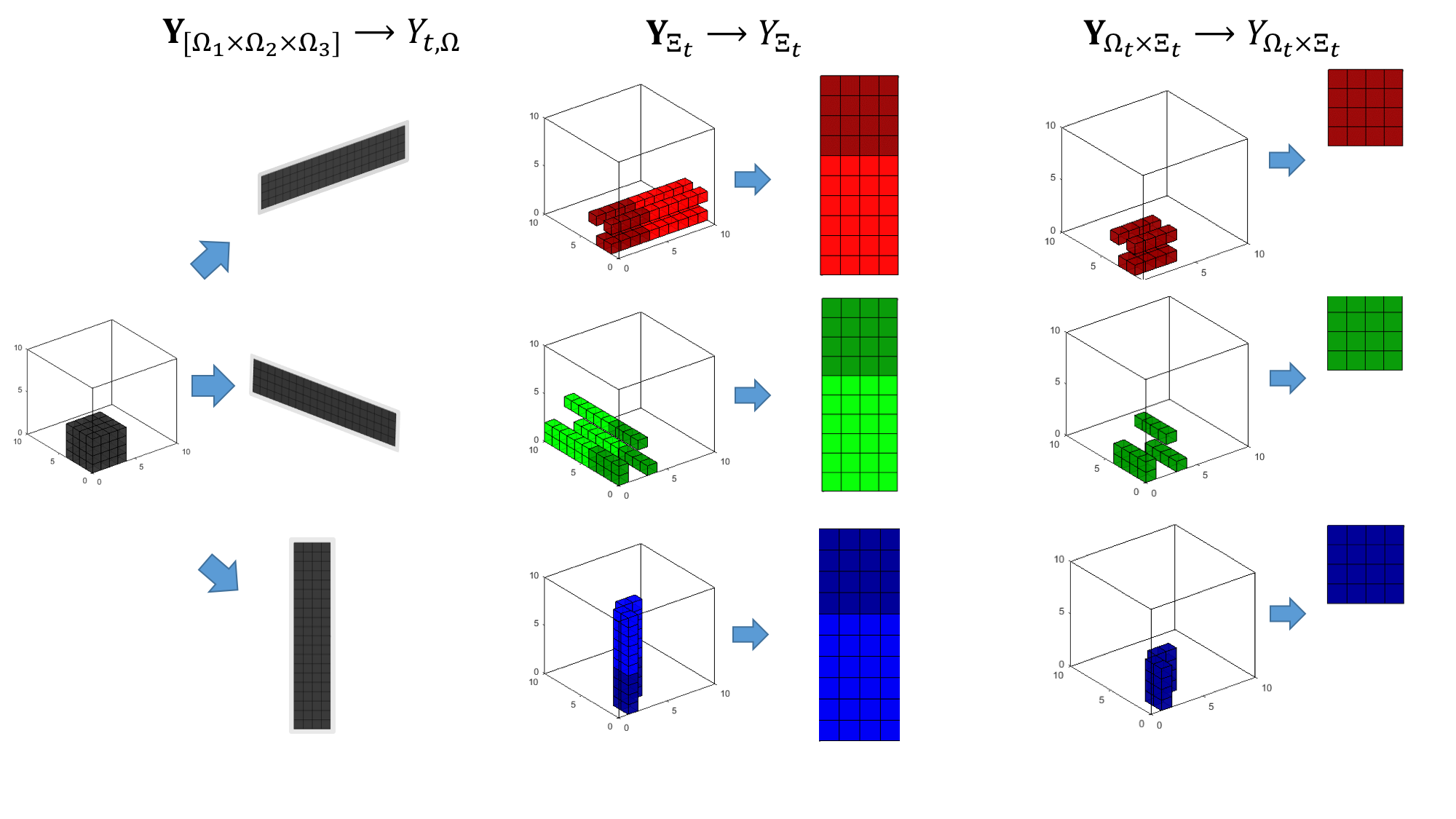}
			\caption{Step 1. All matricizations: $Y_{t,\Omega}, Y_{\Xi_t}$ and $Y_{\Omega_t\times \Xi_t}$}
		\end{subfigure}%
	}\\
	\makebox[\linewidth][c]{%
		\begin{subfigure}[b]{.6\textwidth}
			\centering
			\includegraphics[width = 2.7in,height = 1.75in]{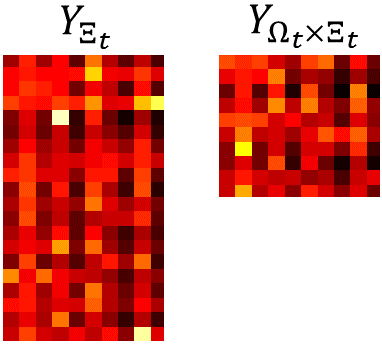}
			\caption{Heatmap illustration of $Y_{\Xi_t}, Y_{\Omega_t\times \Xi_t}$ (darker blocks mean larger absolute values)}
		\end{subfigure}\quad
		\begin{subfigure}[b]{.6\textwidth}
			\centering
			\includegraphics[width = 2.7in,height = 1.75in]{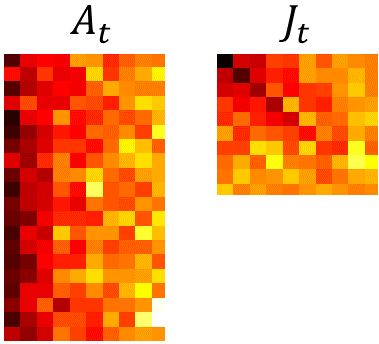}
			\caption{Step 2. We obtain $A_t, J_t$ after rotation\\~}
		\end{subfigure}
	}\\
	\makebox[\linewidth][c]{%
		\begin{subfigure}[b]{.6\textwidth}
			\centering
			\includegraphics[width=2.7in,height=1.75in]{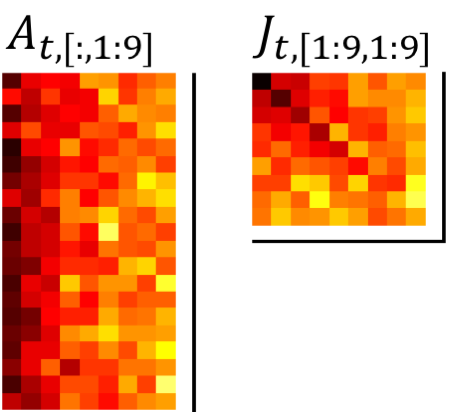}
			\caption{Step 3. Intermediate process of trimming}
		\end{subfigure}\quad
		\begin{subfigure}[b]{.6\textwidth}
			\centering
			\includegraphics[width=2.7in,height=1.75in]{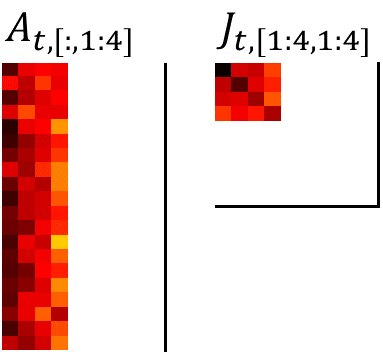}
			\caption{Step 3. Eventually located at $\hat{r}_t = 4$}
		\end{subfigure}
	}
	\caption{Illustration of the proposed procedure in noisy setting}\label{fig:illustration}
\end{figure}

In this section we propose the following procedure for recovery in the noisy setting. The proposed algorithm is divided into four steps and an illustrative example is provided in Figure \ref{fig:illustration} for readers' better understanding.
\begin{itemize}[leftmargin=*]
	\item {\bf (Step 1: Construction of Matricizations)} Same as Algorithm \ref{al:noiseless}, Construct the \emph{arm, body and joint matricizations} as \eqref{eq:arm_measurements} and \eqref{eq:body_matricization} (see Figure \ref{fig:illustration}(a)),
	\begin{equation*}
	\begin{split}
	\text{Arm:} ~ Y_{\Xi_1}, Y_{\Xi_2}, Y_{\Xi_3}; ~ \text{\rm Body:} Y_{1, \Omega}, Y_{2, \Omega}, Y_{3, \Omega};~ \text{Joint:} ~ Y_{\Omega_1\times \Xi_1}, Y_{\Omega_2\times \Xi_2}, Y_{\Omega_3\times \Xi_3}.
	\end{split}
	\end{equation*}	
	\item {\bf (Step 2: Rotation)} For $t=1,2,3$, we calculate the singular value decompositions of $Y_{\Xi_t}$ and $Y_{t, \Omega}$, then store
	\begin{equation}\label{eq:V^A-U^B}
	\begin{split}
	V_t^{(A)} \in \mathbb{O}_{g_t}  & \quad \text{as the right singular vectors of $Y_{\Xi_t}$},\\
	U_t^{(B)} \in \mathbb{O}_{m_t} & \quad \text{as the left singular vectors of $Y_{t, \Omega}$}. 
	\end{split}
	\end{equation}
	Here the superscripts ``(A), (B)" represent arm and body, respectively. We calculate the following rotation for arm and joint matricizations based on SVDs (See Figure \ref{fig:illustration}(b), (c)).
	\begin{equation}\label{eq:A_t-J_t}
	\begin{split}
	A_t = & Y_{\Xi_t} \cdot V_t^{(A)} \in \mathbb{R}^{p_t\times g_t}, \\
	J_t = & (U_t^{(B)})^\top \cdot Y_{\Omega_t \times \Xi_t}\cdot V_t^{(A)} \in \mathbb{R}^{m_t\times g_t}.
	\end{split}
	\end{equation}
	As we can see from Figure \ref{fig:illustration}(c), the magnitude of $A_t$'s columns and $J_t$'s both columns and rows decreases front to back. Therefore, the important factors of $Y_{\Xi_t}$ and $Y_{\Omega_t\times \Xi_t}$ are moved to front rows and columns in this step.
	\item {\bf (Step 3: Adaptive Trimming)} Since $A_t$ and $J_t$ are contaminated with noise, in this step we denoise them 
	by trimming the lower ranking columns of $A_t$ and both lower ranking columns and rows of $J_t$. To decide the number of rows and columns to trim, it will be good to have an estimate for $(r_1, r_2, r_3)$, say $(\hat{r}_1, \hat{r}_2, \hat{r}_3)$. We will show later in theoretical analysis that a good choice of $\hat{r}_t$ should satisfy
	\begin{equation}\label{ineq:condition_check}
	(J_t)_{[1:\hat{r}_t, 1:\hat{r}_t]} \text{ is non-singular} \quad \text{and}  \quad \left\|(A_t)_{[:, 1:\hat{r}_t]}  (J_t)^{-1}_{[1:\hat{r}_t, 1:\hat{r}_t]}\right\|\leq \lambda_t
	\end{equation}
	for $t = 1, 2, 3$. $\lambda_t = c_t\sqrt{p_t/m_t}$ is the tuning parameter here, and the discussion of selection method is provided a little while later. Our final estimator for $r_t$ is the largest $\hat{r}_t$ that satisfies Condition \eqref{ineq:condition_check}, and can be found by verifying \eqref{ineq:condition_check} for all possible $r_t$'s. It is worth mentioning that this step shares similar ideas with structured matrix completion in \cite{cai2016structured}. (See Figure \ref{fig:illustration}(d) and (e)). 
	
	\item {\bf (Step 4: Assembling)} Finally, given $\hat{r}_1, \hat{r}_2, \hat{r}_3$ obtained from Step 3, we calculate
	\begin{equation}\label{eq:hat_R_t}
	\bar{R}_t = (A_t)_{[:, 1:\hat{r}_t]}  (J_t)^{-1}_{[1:\hat{r}_t, 1:\hat{r}_t]} (U_{t, [1:\hat{r}_t, :]}^{(B)})^\top \in  \mathbb{R}^{p_t \times m_t}, \quad t = 1, 2, 3, 
	\end{equation}
	and recover the original low-rank tensor $\X$ by
	\begin{equation}\label{eq:hat_X}
	\hat{\X} =  \Y_{[\Omega_1, \Omega_2, \Omega_3]} \times_1 \bar{R}_1 \times_2 \bar{R}_2 \times_3 \bar{R}_3.
	\end{equation}
\end{itemize}
The procedure is summarized as Algorithm \ref{al:noisy}. It is worth mentioning that both Algorithms \ref{al:noiseless} and \ref{al:noisy} can be easily extended to fourth and higher order tensors.

\begin{algorithm}
	\caption{Noisy Tensor Completion with Cross Measurements}
	\begin{algorithmic}[1]
		\State Input: entries $\Y_{ijk}$, $(i, j ,k) \in \bOmega$ from \eqref{eq:bOmega}, $\lambda_1, \lambda_2, \lambda_3$.
		\State Construct arm, body and joint matricizations as \eqref{eq:arm_measurements} and \eqref{eq:body_matricization},
		$$Y_{\Xi_t}\in \mathbb{R}^{p_t\times g_t}, \quad Y_{\Omega_t\times \Xi_t} \in\mathbb{R}^{m_t\times g_t}, \quad Y_{t, \Omega} \in \mathbb{R}^{m_t \times \left(\prod_{s\neq t}m_s\right)}, \quad t = 1, 2, 3.$$ 
		\State Calculate $U^{(B)}_t$ and $V^{(A)}_t$ via SVDs: 
		\begin{equation*}
		\begin{split}
		U^{(B)}_t \in \mathbb{O}_{m_t},& \text{ as the left singular vectors of } X_{\Xi_t};\\
		V^{(A)}_t \in \mathbb{O}_{g_t},& \text{ as the right singular vectors of } X_{t, \Omega}.
		\end{split}
		\end{equation*}
		\State Rotate the arm and joint measurements as
		$$A_t = X_{\Xi_t}\cdot V^{(A)}\in \mathbb{R}^{p_t\times g_t},\quad J_t = \left(U^{(B)}\right)^\top\cdot X_{\Omega_t\times \Xi_t}\cdot V^{(A)} \in \mathbb{R}^{m_t \times g_t}. $$
		\For {t = 1, 2, 3}
		\For {$s = \min\{g_t, m_t\}: -1: 1$}
		\State {\bf if} $J_{t, [1:s, 1:s]}$ is not singular and $\|A_{t, [:, 1:s]} J_{t, [1:s, 1:s]}^{-1}\| \leq \lambda_t$ {\bf then}
		\State \quad\quad $\hat r_t = s$; {\bf break} from the loop;
		\State {\bf end if}
		\EndFor		
		\State{\bf If} $\hat{r}_t$ is still unassigned {\bf then} $\hat{r}_t = 0$.
		\EndFor
		\State Calculate
		\begin{equation*}
		\bar{R}_t = A_{t, [:, 1:s]} J_{t, [1:s, 1:s]}^{-1} (V_{t, [1:\hat{r}_t, :]}^{(A)})^\top \in \mathbb{R}^{p_t\times m_t}, \quad t = 1, 2, 3.
		\end{equation*}
		\State Compute the final estimator
		$$\hat{\X} =\Y_{[\Omega_1, \Omega_2, \Omega_3]} \times_1 \bar{R}_1 \times_2 \bar{R}_2 \times_3 \bar{R}_3. $$
	\end{algorithmic}\label{al:noisy}
\end{algorithm}

\ \par

\noindent \emph{Selection of tuning parameter:} The tuning parameter $\lambda_t$ is a key factor to the performance of final estimation. Intuitively speaking, a larger value of $\lambda_t$ yields a higher trimming level and a lower-rank estimation. As we will illustrate in the theoretical and numerical analyses, one can simply choose $\lambda_t = 3\sqrt{p_t/m_t}$ in a variety of situations. When more computing power is available, a practical data-driven approach for selecting $c_t$ via $K$-fold subsampling cross-validation can be applied instead. The procedure is described as below, and the detailed numerical analyses for tuning parameter selection is provided in Section \ref{sec.simu}.
	
Suppose the body and arms of $\X$ are observed as \eqref{eq:index_Omega_Xi}--\eqref{eq:Y_omega_observations} and let $T \subseteq [1.5, 4]$ be a grid of candidate values of $c_t$. For $l=1,\ldots, L$, we randomly select subset $\Omega_t^{(train, l)}\subseteq \Omega_t$ with cardinality $|\Omega_t^{(train, l)}|\approx |\Omega_t|\cdot (K-1)/K$. Recall $\Xi_t \subseteq \Omega_{t+1}\times \Omega_{t+2}$, we further denote $\Xi_t^{(train, l)} = \Xi_t \cap (\Omega_{t+1}^{(train, l)}\times \Omega_{t+2}^{(train, l)})$. Apply our proposed procedure based on the Cross measurements 
\begin{equation*}
\begin{split}
\bOmega^{(train, l)} = & \Omega_1^{(train, l)}\times \Omega_2^{(train, l)} \times \Omega_3^{(train, l)}\\
& \cup \big([1:p_1]\times \Xi_1^{(train, l)}\big) \cup \big([1:p_2]\times \Xi_2^{(train, l)}\big) \cup \big([1:p_3]\times \Xi_3^{(train, l)}\big)
\end{split}
\end{equation*}
with each $c_t \in T$, then denote the resulting estimation as $\hat{\X}^l(c_t)$ for $l=1,\ldots, N$. Next, the prediction error is evaluated on the observations outside the training set,
$$\hat{R}(c_t) = \sum_{l=1}^L \sum_{(i, j, k)\in \bOmega \backslash \bOmega^{(train, l)}}\left|(\hat{\X}^l(c_t))_{ijk} - \Y_{ijk}\right|^2,\quad c_t \in T,$$
where $\bOmega$ is defined \eqref{eq:bOmega}. Finally we select $c_t^\ast = \argmin_{c_t\in T}\hat{R}(c_t)$, and apply the proposed procedure again with tuning parameter $\lambda_t = c_t^\ast \sqrt{p_t/m_t}$.

\section{Theoretical Analysis}\label{sec.theory}

In this section, we investigate the theoretical performance for the proposed procedure in the last section. Recall that our goal is to recover $\X$ from $\Y_{\bOmega}$ based on \eqref{eq:bOmega}.
Similarly, one can further define the arm, joint and body matricizations for $\X$, $\Z$, i.e. $X_{\Xi_t}$, $X_{\Omega_t\times \Xi_t}$, $X_{t, \Omega}$, $Z_{\Xi_t}$, $Z_{\Omega_t\times \Xi_t}$ and $Z_{t, \Omega}$ for $t = 1, 2, 3$ in the same fashion as $Y_{\Xi_t}$, $Y_{\Omega_t\times \Xi_t}$ and $Y_{t, \Omega}$ in \eqref{eq:arm_measurements}, \eqref{eq:body_matricization} and \eqref{eq:joint_matricization}.
We first present the following theoretical guarantees for low-rank tensor completion based on noisy observations via Algorithm \ref{al:noisy}.
\begin{Theorem}\label{th:noisy}
	Suppose $\X\in \mathbb{R}^{p_1\times p_2\times p_3}$, rank$(\X)=(r_1, r_2, r_3)$. Assume we observe $\Y_{\bOmega}$ based on Cross tensor measurement scheme \eqref{eq:observations}, where $\X_{\Omega}$ satisfies $\rank(X_{\Omega_t\times \Xi_t}) = r_t$ and 
	\begin{equation}\label{ineq:assumption_gap}
	\sigma_{r_t}\left(X_{\Omega_t\times \Xi_t}\right) > 5\left\|Z_{\Omega_t\times \Xi_t}\right\|,\quad \sigma_{r_t}\left(X_{\Xi_t}\right) > 5\left\|Z_{\Xi_t}\right\|, \quad \sigma_{r_t}\left(X_{t, \Omega}\right) > 5\left\|Z_{t, \Omega}\right\|.
	\end{equation}
	We further define
	\begin{equation}\label{eq:assumption_body_ratio}
	\xi_t = \left\|X^\dagger_{\Omega_t\times \Xi_t} X_{t, \Omega} \right\|, \quad t =1, 2, 3.
	\end{equation}
	Applying Algorithm \ref{al:noisy} with $\lambda_1, \lambda_2, \lambda_3$ satisfying
	\begin{equation}\label{ineq:assumption_arm_body}
	\lambda_t \geq 2\left\|X_{t, \Xi_t} X_{\Omega_t\times \Xi_t}^\dagger \right\|,\quad t = 1, 2, 3,
	\end{equation}
	we have	the following upper bound results for some uniform constant $C$,
	\begin{equation*}
	\begin{split}
	\left\|\hat{\X} - \X \right\|_{\rm HS} \leq & C\lambda_1\lambda_2\lambda_3\|\Z_{[\Omega_1, \Omega_2, \Omega_3]}\|_{\rm HS}\\
	& +  C\lambda_1\lambda_2\lambda_3\left(\sum_{t=1}^3\xi_t\|Z_{\Omega_t \times \Xi_t}\|_F + C\sum_{t=1}^{3}\frac{\xi_t}{\lambda_t}\|Z_{\Xi_t}\|_F\right),
	\end{split}
	\end{equation*}
	\begin{equation*}
	\begin{split}
	\left\|\hat{\X} - \X\right\|_{\rm op} \leq & C\lambda_1\lambda_2\lambda_3\|\Z_{[\Omega_1, \Omega_2, \Omega_3]}\|_{\rm op}\\
	&  +  C\lambda_1\lambda_2\lambda_3\left(\sum_{t=1}^3\xi_t\|Z_{\Omega_t \times \Xi_t}\| + C\sum_{t=1}^{3}\frac{\xi_t}{\lambda_t}\|Z_{\Xi_t}\|\right).
	\end{split}
	\end{equation*}
\end{Theorem}

It is helpful to explain the meanings of the conditions used in Theorem \ref{th:noisy}. The singular value gap condition \eqref{ineq:assumption_gap} is assumed in order to guarantee that signal dominates the noise in the observed blocks. $\lambda_t$ and $\xi_t$ are important factors in our analysis which represent ``arm-joint" and ``joint-body" ratio respectively. These factors roughly indicate how much information is contained in the body and arm measurements and how much impact the noisy terms have on the upper bound, all of which implicitly indicate the difficulty of the problem. Based on $\lambda_t, \xi_t$, we consider
the following classes of low-rank tensors,
the perturbation $\Z$, and indices of observations,
\begin{equation}\label{eq:F_{lambda,Xi}}
\begin{split}
\mathcal{F} = & \mathcal{F}_{\{\lambda_t\}, \{\xi_t\}}\\
= & \left\{(\X, \Z, \Omega_t, \Xi_t): \begin{array}{l}  
\X \in \mathbb{R}^{p_1\times p_2 \times p_3}, \rank(\X) \leq (r_1, r_2, r_3); \\
\left\|X_{\Xi_t} X_{\Omega_t\times \Xi_t}^\dagger\right\| \leq \lambda_t, \left\|X_{\Omega_t\times \Xi_t}^\dagger X_{t, \Omega}\right\| \leq \xi_t.\\
\sigma_{r_t}(X_{\Omega_t\times\Xi_t}) \geq 5 \|Z_{\Omega_t \times \Xi_t}\|, \sigma_{r_t}(X_{\Xi_t}) \geq 5 \|Z_{\Xi_t}\|,\\
\sigma_{r_t}(X_{t, \Omega}) \geq 5\|Z_{t, \Omega}\|; \\
\end{array} \right\}
\end{split}
\end{equation}
and provide the following lower bound result over $\mathcal{F}_{\{\lambda_t\}, \{\xi_t\}}$.
\begin{Theorem}[Lower Bound]\label{th:lower_bound}
	Suppose positive integers $r_t, p_t$ satisfy 
	$4\leq r_t \leq p_t/2$. The arm, body and joint measurement errors are bounded as
	\begin{equation}\label{ineq:Z_upper_bound}
	\|\Z_{[\Omega_1, \Omega_2, \Omega_3]}\|_{\rm HS} \leq C^{(B)},\quad  \|Z_{t, \Xi_t}\|_F \leq C^{(A)}_{t}, \quad \|Z_{t, \Omega_t\times \Xi_t}\|_F \leq C^{(J)}_{t}.
	\end{equation}
	If $C_t^{(J)} \leq \min\{C_t^{(A)}, C^{(B)}\}$, $\xi_t \geq 3, \lambda_t > 1$, then there exists uniform constant $c>0$ such that
	\begin{equation}\label{ineq:lower_bound}
	\inf_{\hat{\X}}\sup_{\substack{(\X, \Z, \Omega_t, \Xi_t) \in \mathcal{F}\\\Z \text{ satisfies \eqref{ineq:Z_upper_bound}}}}\left\|\hat{\X} - \X \right\|_{\rm HS} \geq c\lambda_1\lambda_2\lambda_3C^{(B)} +  c\lambda_1\lambda_2\lambda_3\sum_{t=1}^3 \left(\xi_tC^{(J)}_t + \frac{\xi_t}{\lambda_t}C_t^{(A)}\right).
	\end{equation}
	Similarly, suppose $ C^{(B)}, C_t^{(A)}, C_t^{(J)}$ are the upper bound for arm, body and joint measurement errors in tensor and matrix operator norms respectively, i.e.
	\begin{equation}\label{ineq:Z_upper_bound_op}
	\|\Z_{[\Omega_1, \Omega_2, \Omega_3]}\|_{\rm op} \leq C^{(B)},\quad  \|Z_{t, \Xi_t}\| \leq C^{(A)}_{t}, \quad \|Z_{t, \Omega_t\times \Xi_t}\| \leq C^{(J)}_{t}.
	\end{equation}
	Suppose $C_t^{(J)} \leq \min\{C_t^{(A)}, C^{(B)}\}$, $\xi_t \geq 3, \lambda_t >1$, then
	\begin{equation}
	\inf_{\hat{\X}}\sup_{\substack{(\X, \Z, \Omega_t, \Xi_t) \in \mathcal{F}\\\Z \text{ satisfies \eqref{ineq:Z_upper_bound_op}}}}\left\|\hat{\X} - \X \right\|_{\rm op} \geq c\lambda_1\lambda_2\lambda_3C^{(B)} +  c\lambda_1\lambda_2\lambda_3\sum_{t=1}^3 \left(\xi_tC^{(J)}_t + \frac{\xi_t}{\lambda_t}C_t^{(A)}\right).
	\end{equation}
\end{Theorem}
\begin{Remark}
	Theorems \ref{th:noisy} and \ref{th:lower_bound} together yield the optimal rate of recovery in $\mathcal{F}$ in both Hilbert-Schmidt and operator norms: 
	\begin{equation*}
	\inf_{\hat{\X}}\sup_{\substack{(\X, \Z, \Omega_t, \Xi_t) \in \mathcal{F}\\\Z \text{ satisfies \eqref{ineq:Z_upper_bound}}}}\left\|\hat{\X} - \X \right\|_{\rm HS} \asymp \lambda_1\lambda_2\lambda_3\left\{C^{(B)} +  \sum_{t=1}^3 \left(\xi_tC^{(J)}_t + \frac{\xi_t}{\lambda_t}C_t^{(A)}\right)\right\}.
	\end{equation*}	
	\begin{equation*}
	\inf_{\hat{\X}}\sup_{\substack{(\X, \Z, \Omega_t, \Xi_t) \in \mathcal{F}\\\Z \text{ satisfies \eqref{ineq:Z_upper_bound_op}}}}\left\|\hat{\X} - \X \right\|_{\rm op} \asymp \lambda_1\lambda_2\lambda_3\left\{C^{(B)} +  \sum_{t=1}^3 \left(\xi_tC^{(J)}_t + \frac{\xi_t}{\lambda_t}C_t^{(A)}\right)\right\}.
	\end{equation*}	
\end{Remark}

\begin{Remark}\label{rm:lower-bound}\rm 
	There have been a number of existing lower bound results on the estimation error in matrix/tensor estimation literatures. For example, \cite{negahban2012restricted} considered the setting that one observes uniformly randomly selected entries with noise; \cite{candes2011tight} developed a sharp oracle lower bound when the measurement matrices satisfies restrict isometry property (RIP), \cite{rohde2011estimation,koltchinskii2011nuclear} considered the setting that the measurement matrices/tensors are i.i.d. randomly generated. \cite{raskutti2015convex} considered multi-response regularized tensor regression, auto-regressive regression and interaction model with random Gaussian measurements. As the proposed Cross tensor measurement scheme does not satisfies the assumptions of these existing settings, these previous results cannot be directly applied.
\end{Remark}

As we can see from the theoretical analyses, the choice of $\lambda_t$ is crucial towards the recovery performance of Algorithm \ref{al:noisy}. Theorem \ref{th:noisy} provides a guideline for such a choice depending on the unknown parameter $\left\|X_{\Xi_t}X_{\Omega_t\times \Xi_t}^\dagger\right\|$, which is hard to obtain in practice. However, we can choose $\lambda_t = 3\sqrt{p_t/m_t}$ in a variety of settings. Specifically in the analysis below, we show under random sampling scheme that $\Omega_1, \Omega_2, \Omega_3, \Xi_1, \Xi_2, \Xi_3$ are uniformly randomly selected from $[1:p_1], [1:p_2], [1:p_3]$, $\Omega_2\times \Omega_3$, $\Omega_3\times \Omega_1$, $\Omega_1\times \Omega_2$, Algorithm \ref{al:noisy} with $\lambda_t  = 3\sqrt{p_t/m_t}$ will have guaranteed performance. The choice of $\lambda_t = 3\sqrt{p_t/m_t}$ and the one by cross-validation will be further examined in simulation studies later.

\begin{Theorem}\label{th:random}  
	Suppose $\X$ is with Tucker decomposition $\X = \S\times_1 U_1 \times_2 U_2\times_3 U_3$, where $\S\in \mathbb{R}^{r_1\times r_2\times r_3}, U_1\in \mathbb{O}_{p_1, r_1}, U_2\in \mathbb{O}_{p_2, r_2}, U_3\in \mathbb{O}_{p_3, r_3}$ and $U_1, U_2, U_3, \mathcal{M}_1(\S \times_2 U_2\times_3 U_3), \mathcal{M}_2(\S \times_1 U_1\times_3 U_3),\mathcal{M}_3(\S \times_1 U_1\times_2 U_2)$ all satisfy the matrix incoherence conditions:
	\begin{equation}\label{ineq:incoherence_condition}
	\begin{split}
	& \frac{p_t}{r_t}\max_{1\leq j \leq p_t}\left\|\mathbb{P}_{U_t} e_j^{(p_t)} \right\|_2^2 \leq \rho, \\
	& \frac{\prod_{s\neq t}p_s}{r_t}\max_{1\leq j \leq \prod_{s\neq t}p_s} \left\|\mathbb{P}_{\mathcal{M}_t(\S\times_{(t+1)} U_{t+1} \times_{(t+2)} U_{t+2})^\top} \cdot e_j^{(\prod_{s\neq t} p_{s})}\right\|_2^2 \leq \rho,
	\end{split}
	\end{equation}
	where $e_j^{(p)}$ is the $j$-th canonical basis in $\mathbb{R}^{p}$. Suppose we are given random Cross tensor measurements that $\Omega_t$ and $\Xi_t$ are uniformly randomly chosen $m_t$ and $g_t$ values from $\{1, \ldots, p_t\}$ and $\prod_{s\neq t}\Omega_{s}$, respectively. If for $t = 1, 2, 3$,
	\begin{equation}\label{ineq:th_min_S_condition}
	\begin{split}
	& \sigma_{\min}\left(\mathcal{M}_t(\S)\right)\\
	\geq & \max\left\{10\sqrt{\frac{p_1p_2p_3}{m_tg_t}}\|Z_{\Omega_t\times \Xi_t}\|, 10\sqrt{\frac{p_1p_2p_3}{m_1m_2m_3}}\|Z_{t, \Omega}\|, 19\sqrt{\frac{p_1p_2p_3}{p_tg_t}} \|Z_{\Xi_t}\|\right\},
	\end{split}	
	\end{equation} 
	Algorithm \ref{al:noisy} with $\lambda_t = 3\sqrt{p_t/m_t}$ yields
	\begin{equation*}
	\begin{split}
	\left\|\hat{\X} - \X\right\|_{\rm HS} \leq & C\sqrt{\frac{p_1p_2p_3}{m_1m_2m_3}}\| \Z_{[\Omega_1, \Omega_2, \Omega_3]}\|_{\rm HS}\\
	& + C\sqrt{p_1p_2p_3} \sum_{t=1}^3 \left(\frac{\|Z_{\Omega_t\times \Xi_t}\|_F}{\sqrt{g_tm_t}} + \frac{\|Z_{\Xi_t}\|_F}{\sqrt{g_t p_t}} \right),
	\end{split}
	\end{equation*}
	\begin{equation*}
	\begin{split}
	\left\|\hat{\X} - \X \right\|_{\rm op} \leq & C\sqrt{\frac{p_1p_2p_3}{m_1m_2m_3}} \left\|\Z_{[\Omega_1, \Omega_2, \Omega_3]}\right\|_{\rm op}\\
	& + C\sqrt{p_1p_2p_3} \sum_{t=1}^3 \left(\frac{\|Z_{\Omega_t\times \Xi_t}\|}{\sqrt{g_tm_t}} + \frac{\|Z_{\Xi_t}\|}{\sqrt{g_t p_t}} \right),
	\end{split}
	\end{equation*}
	with probability at least $1 - 2\sum_{t=1}^3 r_t\left\{\exp(-m_t/(16r_t\rho )) + \exp(-g_t/(64 r_t\rho))\right\}$.
\end{Theorem}

\begin{Remark}\rm 
	The incoherence conditions \eqref{ineq:incoherence_condition} are widely used in matrix and tensor completion literature (see, e.g., \cite{candes2010power,recht2011simpler,yuan2016incoherent}). Their conditions basically characterize every entry of $X$ as containing a similar level of information for the whole tensor. Therefore, we should have enough knowledge to recover the original tensor based on the observable entries.
\end{Remark}

\begin{Remark}\rm 
	For a better illustration of the proposed procedure, it is helpful to briefly discuss the matrix counterpart of Cross tensor measurement scheme and recovery algorithm here. Suppose $X$ is a $p_1$-by-$p_2$ unknown low-rank matrix, a row index subset $\Omega_1\subseteq [1:p_1]$ and a column index subset $\Omega_2\subseteq [1:p_2]$ are randomly generated, and one observe the rows $X_{[\Omega_1,:]}$ and columns $X_{[:, \Omega_2]}$. We aim to recover the original low-rank matrix $X$ from observations of $X_{[\Omega_1, \Omega_2]}$, $X_{[\Omega_1, \Omega_2^c]}$ and $X_{[\Omega_1^c, \Omega_2]}$. This problem, which has been studied recently in \cite{wagner2015low} and \cite{cai2016structured} in the context of \emph{row and column matrix completion} and \emph{structured matrix completion}, would be a matrix analogy to tensor completion via Cross tensor measurements. In the noiseless setting, it was shown that the low-rank matrix $X$ can be recovered by the well-regarded Schur complement,
	\begin{equation}\label{eq:schur-complement}
	\hat{X} = X, \quad \text{where}\quad \hat{X} = X_{[:, \Omega_2]}X_{[\Omega_1, \Omega_2]}^\dagger X_{[\Omega_1, :]}.
	\end{equation}
	In the noisy setting, the estimation scheme based on a sequential truncation and MLE-based approach were proposed and analyzed in  \cite{wagner2015low} and \cite{cai2016structured}, respectively.	
	
	Although the Cross tensor measurement scheme shares similarities with the above matrix completion setting, the proposed tensor recovery procedure shows distinct aspects and is much more difficult to analyze in various ways. First, in matrix settings one fully observes an ``L" shaped region including $\X_{[\Omega_1, \Omega_2]}$, $\X_{[\Omega_1^c, \Omega_2]}$ and $\X_{[\Omega_1, \Omega_2^c]}$ \citep{wagner2015low,cai2016structured}. However, the analog of the ``L'' shape in tensor settings, $\X_{[\Omega_1^c, \Omega_2, \Omega_3]}$, $\X_{[\Omega_1, \Omega_2^c, \Omega_3]}$ and $\X_{[\Omega_1, \Omega_2, \Omega_3^c]}$, include $O(p_1m_2m_3 + m_1p_2m_3+ m_1m_2p_3)$ entries in total, which is far more than the level achieved by Cross.
	Such difference also makes it difficult to directly apply the original analysis in matrix setting to Cross. Second, the Cross tensor measurement scheme involves more complicated tensor operations than its matrix counterpart. In particular, the tensor recovery formula \eqref{eq:tensor-Schur-complement} involves seven terms with three inverses including body, arms and joints, making its analysis far more demanding than that of \eqref{eq:schur-complement} where only three submatrices and one inverse are involved.
	Third, the analysis of the proposed tensor completion algorithm relies on tensor terminology and algebra, which are much more complicated than the matrix ones. For example, the $\ell_2$ and Frobenius norms of a matrix can be well characterized by its singular values. However there is no such correspondence for tensors. 
\end{Remark}

\section{Simulation Study}\label{sec.simu}

In this section, we investigate the numerical performance of the proposed procedure in a variety of settings. We repeat each setting 1000 times and record the average relative loss in Hilbert Schmidt norm, i.e., $\|\hat{\X} - \X\|_{\rm HS}/\|\X\|_{\rm HS}$.

We first focus on the setting with i.i.d. Gaussian noise. To be specific, we randomly generate $\X = \S \times_1 E_1 \times_2 E_2 \times E_3$, where $\S \in \mathbb{R}^{r_1\times r_2\times r_3}$, $E_1\in \mathbb{R}^{p_1\times r_1}, E_2\in \mathbb{R}^{p_2\times r_2}, E_3\in \mathbb{R}^{p_3\times r_3}$ are all with i.i.d. standard Gaussian entries. We can verify that $\X$ becomes a rank-$(r_1, r_2, r_3)$ tensor with probably 1 whenever $r_1, r_2, r_3$ satisfy $\max^2(r_1, r_2, r_3) \leq r_1r_2r_3$. Then we generate the Cross tensor measurement $\bOmega$ as in \eqref{eq:bOmega} with $\Omega_t$ including uniformly randomly selected $m_t$ values from $[1:p_t]$ and $\Xi_t$ including uniformly randomly selected $g_t$ values from $\prod_{s\neq t}\Omega_s$, and contaminate $\X_{\bOmega}$ with i.i.d. Gaussian noise: $\Y_{\bOmega} = \X_{\bOmega} + \Z_{\bOmega}$, where $Z_{ijk}\overset{iid}{\sim} (0, \sigma^2)$. Under such configuration, we study the influence of different factors, including $\lambda_t, \sigma, m_t, g_t, p_t$ to the numerical performance.

\begin{figure}
	\makebox[\linewidth][c]{%
		\begin{subfigure}[b]{.5\textwidth}
			\centering
			\includegraphics[height = 1.6in, width=2.4in]{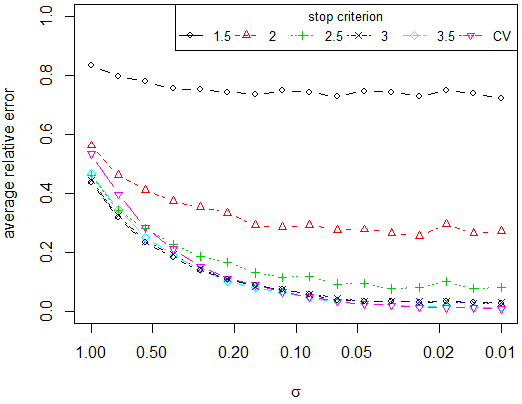}
			\caption{Varying noise level $\sigma$ and stop criteria ($p_t=50$, $m_t=10$)}
		\end{subfigure}
		\begin{subfigure}[b]{.5\textwidth}
			\centering
			\includegraphics[height = 1.6in, width=2.4in]{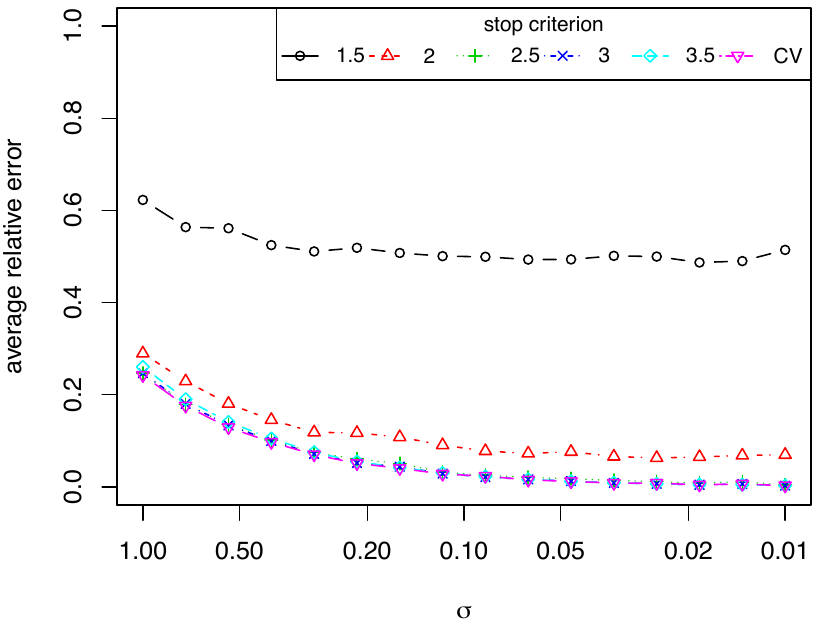}
			\caption{Varying noise level $\sigma$ and stop criteria  ($p_t=80$, $m_t=15$)}
		\end{subfigure}
	}\\
	\makebox[\linewidth][c]{
		\begin{subfigure}[b]{.5\textwidth}
			\centering
			\includegraphics[height = 1.6in,width=2.4in]{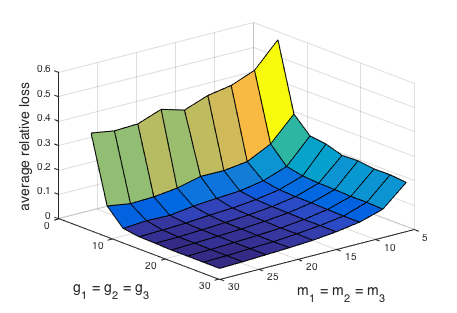}
			\caption{Varying $g_t$ and $m_t$: $\lambda_t = 3\sqrt{p_t/m_t}$}
		\end{subfigure}
	\begin{subfigure}[b]{.5\textwidth}
		\centering
		\includegraphics[height = 1.6in,width=2.4in]{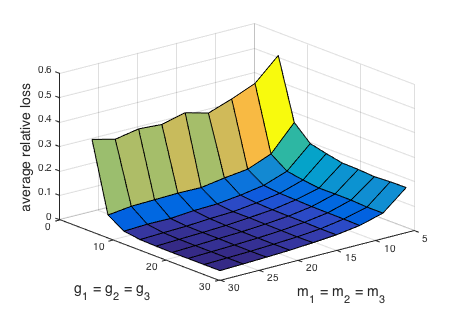}
		\caption{Varying $g_t$ and $m_t$: $\lambda_t$ selected by CV}
		\end{subfigure}
	}\\
	\makebox[\linewidth][c]{%
		\begin{subfigure}[b]{.5\textwidth}
			\centering
			\includegraphics[height = 1.6in,width=2.4in]{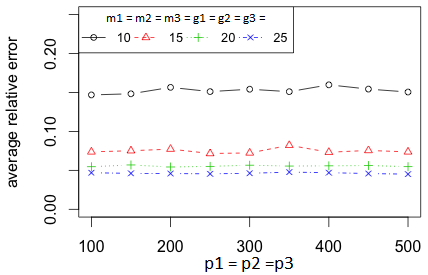}
			\caption{Average relative loss when $p_t, m_t, g_t$ are varying}		\end{subfigure}\quad
		\begin{subfigure}[b]{.5\textwidth}
			\centering
			\includegraphics[height = 1.6in,width=2.4in]{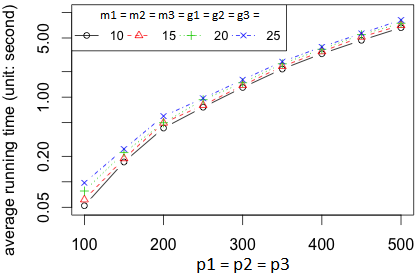}
			\caption{Average time cost when $p_t$, $m_t, g_t$ are varying}
		\end{subfigure}
	}
	\caption{Numerical performance under Gaussian noise settings}\label{fig:simulation_setting_1}
\end{figure}

Under the Gaussian noise setting, we first compare different choices of tuning parameters $\lambda_t$. To be specific, set $p_1 = p_2 = p_3 \in \{50, 80\}$, $m_1 = m_2 = m_3 = g_1 = g_2 = g_3 \in \{10, 15\}$, $r_1 = r_2 = r_3 = 3$ and let $\sigma$ range from 1 to 0.01. We consider both the fixed tuning parameters: $\lambda_t \in \left[1.5\sqrt{p_t/m_t}, 3.5\sqrt{p_t/m_t}\right]$, and the one selected by 5-fold cross-validation. The average relative Hilbert Schmidt norm loss of $\hat{\X}$ from Algorithm \ref{al:noisy} is reported in Figure \ref{fig:simulation_setting_1}. 
It can be seen that the average relative loss decays when the noise level is decreasing. After comparing different choices of $\lambda_t$,  $3\sqrt{p_t/m_t}$ and cross-validation scheme works the best under different $\sigma$, which matches our previous suggestions.

We also compare the effects of $m_t = \left|\Omega_t\right|$ and $g_t = \left|\Xi_t\right|$ in the numerical performance of Algorithm \ref{al:noisy}. We set $p_1 = p_2 = p_3 = 50, r_1 = r_2 = r_3 = 3, \sigma = 0.3$, let $g_t, m_t$ vary from 6 to 30 and let $\lambda_t$ be either fixed as $3\sqrt{p_t/m_t}$ or chosen by 5-fold cross-validation. The average relative Hilbert-Schmidt norm loss are plotted in Figure \ref{fig:simulation_setting_1}(c) and (d). It can be seen that as $g_t, m_t$ grow, namely when more entries are observable, better recovery performance can be achieved. The performance of $\lambda_t = 3\sqrt{p_t/m_t}$ is still similar to the one by cross-validation.

To further study the impact of high-dimensionality to the proposed procedure, we consider the setting where the dimension of $\X$ further grows. Here, $r_1 = r_2 = r_3 = 3, \sigma = 0.3$, $m_1 = m_2 = m_3 = g_1 = g_2 = g_3 \in \{10, 15,20, 25\}$, $p_1, p_2, p_3$ grow from 100 to 500 and $\lambda_t = 3\sqrt{p_t/m_t}$. The average relative loss in Hilbert-Schmidt norm and average running time are provided in Figure \ref{fig:simulation_setting_1}(e) and (f), respectively. Particularly, the recovery of 500-by-500-by-500 tensors involves 125,000,000 variables, but the proposed procedure provides stable recovery within 10 seconds on average by the PC with 3.1 GHz CPU, which demonstrates the efficiency of our proposed algorithm.

\begin{figure}
	\makebox[\linewidth][c]{%
		\begin{subfigure}[b]{.5\textwidth}
			\centering
			\includegraphics[height=1.6in, width=2.4in]{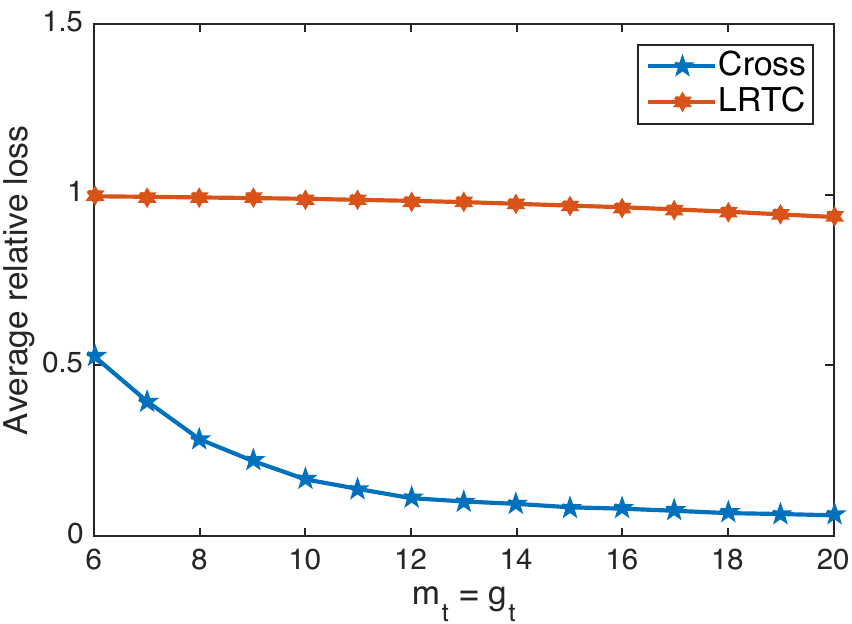}
			\caption{Average relative loss with varying $m_t, g_t \in [6:20]$}
		\end{subfigure}%
		\quad \begin{subfigure}[b]{.5\textwidth}
			\centering
			\includegraphics[height=1.6in,width=2.4in]{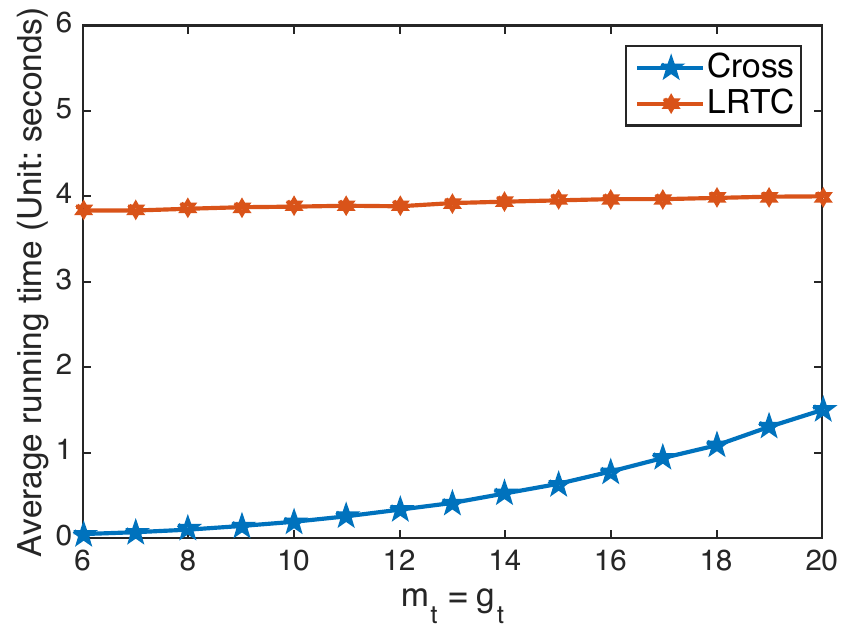}
			\caption{Average running time with varying $m_t, g_t \in [6:20]$}
		\end{subfigure}%
	}
	\makebox[\linewidth][c]{
	\begin{subfigure}[b]{.5\textwidth}
		\centering
		\includegraphics[height=1.6in, width=2.4in]{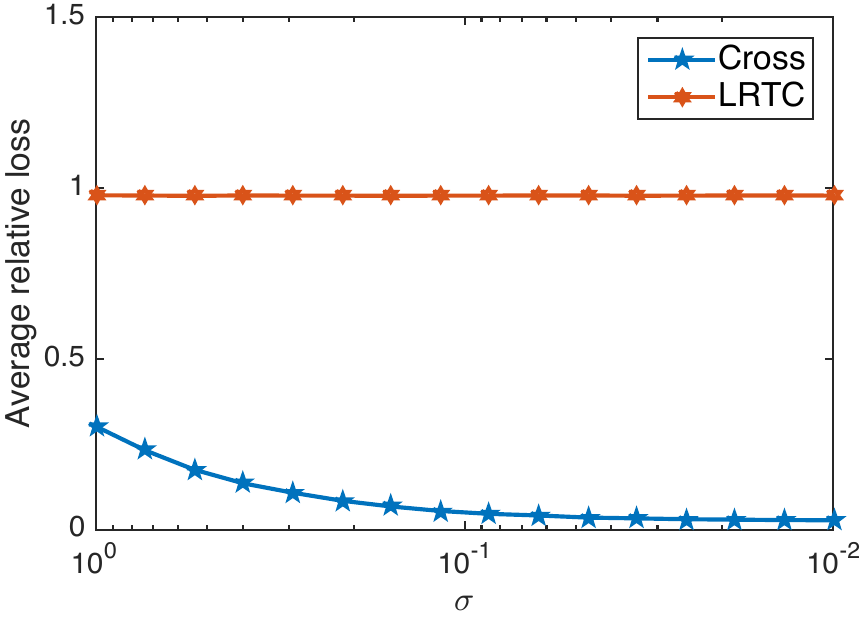}
		\caption{Average relative loss with varying $\sigma^2 \in [0.01, 1]$}
	\end{subfigure}%
	\quad \begin{subfigure}[b]{.5\textwidth}
		\centering
		\includegraphics[height=1.6in,width=2.4in]{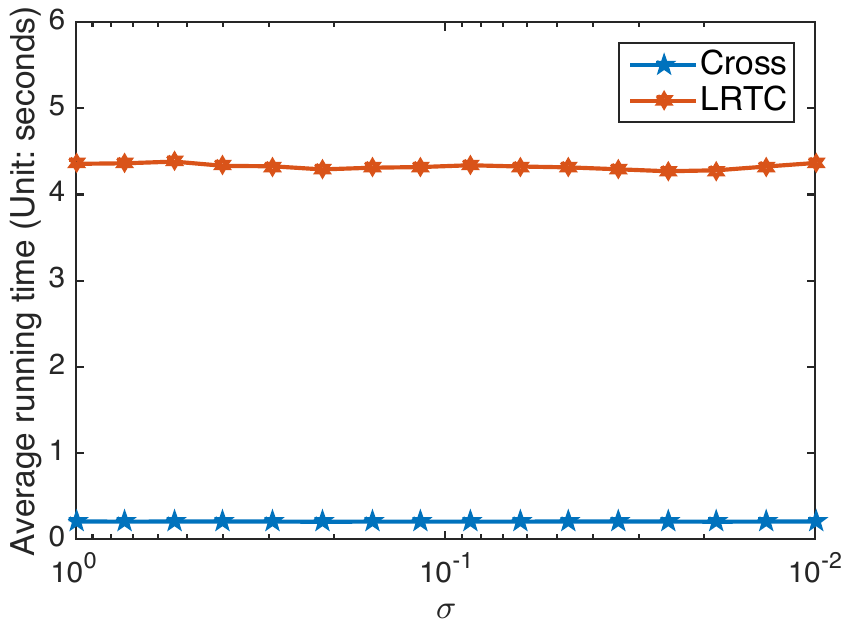}
		\caption{Average running time with varying $\sigma^2 \in [0.01, 1]$}
	\end{subfigure}%
	}
	\caption{Average relative loss and running time for Cross and LRTC.}\label{fig:comparison_LRTC}
\end{figure}

The next simulation setting is designed to compare the proposed Algorithm \ref{al:noisy} with the Low-rank Tensor Completion (LRTC) proposed by \cite{liu2013tensor}. LRTC is a convexified tensor completion method based on matricization nuclear norm minimization. To avoid non-convergence runs of LRTC, we set the maximum number of iterations as 500 and all the other tuning parameters as the default values. Let $p_1 = p_2 = p_3 = 50$, $r_1=r_2=r_3 = 3$, we consider two settings: (i) $\sigma^2 = 0.3$, $m_t, g_t$ vary from 6 to 20; (ii) $m_t = g_t = 10$, $\sigma^2$ varies from 0.01 to 1. We apply both LRTC (with the package downloaded from the authors' website) and our proposed procedure, then present the estimation error in relative Hilbert-Schmidt norm and average running time in Figure \ref{fig:comparison_LRTC}. It is clear that our proposed procedure achieves significantly smaller estimation error in much shorter running time, which substantially outperforms LRTC.

\begin{figure}
	\makebox[\linewidth][c]{%
		\begin{subfigure}[b]{.5\textwidth}
			\centering
			\includegraphics[height = 1.6in, width=2.4in]{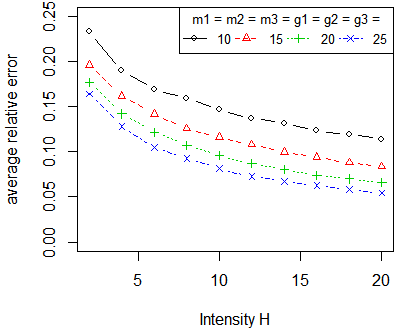}
			\caption{Poisson model with varying $m_t, g_t$ and intensity $H$}
		\end{subfigure}
		\begin{subfigure}[b]{.5\textwidth}
			\centering
			\includegraphics[height = 1.6in,width=2.4in]{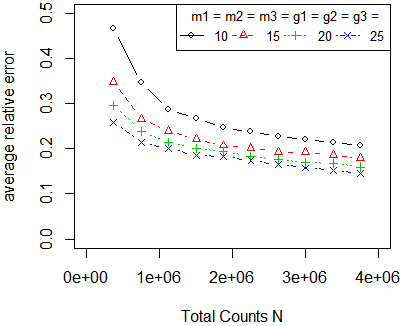}
			\caption{Multinomial model with varying $m_t, g_t$, total count $N$}
		\end{subfigure}
	}
	\caption{Average relative loss in HS norm based on Poisson and multinomial observations. Here, $p_1 = p_2 = p_3 =50, r_1 = r_2 = r_3 = 3$.}\label{fig:simulation-poisson}
\end{figure}

Then we move on to the setting where observations take discrete random values. High-dimensional count data commonly appear in a wide range of applications, including fluorescence microscopy, network flow, and microbiome (see, e.g., \cite{nowak2000statistical,jiang2015minimax,cao2016poisson,cao2017microbial}, etc.), where Poisson and multinomial distributions are often used in modeling the counts. In this simulation study, we generate $\S\in \mathbb{R}^{r_1\times r_2 \times r_3}, E_t\in \mathbb{R}^{p_t\times r_t}$ as absolute values of i.i.d. standard normal random variables, and calculate
$$ \X = \frac{\S \times_1 E_1 \times_2 E_2 \times_3E_3}{\sum_{i=1}^{p_1}\sum_{j=1}^{p_2}\sum_{k=1}^{p_3} \left(\S \times_1 E_1 \times_2 E_2 \times_3E_3\right)_{ijk}}.$$ 
$\Omega_t, \Xi_t$ are generated similarly as before, $p_1 = p_2 = p_3 = 50$, $r_1 = r_2 = r_3 = 3$, $m_1 = m_2 = m_3 = g_1 = g_2 = g_3 \in\{10, 15, 20, 25\}$, and $\Y = (\Y_{ijk})_{1\leq i \leq p_1, 1\leq j \leq p_2, 1\leq k \leq p_3}$ are Poisson or multinomial distributed:
$$\Y_{ijk} \sim \text{Poisson}(H \X_{ijk}), \quad \text{or}\quad \Y_{ijk} \sim \text{Multinomial}(N; \X).$$
Here $H$ is a known intensity parameter in Poisson observations and $N$ is the total count parameter in multinomial observations. As shown in Figure \ref{fig:simulation-poisson}, the proposed Algorithm \ref{al:noisy} performs stably for these two types of noisy structures.

\begin{figure}
	\makebox[\linewidth][c]{%
		\begin{subfigure}[b]{.5\textwidth}
			\centering
		\includegraphics[height=1.6in, width=2.4in]{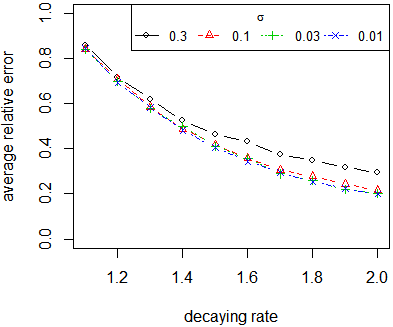}
		\caption{Fixed $m_t = g_t = 10$, varying singular value decaying rate $\alpha$ and noise level $\sigma$.}
		\end{subfigure}%
		\quad \begin{subfigure}[b]{.5\textwidth}
			\centering
			\includegraphics[height=1.6in,width=2.4in]{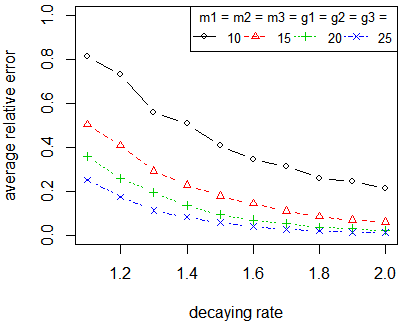}
			\caption{Fixed $\sigma = 0.3$, varying singular value decaying rate $\alpha$ and $m_t, g_t$.}
		\end{subfigure}%
	}
	\caption{Average relative loss for approximate low-rank tensors.}\label{fig:simulation_approx}
\end{figure}

Although $\X$ is assumed to be exactly low-rank in all theoretical studies, it is not necessary in practice. In fact, our simulation study shows that Algorithm \eqref{al:noisy} performs well when $\X$ is only approximately low-rank. Specifically, we fix $p_1 = p_2 = p_3 =50$, generate $\W\in \mathbb{R}^{p_1\times p_2\times p_3}$ from i.i.d. standard normal, set $U_1\in \mathbb{O}_{p_1}, U_2\in \mathbb{O}_{p_2}, U_3\in \mathbb{O}_{p_3}$ as uniform random orthogonal matrices, and $E_t = \diag(1, 1, 1^{-\alpha}, \cdots, (p_t - 2)^{-\alpha})$. $\X$ is then constructed as
\begin{equation*}
\X = \W \times_1 \left(E_1U_1\right) \times_2 \left(E_2U_2\right) \times_3 \left(E_3U_3\right).
\end{equation*}
Here, $\alpha$ measures the decaying rate of singular values of each matricization of $\X$ and $\X$ becomes exactly rank-(3, 3, 3) when $\alpha = \infty$. We consider different decay rates $\alpha$, noise levels $\sigma$, and observation set sizes $m_t$ and $g_t$.
The corresponding average relative Hilbert-Schmidt norm loss is reported in Figure \ref{fig:simulation_approx}. It can be seen that although $\X$ is not exactly low rank, as long as the singular values of each matricization of $\X$ decay sufficiently fast, a desirable completion of $\X$ can still be achieved, which again demonstrates the robustness of the proposed procedure.

\section{Real Data Illustration}\label{sec.real_data}

In this section, we apply the proposed Cross tensor measurements scheme to a real dataset on attention hyperactivity disorder (ADHD) available from ADHD-200 Sample Initiative ( \url{http://fcon_1000.projects.nitrc.org/indi/adhd200/}). ADHD is a common disease that affects at least 5-7\% of school-age children and may accompany patients throughout their life with direct costs of at least \$36 billion per year in the United States. Despite being the most common mental disorder in children and adolescents, the cause of ADHD is largely unclear. To investigate the disease, the ADHD-200 study covered 285 subjects diagnosed with ADHD and 491 control subjects. After data cleaning, the dataset contains 776 tensors of dimension 121-by-145-by-121: $\Y_i,i=1,\dots, 776$. The storage space for these data through naive format is $121\times 145\times 121\times 776 \times 4$B $\approx$ 6.137 GB, which makes it difficult and costly for sampling, storage and computation. Therefore, we hope to reduce the sampling size for ADHD brain imaging data via the proposed Cross tensor measurement scheme. 

\begin{figure}
	\includegraphics[height = 1.6in,width=1.6in]{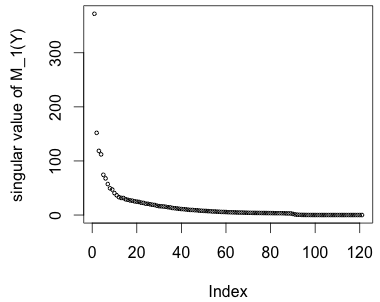}
	\includegraphics[height = 1.6in,width=1.6in]{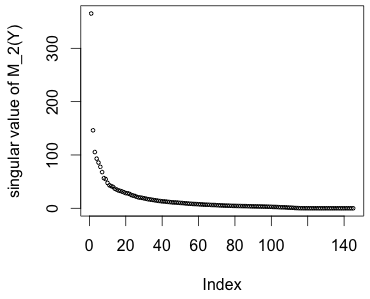}
	\includegraphics[height = 1.6in,width=1.6in]{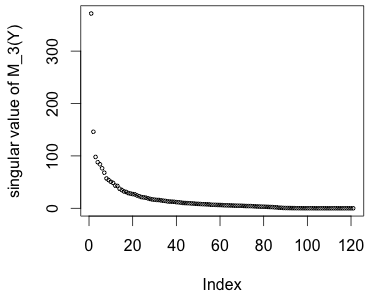}
	\caption{Singular value decompositions for each matricization of $\Y$}\label{fig:real_data_matricization}
\end{figure}

Figure \ref{fig:real_data_matricization} shows the singular values of each matricization of a randomly selected $\Y_i$. We can see that $\Y_i$ is approximately Tucker low-rank. 
Similar to previous simulation settings, we uniformly randomly select $\Omega_t \subseteq [1:p_t]$, $\Xi_t \subseteq \prod_{s\neq t}\Omega_{s}$ such that $|\Omega_t|=m_t, |\Xi_t| = g_t$. Particularly, we choose $m_t = \text{round}(\rho \cdot p_t), \quad g_t = \text{round}(m_1m_2m_3/p_t)$,
where $\rho$ varies from 0.1 to 0.5 and round($\cdot$) is the function that rounds its input to the nearest integer. After observing partial entries of each tensor, we apply Algorithm \ref{al:noisy} with $\lambda_t = 3\sqrt{p_t/m_t}$ to obtain $\hat{\X}$. Different from some of the previous studies (e.g. \cite{zhou2013tensor}), our algorithm is adaptive and tuning-free so that we do not need to subjectively specify the rank of the target tensors beforehand. 

Suppose $\rank(\hat{\X}) =  (\hat{r}_1, \hat{r}_2, \hat{r}_3)$, $\hat{U}_1\in \mathbb{O}_{p_1, r_1}, \hat{U}_2\in \mathbb{O}_{p_2, r_2}, \hat{U}_3\in \mathbb{O}_{p_3, r_3}$ are the left singular vectors of $\mathcal{M}_1(\hat{\X}), \mathcal{M}_2(\hat{\X}), \mathcal{M}_3(\hat{\X})$, respectively. We are interested in investigating the performance of $\hat\X$, but the absence of the true tank of the underlying tensor $\X$ makes it difficult to directly compare $\hat{\X}$ and $\X$. Instead, we compare $\hat{\X}$ with $\tilde{\X}$, where $\tilde{\X}$ is the rank-$(\hat{r}_1, \hat{r}_2, \hat{r}_3)$ tensor obtained through the high-order singular value decomposition (HOSVD) (see e.g. \cite{kolda2009tensor}) based on all observations in $\Y$:
\begin{equation}
\tilde{\X} = \Y \times_1 \mathbb{P}_{\tilde{U}_{1}} \times _2 \mathbb{P}_{\tilde{U}_{2}} \times_3\mathbb{P}_{\tilde{U}_{3}}.
\end{equation}
Here $\tilde{U}_1\in \mathbb{O}_{p_1, r_1}, \tilde{U}_2\in \mathbb{O}_{p_2, r_2}, \tilde{U}_3\in \mathbb{O}_{p_3, r_3}$ are the first $r_1, r_2$ and $r_3$ left singular vectors of $\mathcal{M}_1(\Y)$, $\mathcal{M}_2(\Y)$ and $\mathcal{M}_3(\Y)$, respectively. 
\begin{table}[h]
	\begin{center}
		\begin{tabular}{cccccc}
			\hline
			$m_t$ & $\substack{\text{Sampling Ratio}\\\text{(See \eqref{eq:sampling_ratio})}}$ & $\frac{\|\hat{\X} - \Y\|_{\rm HS}}{\|\tilde{\X} - \Y\|_{\rm HS}}$  &
			$\frac{1}{\sqrt{\hat{r}_1}}\|\hat{U}_1^\top\tilde{U}_1\|_F$ & $\frac{1}{\sqrt{\hat{r}_2}}\|\hat{U}_2^\top\tilde{U}_2\|_F$ & $\frac{1}{\sqrt{\hat{r}_3}}\|\hat{U}_3^\top\tilde{U}_3\|_F$ \\\hline
			$\text{round}(.1p_t)$ & 0.0035 & 1.5086 & 0.8291 & 0.8212 & 0.8318\\
			$\text{round}(.2p_t)$ & 0.0267 & 1.2063 & 0.9352 & 0.9110 & 0.9155\\
			$\text{round}(.3p_t)$ & 0.0832 & 1.0918 & 0.9650 & 0.9571 & 0.9634\\
			$\text{round}(.4p_t)$ & 0.1766 & 1.0506 & 0.9769 & 0.9745 & 0.9828\\
			$\text{round}(.5p_t)$ & 0.3066 & 1.0312 & 0.9832 & 0.9840 & 0.9905\\\hline\hline
		\end{tabular}
	\end{center}
	\caption{Comparison between $\hat{\X}$ and $\tilde{\X}$ for ADHD brain imaging data.}\label{tb:ADHD_result}
\end{table}
Particularly, we compare $\|\hat{\X} - \Y\|_{\rm HS}$ and $\|\tilde{\X}-\Y\|_{\rm HS}$, i.e. the rank-$(r_1, r_2, r_3)$ approximation based on limited number of Cross tensor measurements and the approximation based on all measurements. We also compare $\hat{U}_1, \hat{U}_2, \hat{U}_3$ and $\tilde{U}_1, \tilde{U}_2, \tilde{U}_3$ by calculating $\frac{1}{\sqrt{\hat{r}_t}} \|\hat{U}_t^\top \tilde{U}_t\|_F$. The study is performed on 10 randomly selected images and repeated 100 times for each of them. We can immediately see from the result in Table \ref{tb:ADHD_result} that on average, $\|\hat{\X}-\Y\|_{\rm HS}$, i.e. rank-$(r_1, r_2, r_3)$ approximation error with limited numbers of Cross tensor measurements, can get very close to $\|\tilde{\X} - \Y\|_{\rm HS}$, i.e. rank-$(r_1, r_2, r_3)$ approximation error with the whole tensor $\Y$. Besides, $\frac{1}{\sqrt{\hat{r}_t}} \|\hat{U}_t^\top \tilde{U}_t\|_F$ is close to 1, which means the singular vectors calculated from limited numbers of Cross tensor measurements are not too far from the ones calculated from the whole tensor. 

Therefore, by the proposed Cross Tensor Measurement Scheme and a small fraction of observable entries, we can approximate the leading principle component of the original tensor just as if we have observed all voxels. This illustrates the power of the proposed algorithm.

	\section{Discussions: Extensions to Fourth and higher order Tensors}\label{sec.discussion}

In this paper, we propose a novel tensor measurement scheme called Cross and the corresponding low-rank tensor completion algorithm. The theoretical analyses are provided for the proposed procedure to guarantee the optimality in both the sample size requirement and the completion error. The proposed procedure is efficient and easy to implement even for large-scale dataset.

Throughout the paper, we focus our presentations and analyses on order-3 tensors. Moreover, the proposed methods can be easily extended for fourth or higher order tensors. Suppose we aim to complete an unknown, order-$d$, and rank-$(r_1, \ldots, r_d)$ tensor: $\X \in \mathbb{R}^{p_1\times \cdots\times p_d}$. Similarly, we introduce the order-$d$ Cross Tensor Measurement Scheme as
\begin{equation*}
\Omega_t \subseteq [1:p_t], \quad \text{where}\quad |\Omega_t| = m_t,\quad t = 1,\ldots, d,
\end{equation*}
\begin{equation*}
\Xi_t = \prod_{s \neq t} \Omega_s,\quad \text{where} \quad |\Xi_t| = g_t, \quad t = 1,\ldots, d,
\end{equation*}
\begin{equation*}
\bOmega = \left(\prod_{t = 1}^d \Omega_t\right)  \bigcup\limits_{t=1}^{d} \left([1:p_t]\times \Xi_t\right).
\end{equation*}
By observing $\Y_{\bOmega} = \X_{\bOmega}+\Z_{\bOmega}$, we can construct the body, arm, and joint matricizations as
\begin{equation*}
Y_{t, \Omega} = \mathcal{M}_t\left(\Y_{[\Omega_1,\ldots, \Omega_d]}\right),\quad Y_{\Xi_t} = \mathcal{M}_t\left(\Y_{[1:p_t]\times \Xi_t}\right),\quad Y_{\Omega_t\times \Xi_t} = \mathcal{M}_t\left(\Y_{\Omega_t\times \Xi_t}\right).
\end{equation*}
Similarly to Theorem \ref{th:noiseless}, we can prove that $\X$ can be recovered by
\begin{equation*}
\hat{\X} = \Y_{[\Omega_1, \ldots, \Omega_d]} \times_1 R_1 \times_2 \cdots \times_d R_d,\quad \text{where}\quad R_t = Y_{\Xi_t}Y_{\Omega_t\times \Xi_t}^{\dagger} , \quad t = 1,\ldots, d,
\end{equation*}
in the noiseless setting, provided that $\min\{m_t, g_t\}\geq r_t$ (so that $|\bOmega| \geq \prod_{t=1}^d r_t +\sum_{t=1}^d r_t(p_t-r_t)$) and some other mild assumptions holds. This result achieves the optimal sampling requirement since the degrees of freedom for rank-$(r_1, \ldots, r_d)$ tensors in $\mathbb{R}^{p_1\times \cdots\times p_d}$ is exactly $\prod_{t=1}^d r_t +\sum_{t=1}^d r_t(p_t-r_t)$ (see Proposition \ref{pr:degree_of_freedom} and Remark \ref{rm:dof}). Additionally, the procedure for order-$d$ tensor completion with noisy Cross measurements essentially follow from the proposed procedure in Algorithm \ref{al:noisy}, as long as we replace ``$t= 1, 2, 3$" by ``$t = 1,\ldots, d$". An interesting problem for further exploration is on how to select the tuning parameter $\lambda_t$ for the general order-$d$ tensor completion. 

	The main results on Cross tensor measurements can be further extended from the entry-wise observations to the more general projection settings. Suppose $P_t \in \mathbb{O}_{p_t, m_t}$ and $Q_t \in \mathbb{O}_{m_{t+1}m_{t+2}, g_t}$ are orthogonal matrices for $t=1,2,3$. We observe the following \emph{body, arm, and joint projections} of $\X$,
\begin{equation}\label{eq:projection-measurements}
\begin{split}
\X^{(B)} = & \X \times_1 P_1^\top \times_2 P_2^\top \times_3 P_3^\top \in \mathbb{R}^{m_1\times m_2\times m_3},\\
X^{(A)}_t = & \mathcal{M}_1\left(\X \times_2 P_2^\top \times_3 P_3^\top \right)\cdot Q_1 \in \mathbb{R}^{p_t\times g_t},\quad t=1,2,3,\\
X^{(J)}_t = & P_1^\top \cdot \mathcal{M}_1\left(\X \times_2 P_2^\top \times_3 P_3^\top \right)\cdot Q_1 \in \mathbb{R}^{m_t\times g_t},\quad t= 1,2,3.
\end{split}
\end{equation}
The regular Cross tensor measurements discussed in Section \ref{sec.Cross_scheme} can be seen as a special case of \eqref{eq:projection-measurements}, where
\begin{equation*}
\begin{split}
(P_t)_{ij} = & 1_{\{i = \Omega_t(j)\}}, \quad 1\leq i\leq p_t,\quad 1\leq j \leq m_t;\\
(Q_t)_{ij} = & 1_{\{i = \Xi_t(j)\}}, \quad 1\leq i \leq m_{t+1}m_{t+2},\quad 1\leq j \leq g_t.
\end{split}
\end{equation*}
When $\X^{(B)}, X_t^{(A)}, X_t^{(J)}$ are observed without noise, by the similar argument of Theorem \ref{th:noiseless}, one can show that 
$$\X = \X^{(B)}\times_1 R_1 \times R_2\times_3 R_3, \quad R_t = X^{(A)}_t (X^{(J)}_t)^\dagger, \quad t=1,2,3.$$
In the noisy setting, one can similarly apply the proposed Algorithm \ref{al:noisy} to recover $\X$. Suppose the additive observation noises are
\begin{equation*}
\begin{split}
\Z^{(B)} = & \Y^{(B)} - \X^{(B)} \in \mathbb{R}^{m_1\times m_2\times m_3},\\
Z_t^{(A)} = & Y_t^{(A)} - X_t^{(A)} \in \mathbb{R}^{p_t\times g_t}, \quad t=1,2,3,\\
Z_t^{(J)} = & Y_t^{(J)} - X_t^{(J)} \in \mathbb{R}^{m_t\times g_t}, \quad t=1,2,3.
\end{split}
\end{equation*}
If $\hat{\X}$ is the output from Algorithm 2, by similar procedure of Theorem \ref{th:noisy}, one can show that
\begin{equation}\label{ineq:projection-error-bound}
\begin{split}
\left\|\hat{\X} - \X \right\|_{\rm HS} & \leq C\lambda_1\lambda_2 \lambda_3 \|\Z^{(B)}\|_{\rm HS} \\
& + C\lambda_1\lambda_2\lambda_3 \left(\sum_{t=1}^3\xi_t \|Z_t^{(J)}\|_F + C\sum_{t=1}^3\frac{\xi_t}{\lambda_t}\|Z_t^{(A)}\|_F\right).
\end{split}
\end{equation}
where $\lambda_t \geq 2\|X^{(A)}(X^{(J)}_t)^\dagger \|, \xi_t = \|(X^{(J)}_t)^\dagger \mathcal{M}_t(\X^{(B)})\|$.
This extension yields a possible application of Cross to general tensor estimation problems. Suppose one aims to recover low-rank tensor $\X$ from limited number of (not necessarily entry-wise) measurements. If one can obtain reasonable estimations for the following low-dimensional projections of $\X$,
\begin{equation*}
\begin{split}
& \X^{(B)} = \X\times_1 P_1 \times P_2\times P_3 \in \mathbb{R}^{m_1\times m_2\times m_3},\\ ~~\text{and}~~ & X^{(A)}_t = \mathcal{M}_t\left(\X \times_{t+1} P_{t+1}^\top \times_{t+2} P_{t+2}^\top\right)Q_t \in \mathbb{R}^{p_t\times g_t},
\end{split}
\end{equation*}
the proposed Algorithm \ref{al:noisy} yields an efficient estimation for $\X$ with guaranteed performance \eqref{ineq:projection-error-bound}. It would be an interesting future topic to apply Cross tensor measurement scheme to develop efficient algorithm for other low-rank tensor estimation problems, such as tensor completion with uniform random measurements, tensor regression and tensor denoising.

\section*{Acknowledgments}

The author would like to thank Lexin Li for sharing the ADHD dataset and helpful discussion. The author would also like to thank the editor, associate editor, and anonymous referees for valuable suggestions on improving the paper.


\bibliographystyle{apa}
\bibliography{reference}

\newpage

\setcounter{page}{1}
\setcounter{section}{0}
\appendix

\begin{center}
	{\LARGE Supplement to ``Cross: Efficient Low-rank Tensor Completion"}
	
	\bigskip
	
	Anru Zhang\footnote{E-mail: anruzhang@stat.wisc.edu}\ \par University of Wisconsin-Madison
\end{center}
	\medskip
\begin{center}
	{\bf Abstract}
\end{center}
\begin{changemargin}{3em}{3em} 
	\qquad In this Supplement, we provide proofs for the main results and technical lemmas. For better presentation for the long proof of Theorem \ref{th:noisy}, we also provide a table of used notations to Table \ref{tb:sympbols}.
\end{changemargin} 	


\bigskip

\section{Proofs}


We collect the proofs for the main results in this section.

\subsection{Proof of Theorem \ref{th:noiseless}} 
First, we shall note that $\Y = \X$ in the exact low-rank and noiseless setting. For each $t = 1, 2, 3$, according to definitions, $X_{\Xi_t}$ is a collection of columns of $\mathcal{M}_t(\X)$ and $X_{\Omega_t\times \Xi_t}$ is a collection of rows of $X_{\Xi_t}$. Then we have $\rank(X_{\Omega_t \times \Xi_t}) \leq  \rank(X_{\Xi_t}) \leq \rank(\mathcal{M}_t(\Y))$. Given the assumptions, we have $\rank(X_{\Omega_t\times \Xi_t}) = \rank(\mathcal{M}_t(\X)) = r_t$. Thus, $$\rank(X_{\Omega_t\times \Xi_t}) = \rank(X_{\Xi_t}) = \rank(\mathcal{M}_t(\X)) = r_t .$$
Then $\mathcal{M}_t(\X)$ and $X_{\Xi_t}$ share the same column subspace and there exists a matrix $W_t\in \mathbb{R}^{g_t\times \prod_{s\neq t}p_s}$ such that $\mathcal{M}_t(\X) = X_{\Xi_t} \cdot W_t.$ On the other hand, since $X_{\Xi_t}$ and $X_{\Omega_t\times \Xi_t}$ have the same row subspace, we have
\begin{equation}\label{eq:theorem_noiseless_1}
X_{\Xi_t} = X_{\Xi_t} \cdot (X_{\Omega_t\times \Xi_t})^\dagger \cdot X_{\Omega_t\times \Xi_t} = R_t \cdot X_{\Omega_t\times \Xi_t}.
\end{equation}
Additionally, $X_{\Xi_t}$ and $X_{\Omega_t\times \Xi_t}$ can be factorized as $X_{\Xi_t} = P_1 Q$, $X_{\Omega_t\times \Xi_t} = P_2Q$, where $P_1\in \mathbb{R}^{p_t \times r_t}, P_2\in \mathbb{R}^{m_t\times r_t}$, $Q\in \mathbb{R}^{r_t\times g_t}$ and $\rank(P_1) = \rank(P_2) = \rank(Q) = r_t$. In this case, for any matrices $\tilde{M} \in \mathbb{R}^{m_t\times r_t}, \tilde{N}\in \mathbb{R}^{g_t\times r_t}$ with $\rank(\tilde{M}^\top X_{\Omega_t\times \Xi_t}\tilde{N}) = r_t$, we have 
\begin{equation}\label{eq:theorem_noiseless_2}
\begin{split}
\tilde{R}_t\cdot X_{\Omega_t\times \Xi_t} = & X_{\Xi_t} \tilde{N}\left(\tilde{M}^\top X_{\Omega_t\times \Xi_t}\tilde{N}\right)^{-1} \tilde{M}^\top \cdot X_{\Omega_t\times \Xi_t}\\
= & P_1Q\tilde{N} \left(\tilde{M}^\top P_2 Q \tilde{N}\right)^{-1} \tilde{M}^\top P_2Q\\
= & P_1 Q\tilde{N} (Q\tilde{N})^{-1} (\tilde{M}^\top P_2)^{-1}\tilde{M}^\top P_2 Q = P_1Q = X_{\Xi_t}.
\end{split}
\end{equation}
Right multiplying $W_t$ to \eqref{eq:theorem_noiseless_1} and \eqref{eq:theorem_noiseless_2}, we obtain
\begin{equation*}
\begin{split}
\mathcal{M}_t(\X) = X_{\Xi_t} \cdot W_t = & R_t \cdot X_{\Omega_t\times \Xi_t} \cdot W_t =  R_t \cdot \left(\mathcal{M}_t(\Y)\right)_{[\Omega_t, :]},\\
= & \tilde{R}_t \cdot X_{\Omega_t\times \Xi_t} \cdot W_t =  \tilde{R}_t \cdot \left(\mathcal{M}_t(\Y)\right)_{[\Omega_t, :]}.
\end{split}
\end{equation*}
By folding $\mathcal{M}_t(\X)$ back to the tensors and Lemma 3, we obtain
\begin{equation*}
\begin{split}
\X = & \X_{[\Omega_1, :, :]} \times_1 R_1 = \X_{[:, \Omega_2, :]} \times_2 R_2 =  \X_{[:, :, \Omega_3]} \times_3 R_3\\
= & \X_{[\Omega_1, :, :]} \times_1 \tilde{R}_1 = \X_{[:, \Omega_2, :]}\times_2 \tilde{R}_2 = \X_{[:, :, \Omega_3]} \times_3 \tilde{R}_3.
\end{split}
\end{equation*}
By the equation above,
\begin{equation*}
\begin{split}
& \X_{[\Omega_1, \Omega_2, \Omega_3]} \times_1 R_1= \X_{[\Omega_1, \Omega_2, \Omega_3]} \times_1 \tilde{R}_1 = \X_{[:, \Omega_2, \Omega_3]}, \\
& \X_{[:, \Omega_2, \Omega_3]} \times_2 R_2 = \X_{[:, \Omega_2, \Omega_3]} \times_2 \tilde{R}_2 =\X_{[:, :, \Omega_3]},\\
& \X_{[:, :, \Omega_3]} \times_3 R_3 = \X_{[:, :, \Omega_3]} \times_3 \tilde{R}_3 = \X.
\end{split}
\end{equation*}
Therefore,
\begin{equation*}
\X_{[\Omega_1, \Omega_2, \Omega_3]} \times_1 R_1 \times_2 R_2 \times_3 R_3 = \X_{[\Omega_1, \Omega_2, \Omega_3]} \times_1 \tilde{R}_1 \times_2 \tilde{R}_2 \times_3 \tilde{R}_3 = \X,
\end{equation*}
which concludes the proof of Theorem \ref{th:noiseless}. \quad $\square$

\subsection{Proof of Theorem \ref{th:noisy}} Based on the proof for Theorem \ref{th:noiseless}, we know
\begin{equation*}
\X = \X_{[\Omega_1, \Omega_2, \Omega_3]} \times_1 \tilde{R}_1 \times_2 \tilde{R}_2 \times_3 \tilde{R}_3, 
\end{equation*}
where 
\begin{equation}\label{eq:def_tilde_R}
\tilde{R}_t = X_{\Xi_t}\tilde{N}\left(\tilde{M}^\top X_{\Omega_t\times \Xi_t}\tilde{N}\right)^{-1}\tilde{M}^\top \in \mathbb{R}^{p_t\times m_t},
\end{equation} for any $\tilde{M}\in \mathbb{R}^{m_t\times r_t}, \tilde{N}\in \mathbb{R}^{g_t\times r_t}$ satisfying $\tilde{M}^\top X_{\Omega_t\times \Xi_t}\tilde{N}$ is non-singular. The proof for Theorem \ref{th:noisy} is relatively long. For better presentation, we divide the proof into steps. Before going into detailed discussions, we list all notations with the definitions and possible simple explanations in Table 2 in the supplementary materials. 

\ \par

\noindent (\emph{Step 1.}) Denote 
\begin{equation}\label{eq:N_t,M_t}
\begin{split}
\hat{N}_t, N_t \in \mathbb{O}_{g_t, r_t} \text{ as the first $r_t$ right singular vectors of $Y_{\Xi_t}$, $X_{\Xi_t}$}; \\
\hat{M}_t, M_t\in \mathbb{O}_{m_t, r_t} \text{ as the first $r_t$ left singular vectors of $Y_{t, \Omega}$, $X_{t, \Omega}$}.
\end{split}
\end{equation} 
It is easy to see that alternate characterization for $\hat{N}_t$ and $\hat{M}_t$ are
\begin{equation}\label{eq:hat_N_t_hat_M_t_U_V_related}
\hat{N}_t = V^{(A)}_{[:, 1:r_t]},\quad \hat{M}_t = U^{(B)}_{[:, 1:r_t]}.
\end{equation} 
Denote $\tau  = 1/5$. In this step, we prove that $\hat{M}_t, M_t$; $\hat{N}_t, N_t$ are close by providing the upper bounds on singular subspace perturbations,
\begin{equation}\label{ineq:hatMM_hatNN}
\|\sin\Theta(\hat{M}_t, M_t)\| \leq \tau^2/(1-2\tau), \quad \|\sin\Theta(\hat{N}_t, N_t)\|\leq \tau^2/(1-2\tau);
\end{equation}
as well as the inequality which ensures that $J_t$ is bounded away from being singular,
\begin{equation}\label{ineq:sigma_min-MXN}
\begin{split}
\sigma_{\min}(\hat{M}_t^\top X_{\Omega_t\times \Xi_t}\hat{N}_t) \geq & \left(1 - \tau^4/(1-2\tau)^2\right)\sigma_{\min}(X_{\Omega_t\times \Xi_t}),\\
\sigma_{\min}(\hat{M}_t^\top Y_{\Omega_t\times \Xi_t}\hat{N}_t) \geq & \left(1 - \tau^4/(1-2\tau)^2 - \tau\right)\sigma_{\min}(X_{\Omega_t\times \Xi_t}).
\end{split}
\end{equation}
Actually, based on Assumption \eqref{ineq:assumption_gap} , we have
\begin{equation*}
\begin{split}
& \sigma_{r_t}(Y_{\Xi_t}N_t) \geq \sigma_{r_t}(X_{\Xi_t}N_t) - \|Z_{\Xi_t}N_t\|\\
\geq & \sigma_{r_t}(X_{\Xi_t}) - \tau\sigma_{r_t}(X_{\Xi_t}) = (1-\tau)\sigma_{r_t}(A_{\Xi_t});
\end{split}
\end{equation*}
\begin{equation*}
\sigma_{r_t+1}(X_{\Xi_t}) \leq \sigma_{r+1}(X_{\Xi_t}) + \|Z_{\Xi_t}\| \leq \tau \sigma_{r_t}(A_{\Xi_t});
\end{equation*}
\begin{equation*}
\begin{split}
& \left\|\mathbb{P}_{X_{\Xi_t}N_t}X_{\Xi_t}(N_t)_{\perp}\right\| \leq \left\|X_{\Xi_t}(N_t)_{\perp}\right\| = \left\|A_{\Xi_t}(N_t)_{\perp} + Z_{\Xi_t}(N_t)_{\perp}\right\|\\
\leq & \|Z_{\Xi_t}\| \leq \tau\sigma_{r_t}(A_{\Xi_t}).
\end{split}
\end{equation*}
Here $(N_t)_{\perp}$ is the orthogonal complement matrix, i.e. $[N_t ~ (N_t)_{\perp}] \in \mathbb{O}_{g_t}$. By setting $A = X_{\Xi_t}$, $W = N_t$ in the scenario of the unilateral perturbation bound (Proposition 1 in \cite{cai2016rate}), we then obtain
\begin{equation*}
\begin{split}
& \left\|\sin\Theta(\hat{N}_t, N_t)\right\| \leq \frac{\sigma_{r_t+1}(X_{\Xi_t}) \left\|\mathbb{P}_{(X_{\Xi_t}N_t)} Z_{\Xi_t}(N_t)_{\perp}\right\|}{\sigma_{r_t}^2(X_{\Xi_t}N_t) - \sigma_{r_t+1}^2(X_{\Xi_t})}\\
\leq & \frac{\tau\cdot \tau}{(1-\tau)^2 - \tau^2} = \tau^2/(1-2\tau).
\end{split}
\end{equation*} 
Here $\|\sin\Theta(\cdot,\cdot)\|$ is a commonly used distance between orthogonal subspaces. Similarly, based on the assumption that $\tau\sigma_{r_t}(A_{t, \Omega}) \geq \|Z_{t, \Omega}\|$, we can derive
\begin{equation*}
\left\|\sin\Theta(\hat{M}_t, M_t)\right\| \leq \tau^2/(1-2\tau),
\end{equation*}
which proves \eqref{ineq:hatMM_hatNN}. Based on the property for $\sin\Theta$ distance (Lemma 1 in \cite{cai2016rate}),
\begin{equation*}
\begin{split}
\sigma_{\min}(\hat{N}_t^\top N_t) = \sqrt{1 - \|\sin\Theta(\hat{N}_t, N_t)\|^2} \geq \sqrt{1-\tau^4(1-2\tau)^2};\\
\sigma_{\min}(\hat{M}_t^\top M_t) = \sqrt{1 - \|\sin\Theta(\hat{M}_t, M_t)\|^2} \geq \sqrt{1-\tau^4/(1-2\tau)^2}.
\end{split}
\end{equation*}
Also, since $M_t, N_t$ coincide with the left and right singular subspaces of $X_{\Omega_t\times \Xi_t}$ respectively, we have
\begin{equation*}
\begin{split}
& \sigma_{\min}(\hat{M}_t^\top X_{\Omega_t\times \Xi_t}\hat{N}_t)\\
= & \sigma_{\min}(\hat{M}_t^\top \mathbb{P}_{M_t}X_{\Omega_t\times \Xi_t}\mathbb{P}_{N_t}\hat{N}_t) = \sigma_{\min}(\hat{M}_t^\top M_t M_t^\top X_{\Omega_t\times \Xi_t}N_tN_t^\top\hat{N}_t)\\
\geq & \sigma_{\min}(\hat{M}_t^\top M_t) \sigma_{\min} (M_t^\top X_{\Omega_t\times \Xi_t}N_t)\sigma_{\min}(N_t^\top\hat{N}_t)\\ 
\geq & (1-\tau^4/(1-\tau)^2)\sigma_{\min}(X_{\Omega_t\times \Xi_t})
\end{split}
\end{equation*}
\begin{equation*}
\begin{split}
& \sigma_{\min}(\hat{M}_t^\top Y_{\Omega_t\times \Xi_t}\hat{N}_t) \geq \sigma_{\min}(\hat{M}_t^\top X_{\Omega_t\times \Xi_t}\hat{N}_t) - \|Z_{\Omega_t\times \Xi_t}\|\\
\geq & (1- \tau^4/(1-2\tau)^2 -\tau)\sigma_{\min}(X_{\Omega_t\times \Xi_t}),
\end{split}
\end{equation*}
which has finished the proof for \eqref{ineq:sigma_min-MXN}.

\ \par

\noindent(\emph{Step 2.})  In this step, we prove that under the given setting, $\hat{r}_t \geq r_t$ for $t = 1, 2, 3$. We only need to show that for each $t =1, 2, 3$, the stopping criterion holds when $s = r_t$, i.e.
\begin{equation}\label{ineq:thm2_step2_target}
\left\|(A_t)_{[:, 1:r_t]} (J_t)_{[1:r_t, 1:r_t]}^{-1}\right\| \leq \lambda_t.
\end{equation}
According to the definitions in \eqref{eq:A_t-J_t}, $A_t, J_t$ and $\hat{M}_t, \hat{N}_t$ can be related as
\begin{equation*}
(A_t)_{[:, 1:r_t]} = Y_{\Xi_t}V^{(A)}_{[:, 1:r_t]} \overset{\eqref{eq:hat_N_t_hat_M_t_U_V_related}}{=} Y_{\Xi_t} \hat{N}_t = (X_{\Xi_t}+Z_{\Xi_t})\hat{N}_t,
\end{equation*}
\begin{equation*}
\begin{split}
(J_t)_{[1:r_t, 1:r_t]} = & (U^{(B)}_{[:, 1:r_t]})^\top Y_{\Omega_t\times \Xi_t} V^{(A)}_{[:, 1:r_t]} \overset{\eqref{eq:hat_N_t_hat_M_t_U_V_related}}{=} \hat{M}_t^\top Y_{\Omega\times \Xi_t}\hat{N}_t\\
= & \hat{M}_t^\top(X_{\Omega\times \Xi_t} + Z_{\Omega\times \Xi_t})\hat{N}_t. 
\end{split}
\end{equation*}
Therefore, in order to show \eqref{ineq:thm2_step2_target}, we only need to prove that $\hat{M}_t^\top Y_{\Omega_t\times \Xi_t} \hat{N}_t$ is non-singular and
\begin{equation}
\left\|(X_{\Xi_t} + Z_{\Xi_t})\hat{N}_t\cdot \left(\hat{M}_t^\top (X_{\Omega_t\times \Xi_t} + Z_{\Omega_t\times \Xi_t}) \hat{N}_t\right)^{-1}\right\| \leq \lambda_t.
\end{equation}
Recall that $U_t\in \mathbb{O}_{p_t, r_t}$ is the singular subspace for $\mathcal{M}_t(\X)$, so there exists another matrix $Q\in \mathbb{R}^{r_t\times g_t}$ such that $X_{\Xi_t}$, a set of columns of $\mathcal{M}_t(\X)$, can be written as 
\begin{equation}\label{eq:Q-U-X_Xi-association}
X_{\Xi_t} = U_t \cdot Q, \quad \text{ then }\quad X_{\Omega_t\times \Xi_t} = U_{t, \Omega}\cdot Q.
\end{equation}
Here $U_{t, \Omega} = (U_t)_{[\Omega_t, :]}$ is a collection of rows from $U_t$, and
$$X_{\Xi_t}X_{\Omega_t\times \Xi_t}^\dagger = U_tQ\left(U_{t, \Omega} Q\right)^\dagger = U_tU_{t, \Omega}^\dagger, \quad \text{then}\quad \left\|X_{\Xi_t}X_{\Omega_t\times \Xi_t}^\dagger\right\| = \sigma_{\min}^{-1}(U_{t, \Omega}).$$
For convenience, we denote 
\begin{equation}\label{ineq:bar_lambda_t}
\bar{\lambda}_t  = \frac{1}{\sigma_{\min}(U_{t, \Omega})} = \left\|X_{\Xi_t}X_{\Omega_t\times \Xi_t}^\dagger \right\| \overset{\eqref{ineq:assumption_arm_body}}{\leq} \frac{1}{2} \lambda_t.
\end{equation}
In this case,
\begin{equation}\label{ineq:X_leq_bar_lambda_t}
\begin{split}
& \left\|X_{\Xi_t} \hat{N}_t (\hat{M}_t^\top X_{\Omega_t\times \Xi_t} \hat{N}_t)^{-1} \right\| = \left\|U_t Q\hat{N}_t (\hat{M}_t^\top U_{t, \Omega} Q \hat{N}_t)^{-1}\right\|\\
= & \left\|U_t \left(\hat{M}_t^\top U_{t, \Omega}\right)^{-1}\right\| = \left\|\left(\hat{M}_t^\top U_{t, \Omega}\right)^{-1}\right\|\leq \frac{1}{\sigma_{\min}(U_{t, \Omega})} = \bar{\lambda}_t.
\end{split}
\end{equation}
Furthermore,
\begin{equation}\label{ineq:sigma_rX_ratio}
\begin{split}
&\sigma_{r_t}(X_{\Omega\times \Xi_t}) = \sigma_{\min}(U_{t, \Omega}\cdot Q) \geq \sigma_{\min}(U_{t, \Omega})\cdot \sigma_{\min}(Q)\\
= & \sigma_{\min}(U_{t, \Omega})\cdot \sigma_{\min}(U_tQ) =  \sigma_{\min}(U_{t, \Omega}) \cdot \sigma_{r_t}(X_{\Xi_t}).
\end{split}
\end{equation}
Therefore,
\begin{equation*}
\begin{split}
& \left\|(X_{\Xi_t}+Z_{\Xi_t})\hat{N}_t\cdot \left(\hat{M}_t^\top \left(X_{\Omega_t\times \Xi_t}+Z_{\Omega_t\times \Xi_t}\right) \hat{N}_t\right)^{-1}\right\|\\
\leq & \left\|X_{\Xi_t}\hat{N}_t\cdot \left(\hat{M}_t^\top \left(X_{\Omega_t\times \Xi_t}+Z_{\Omega_t\times \Xi_t}\right) \hat{N}_t\right)^{-1}\right\|\\
& + \left\|Z_{\Xi_t}\hat{N}_t\cdot \left(\hat{M}_t^\top \left(X_{\Omega_t\times \Xi_t}+Z_{\Omega_t\times \Xi_t}\right) \hat{N}_t\right)^{-1}\right\|\\
\overset{\text{Lemma 5}}{\leq} & \Big\|X_{\Xi_t}\hat{N}_t\cdot \left(\hat{M}_t^\top X_{\Omega_t\times \Xi_t}\hat{N}_t\right)^{-1} \left(\hat{M}_t^\top Z_{\Omega_t\times \Xi_t} \hat{N}_t\right)\\
& \quad\quad\quad\cdot \left(\hat{M}^\top (X_{\Omega_t\times \Xi_t} + Z_{\Omega_t\times \Xi_t})\hat{N}_t\right)^{-1}\hat{M}_t\Big\|\\
& + \left\|X_{\Xi_t}\hat{N}_t\cdot \left(\hat{M}_t^\top X_{\Omega_t\times \Xi_t} \hat{N}_t\right)^{-1}\right\| + \frac{\|Z_{\Xi_t}\|}{\sigma_{\min}(\hat{M}_t^\top Y_{\Omega_t\times \Xi_t}\hat{N}_t)}\\
\overset{\eqref{ineq:sigma_min-MXN} \eqref{ineq:X_leq_bar_lambda_t}}{\leq} & \bar{\lambda}_t + \bar{\lambda}_t \frac{\|Z_{\Omega_t\times \Xi_t}\|}{\left(1 - \tau^4/(1-2\tau)^2-\tau\right)\sigma_{\min}(X_{\Omega_t\times \Xi_t})}\\ 
& + \frac{\|Z_{\Xi_t}\|}{\left(1 - \tau^4/(1-2\tau)^2-\tau\right)\sigma_{\min}(X_{\Omega_t\times \Xi_t})}\\
\overset{\eqref{ineq:assumption_gap}}{\leq} & \bar{\lambda}_t + \frac{\bar{\lambda}_t\tau}{1 - \tau^4/(1-2\tau)^2-\tau}\\
& + \frac{\tau\sigma_{r_t}(X_{\Xi_t})}{\left(1 - \tau^4/(1-2\tau)^2-\tau\right)\sigma_{\min}(X_{\Omega_t\times \Xi_t})}\\
\overset{\eqref{ineq:sigma_rX_ratio}}{\leq} & \bar{\lambda}_t\left(1 + \frac{2\tau}{1 - \tau^4/(1-2\tau)^2 - \tau}\right)
\overset{\eqref{ineq:bar_lambda_t}}{\leq} \lambda_t,
\end{split}
\end{equation*}	
which has proved our claim that $\hat{r}_t \geq r_t$ for $t =1, 2, 3$.

\ \par

\noindent(\emph{Step 3.})  In this step, we provide an important decomposition of $\hat{\X}$ under the scenario that $\hat{r}_t \geq r_t$. One major difficulty is measuring the difference between $J_{t, [1:\hat{r}_t, 1:\hat{r}_t]}^{-1}$ and $J_{t, [1:r_t, 1:r_t]}^{-1}$. For convenience, we further introduce the following notations,
\begin{equation}\label{eq:J_hat_r_t}
J_{\hat{r}_t} \in \mathbb{R}^{\hat{r}_t\times \hat{r}_t},\quad J_{\hat{r}_t} := J_{t, [1:\hat{r}_t, 1:\hat{r}_t]} = U^{(B)\top}_{t, [:, 1:\hat{r}_t]}Y_{\Omega_t\times \Xi_t} V_{t, [:, 1:\hat{r}_t]}^{(A)}, 
\end{equation}
\begin{equation}\label{eq:J_hat_r_t^X^Z}
\begin{split}
J_{\hat{r}_t}^{(X)} := & U^{(B)\top}_{t, [:, 1:\hat{r}_t]}X_{\Omega_t\times \Xi_t} V_{t, [:, 1:\hat{r}_t]}, \quad J_{\hat{r}_t}^{(Z)} := U^{(B)\top}_{t, [:, 1:\hat{r}_t]}Z_{\Omega_t\times \Xi_t} V_{t, [:, 1:\hat{r}_t]},\\ 
& \text{then} \quad J_{\hat{r}_t} = J_{\hat{r}_t}^{(X)} + J_{\hat{r}_t}^{(Z)}.
\end{split}
\end{equation}
\begin{equation}\label{eq:A_hat_r_t}
A_{\hat{r}_t} \in \mathbb{R}^{p_t\times \hat{r}_t},\quad A_{\hat{r}_t} := A_{t, [:, 1:\hat{r}_t]} = Y_{\Xi_t} V_{t, [:, 1:\hat{r}_t]}^{(A)}, 
\end{equation}
\begin{equation}\label{eq:A_hat_r_t^X^Z}
A_{\hat{r}_t}^{(X)} := X_{\Xi_t} V^{(A)}_{t, [:, 1:\hat{r}_t]}, \quad A_{\hat{r}_t}^{(Z)} := Z_{\Xi_t} V^{(A)}_{t, [:, 1:\hat{r}_t]},\quad \text{then} \quad A_{\hat{r}_t} = J_{\hat{r}_t}^{(X)} + J_{\hat{r}_t}^{(Z)}.
\end{equation}
Let the singular value decompositions of $J_{\hat{r}_t}$ be
\begin{equation}\label{eq:K_t1_Lt1}
J_{\hat{r}_t} = K_t \Lambda_t  L_t^\top = \begin{bmatrix}
K_{t1} & K_{t2}
\end{bmatrix}\cdot \begin{bmatrix}
\Lambda_{t1} & \\
& \Lambda_{t2}
\end{bmatrix} \cdot \begin{bmatrix}
L_{t1}^\top\\
L_{t2}^\top
\end{bmatrix}, \quad t =1, 2, 3.
\end{equation}
Here $K_t, L_t\in \mathbb{O}_{\hat{r}_t}$, $K_{t1}, L_{t1} \in \mathbb{O}_{\hat{r}_t, r_t}, K_{t2}, L_{t2} \in \mathbb{O}_{\hat{r}_t, r_t}$ are the singular vectors, $\Lambda_t \in \mathbb{R}^{\hat{r}_t\times \hat{r}_t}$, $\Lambda_{t1} \in \mathbb{R}^{r_t\times r_t}$ and $\Lambda_{t2} \in \mathbb{R}^{(\hat{r}_t-r_t)\times (\hat{r}_t-r_t)}$ are the singular values. Based on these SVDs, we can correspondingly decompose $\bar{R}_t$ (defined in \eqref{eq:hat_R_t}) as
\begin{equation}\label{eq:decomposition_hat_R}
\begin{split}
\bar{R}_t = & A_{t, [:, 1:\hat{r}_t]} J_{\hat{r}_t}^{-1} U^{(B)}_{t, [:, 1:\hat{r}_t]} = A_{t, [:, 1:\hat{r}_t]} \left(K_{t1}\Lambda_{t1} L_{t1}^\top + K_{t2}\Lambda_{t2} L_{t2}^\top\right)^{-1}U^{(B)}_{t, [:, 1:\hat{r}_t]}\\
= & A_{t, [:, 1:\hat{r}_t]}\left(L_{t1}\Lambda_{t1}^{-1} K_{t1}^\top + L_{t2} \Lambda_{t2}^{-1} K_{t2}^\top \right) U^{(B)}_{t, [:, \hat{r}_t]}\\
= & A_{t, [:, 1:\hat{r}_t]}L_{t1}\Lambda_{t1}^{-1} K_{t1}^\top U^{(B)}_{t, [:, 1: \hat{r}_t]} + A_{t, [:, 1:\hat{r}_t]}L_{t2} \Lambda_{t2}^{-1} K_{t2}^\top U^{(B)}_{t, [:, 1:\hat{r}_t]}\\
= & Y_{\Xi_t} V_{t, [:, 1:\hat{r}_t]}^{(A)} L_{t1} \left(K_{t1}^\top J_{\hat{r}_t} L_{t1}\right)^{-1} K_{t1}^\top U^{(B)}_{t, [:, 1:\hat{r}_t]}\\
& + Y_{\Xi_t} V_{t, [:, 1:\hat{r}_t]}^{(A)} L_{t2} \left(K_{t2}^\top J_{\hat{r}_t}L_{t2}\right)^{-1} K_{t2}^\top U^{(B)}_{t, [:, 1:\hat{r}_t]}.
\end{split}
\end{equation}
Here, the first term above associated with $K_{t1}, L_{t1} ...$ stands for the major part in $\bar{R}_t$ while the second associated with $K_{t2}, K_{t2}$ stands for the minor part. Based on this decomposition, we introduce the following notations: for any $t=1,2,3$ (indicating Mode-1, 2 or 3) and $s = 1, 2$ (indicating the major or minor parts),
\begin{equation}\label{eq:A_ts^X^Z}
A_{ts}^{(X)}:= A_{\hat{r}_t}^{(X)} L_{ts} = X_{\Xi_t}V_{t,[:,1:\hat{r}_t]}^{(A)}L_{ts},\quad A_{ts}^{(Z)} := A_{\hat{r}_s}^{(Z)} L_{ts} = Z_{\Xi_t}V_{t,[:,1:\hat{r}_t]}^{(A)}L_{ts}
\end{equation}
\begin{equation}\label{eq:A_ts}
A_{ts} := A_{\hat{r}_t} L_{ts} =  Y_{\Xi_t}V_{t,[:,1:\hat{r}_t]}^{(A)}L_{ts} =  A_{ts}^{(X)} + A_{ts}^{(Z)}.
\end{equation}
\begin{equation}\label{eq:J_ts^X^Z}
\begin{split}
& J_{ts}^{(X)}:= K_{ts}^\top J_{\hat{r}_t} L_{ts} = K_{ts}^\top J_{\hat{r}_t}^{(X)}L_{ts} = K_{ts}^\top U^{(B)\top}_{t, [:, 1:\hat{r}_t]} X_{\Omega_t\times \Xi_t}V_{t, [:, 1:\hat{r}_t]}^{(A)} L_{ts},\\ 
& J_{ts}^{(Z)}:= K_{ts}^\top J_{\hat{r}_t} L_{ts} = K_{ts}^\dagger A_{\hat{r}_t}L_{ts} =  K_{ts}^\top J_{\hat{r}_t}^{(A)} L_{ts}\\
&\quad\quad ~ = K_{ts}^\top U^{(B)\top}_{t, [:, 1:\hat{r}_t]} Z_{\Omega_t\times \Xi_t}V_{t, [:, 1:\hat{r}_t]}^{(A)} L_{ts},
\end{split}
\end{equation}
\begin{equation}\label{eq:J_ts}
J_{ts}:=J_{ts}^{(X)} + J_{ts}^{(Z)} = K_{ts}^\top U^{(B)\top}_{t, [:, 1:\hat{r}_t]} Y_{\Omega_t\times \Xi_t}V_{t, [:, 1:\hat{r}_t]}^{(A)} L_{ts}.
\end{equation}
Also, for $s_1, s_2, s_3 \in \{1, 2\}^3$, we define the projected body measurements
\begin{equation}\label{eq:B_s1s2s3^X}
\B^{(X)}_{s_1s_2s_3} = \X_{[\Omega_1, \Omega_2, \Omega_3]}\times_{t=1}^3 \left(K_{ts_t}^\top U_{t, [1:\hat{r}_t]}^{(B)}\right)
\end{equation}
\begin{equation}\label{eq:B_s1s2s3^Z}
\B^{(Z)}_{s_1s_2s_3} = \Z_{[\Omega_1, \Omega_2, \Omega_3]}\times_{t=1}^3 \left(K_{ts_t}^\top U_{t, [1:\hat{r}_t]}^{(B)}\right)
\end{equation}
\begin{equation}\label{eq:B_s1s2s3}
\begin{split}
\B_{s_1s_2s_3} = & \Y_{[\Omega_1, \Omega_2, \Omega_3]}\times_{t=1}^3 \left(K_{ts_t}^\top U_{t, [1:\hat{r}_t]}^{(B)}\right)= \B^{(X)}_{s_1s_2s_3} + \B^{(Z)}_{s_1s_2s_3}.
\end{split}
\end{equation}
Combining \eqref{eq:decomposition_hat_R}-\eqref{eq:B_s1s2s3}, we can write down the following decomposition for $\hat{\X}$.
\begin{equation}\label{eq:decomposition_hat_X}
\begin{split}
& \hat{\X} = \Y_{[\Omega_1, \Omega_2, \Omega_3]} \times_1 \bar{R}_1\times_2\bar{R}_2\times_3\bar{R}_3  \\
= & \left(\sum_{s_1, s_2, s_3 = 1}^2 \B_{s_1s_2s_3}\right) \times_1 \left(A_{11}J_{11}^{-1} + A_{12}J_{12}^{-1}\right) \times_2 \left(A_{21}J_{21}^{-1} + A_{22}J_{22}^{-1}\right)\\
& \quad\quad\quad\quad\quad\quad\quad\quad\quad \times_3 \left(A_{31}J_{31}^{-1} + A_{32}J_{32}^{-1}\right)\\
= & \sum_{s_1, s_2, s_3=1}^2  
\B_{s_1s_2s_3} \times_1 A_{1 s_1}\left(J_{1 s_1}\right)^{-1} \times_2 A_{2 s_2}\left(J_{2 s_2}\right)^{-1} \times_3 A_{3 s_3}\left(J_{3 s_3}\right)^{-1}.
\end{split}
\end{equation}
This form is very helpful in our analysis later.

\ \par

\noindent(\emph{Step 4.})  In this step, we derive a few formulas for the terms in \eqref{eq:decomposition_hat_X}. To be specific, we shall prove the following results.
\begin{itemize}[leftmargin=*]
	\item Lower bound for singular value of the joint major part: for $t = 1, 2, 3$,
	\begin{equation}\label{ineq:sigma_min_J_t1}
	\begin{split}\sigma_{\min}\left(J_{t1}^{(X)}\right) \geq & (1-\tau_1^2)(1-\tau^4/(1-2\tau)^2)\sigma_{\min}(X_{\Omega_t\times \Xi_t}),\\
	\tau_1 = & \frac{\tau^2}{(1-\tau^4/(1-2\tau)^2-\tau)^2 - \tau^2}.
	\end{split}
	\end{equation}
	In fact, according to definitions, $\hat{M}_t^\top X_{\Omega_t\times \Xi_t} \hat{N}_t=(J_{\hat{r}_t}^{(X)})_{[1:r_t, 1:r_t]}$, i.e., $\hat{M}_t^\top X_{\Omega_t\times \Xi_t} \hat{N}_t$ is a sub-matrix of $J_{\hat{r}_t}^{(X)}$, we have
	\begin{equation*}
	\sigma_{r_t}(J_{\hat{r}_t}^{(X)}) \geq  \sigma_{r_t}\left(\hat{M}^\top X_{\Omega_t\times \Xi_t} \hat{N}\right) \overset{\eqref{ineq:sigma_min-MXN}}{=} (1-\tau^4/(1-2\tau)^2) \sigma_{\min}(X_{\Omega_t\times \Xi_t}).
	\end{equation*}	
	Let 
	\begin{equation}\label{eq:def_K_t_L_t}
	\begin{split}
	& \bar{K}_t \in \mathbb{O}_{\hat{r}_t, r_t},  \bar{L}_t\in \mathbb{O}_{\hat{r}_t, r_t} \text{ be the left and right singular vectors}\\ 
	& \text {for rank-}r_t \text{ matrix }  J_{\hat{r}_t}^{(X)}.
	\end{split}
	\end{equation} 
	We summarize some facts here: 
	\begin{itemize}
		\item $J_{\hat{r}_t} = J_{\hat{r}_t}^{(X)} + J_{\hat{r}_t}^{(Z)}$; 
		\item $\bar{K}_t, \bar{L}_t$ are the left and right singular vectors of rank-$r_t$ matrix $J_{\hat{r}_t}^{(X)}$; 
		\item $K_{t1}, L_{t1}$ are the first $r_t$ left and right singular vectors of $J_{\hat{r}_t}$. 
		\item $\sigma_{r_t+1}(J_{\hat{r}_t}) \overset{\text{ \cite{weyl1912asymptotische}}}{\leq} \sigma_{r_t+1}(J_{\hat{r}_t}^{(X)}) + \|J_{\hat{r}_t}^{(Z)}\| \leq \tau \sigma_{\min}(X_{\Omega_t\times \Xi_t})$.
		\item $\sigma_{r_t}(J_{\hat{r}_t} \bar{L}_t) \overset{\text{ \cite{weyl1912asymptotische}}}{\geq} \sigma_{r_t}(J_{\hat{r}_t}^{(X)} \bar{L}_t) - \|J_{\hat{r}_t}^{(Z)}\| = \sigma_{r_t}(J_{\hat{r}_t}^{(X)}) - \|J_{\hat{r}_t}^{(Z)}\|$\\ 
		$\geq (1-\tau^4/(1-2\tau)^2-\tau) \sigma_{\min}(X_{\Omega_t\times \Xi_t})$.
	\end{itemize}
	Then by the unilateral perturbation bound result (Proposition 1 and Lemma 1 in \cite{cai2016rate}) and the facts above,
	\begin{equation*}
	\begin{split}
	\left\|\sin\Theta(\bar{L}_t, L_{t1})\right\| \leq & \frac{\sigma_{r_t+1}(J_{\hat{r}_t})\cdot\|\mathbb{P}_{(J_{\hat{r}_t}\bar{L}_t)} J_{\hat{r}_t}^{(Z)}\|}{\sigma_{r_t}^2(J_{\hat{r}_t}\bar{L}_t) - \sigma^2_{r_t+1}(J_{\hat{r}_t})}\\
	\overset{\eqref{ineq:J_t1_Z}}{\leq} & \frac{\tau^2}{( 1-\tau^4/(1-2\tau)^2-\tau)^2 - \tau^2}:= \tau_1.
	\end{split}
	\end{equation*}
	\begin{equation*}
	\sigma_{\min}(L_{t1}^\top \bar{L}_t) \geq \sqrt{1 - \tau_1^2}.
	\end{equation*}
	Similarly, $\sigma_{\min}(K_{t1}^\top \bar{K}_t) \geq \sqrt{1-\tau_1^2}$. Therefore,
	\begin{equation*}
	\begin{split}
	\sigma_{\min}\left(J_{t1}^{(X)}\right) = & \sigma_{\min}\left(K_{t1}^\top J_{\hat{r}_t}^{(X)} L_{t1}\right) \overset{\eqref{eq:def_K_t_L_t}}{=} \sigma_{\min}\left(K_{t1}^\top \mathbb{P}_{\bar{K}_t} J_{\hat{r}_t}^{(X)} \mathbb{P}_{\bar{L}_t} L_{t1}\right)\\
	= & \sigma_{\min}\left(K_{t1}^\top \bar{K}_t\bar{K}_t^\top J_{\hat{r}_t}^{(X)} \bar{L}_t\bar{L}_t^\top L_{t1}\right)\\
	\geq & \sigma_{\min}(K_{t1}^\top \bar{K}_t)\sigma_{\min}\left(\bar{K}_t^\top J_{\hat{r}_t}^{(X)}\bar{L}_t\right) \sigma_{\min}(L_{t1}^\top \bar{L}_t)\\
	\geq & (1-\tau_1^2)\sigma_{r_t}(J_{\hat{r}_t}^{(X)})\\
	\geq & (1-\tau_1^2)(1-\tau^4/(1-2\tau)^2)\sigma_{\min}(X_{\Omega_t \times \Xi_t}).
	\end{split}
	\end{equation*}
	which has finished the proof for \eqref{ineq:sigma_min_J_t1}.
	\item Upper bound for all terms related to perturbation ``$Z$:" for example,
	\begin{equation}\label{ineq:J_t1_Z}
	\|J_{t1}^{(Z)}\| \leq \|Z_{\Omega_t\times \Xi_t}\| \leq \tau \sigma_{\min}(X_{\Omega_t\times \Xi_t}),
	\end{equation}
	\begin{equation*}
	\|A_{t1}^{(Z)}\| \leq \|Z_{\Xi_t}\| \leq \tau \sigma_{\min}(X_{\Xi_t}).
	\end{equation*}
	\begin{equation}\label{ineq:B^Z}
	\left\|\B_{s_1s_2s_3}^{(Z)}\right\|_{\rm HS} \leq \|\Z_{[\Omega_1, \Omega_2, \Omega_3]}\|_{\rm HS}, \quad \left\|\B_{s_1s_2s_3}^{(Z)}\right\|_{\rm op} \leq \|\Z_{[\Omega_1, \Omega_2, \Omega_3]}\|_{\rm op}.
	\end{equation}
	Since all these terms related to perturbation ``$(Z)$" are essentially projections of $Z_{\Xi_t}$, $Z_{\Omega_t\times \Xi_t}$, $Z_{t, \Omega}$ or $\Z_{[\Omega_1, \Omega_2, \Omega_3]}$, they can be derived easily.
	\item Upper bounds in spectral norm for ``arm $\cdot$ joint$^{-1}$" and ``joint$^{-1}$ $\cdot$ body":
	\begin{equation}\label{ineq:A_t1^-1J_t1}
	\left\|A_{t1}^{(X)} (J_{t1}^{(X)})^{-1}\right\| \leq \bar{\lambda}_t; 
	\end{equation}
	\begin{equation}\label{ineq:J_t1^-1B_111}
	\text{if } (s_1, s_2, s_3)\in \{1, 2\}^3, t \in \{1, 2, 3\}, \quad \left\|(J^{(X)}_{ts_t})^{-1} \mathcal{M}_t(\B_{s_1s_2s_3}^{(X)})\right\| \leq \xi_t;
	\end{equation}
	\begin{equation}\label{ineq:A^X+A^Z_J^X+J^Z}
	\left\|A_{t1} \left(J_{t1}\right)^{-1}\right\| \leq \bar\lambda_t + \frac{2\bar\lambda_t \tau}{(1-\tau_1^2)(1-\tau^4/(1-2\tau)^2) - \tau};
	\end{equation}	
	\begin{equation}\label{ineq:A_t2_J^X_t2^-1}
	\left\|A_{t2}\left(J_{t2}\right)^{-1}\right\| \leq \lambda_t + \bar{\lambda}_t + \frac{2\bar{\lambda}_t\tau}{(1-\tau_1^2)(1-\tau^4/(1-2\tau)^2) - \tau}.
	\end{equation}
	Recall \eqref{eq:Q-U-X_Xi-association} that $X_{\Xi_t} = U_tQ$, $X_{\Omega_t\times \Xi_t} = U_{t, \Omega}Q$, the definition for $A_{t1}^{(X)}$, $J_{t1}^{(X)}$ and the fact that $\sigma_{\min}(J_{t1}^{(X)})>0$, we have
	\begin{equation*}
	\begin{split}
	& \left\|A_{t1}^{(X)}(J_{t1}^{(X)})^{-1} \right\|\\ 
	= & \left\|\left(X_{\Xi_t}V^{(A)}_{t, [1:\hat{r}_t]}L_{t1}\right) \left(K_{t1}^\top U_{t, [:, 1:\hat{r}_t]}^{(B)\top} X_{\Omega_t\times \Xi_t} V_{t, [:, 1:\hat{r}_t]}^{(A)\top}L_{t1}\right)^{-1}\right\|\\
	= &\left\|U_t Q V^{(A)}_{t, [1:\hat{r}_t]}L_{t1} \left(K_{t1}^\top U_{t, [:, 1:\hat{r}_t]}^{(B)\top} U_{t, \Omega} Q V_{t, [:, 1:\hat{r}_t]}^{(A)\top}L_{t1}\right)^{-1}\right\|\\
	= & \left\|U_t Q V^{(A)}_{t, [1:\hat{r}_t]}L_{t1} \left(Q V_{t, [:, 1:\hat{r}_t]}^{(A)\top}L_{t1}\right)^{-1}\left(K_{t1}^\top U_{t, [:, 1:\hat{r}_t]}^{(B)\top} U_{t, \Omega}\right)^{-1}\right\| \\
	= & \frac{1}{\sigma_{\min}\left(K_t^\top U^{(B)\top}_{t,[:, 1:\hat{r}_t]} U_{t, \Omega}\right)} \leq \frac{1}{\sigma_{\min}\left(U_{t, \Omega}\right)} = \bar{\lambda}_t,
	\end{split}
	\end{equation*}
	which has proved \eqref{ineq:A_t1^-1J_t1}. The proof for \eqref{ineq:J_t1^-1B_111} is similar. Since $X_{\Omega_t\times \Xi_t}$ is a collection of columns of $X_{t, \Omega}$ and $\rank(X_{\Omega_t\times \Xi_t}) = \rank(\X_{t, \Omega}) = r_t$, these two matrices share the same column subspace. In this case, 
	\begin{equation*}
	X_{\Omega_t\times \Xi_t} \left(X_{\Omega_t\times \Xi_t}^\dagger X_{t, \Omega}\right) = \mathbb{P}_{X_{\Omega_t\times \Xi_t}} X_{t, \Omega} = X_{t, \Omega}.
	\end{equation*} 
	Given the assumption \eqref{ineq:assumption_arm_body} that $\|X_{\Omega_t\times \Xi_t}^\dagger X_{t, \Omega}\|\leq \xi_t$, the rest of the proof for \eqref{ineq:J_t1^-1B_111} essentially follows from the proof for \eqref{ineq:A_t1^-1J_t1}.
	
	The proof for \eqref{ineq:A^X+A^Z_J^X+J^Z} is relatively more complicated. Note that
	\begin{equation}\label{ineq:J_t1^Z, A_t1^Z leq}
	\begin{split}
	\|J_{t1}^{(Z)}\| \leq & \|Z_{\Omega_t\times \Xi_t}\| \leq \tau\sigma_{\min}(X_{\Omega_t\times \Xi_t}), \\
	\|A_{t1}^{(Z)}\| \leq & \|Z_{\Xi_t}\| \leq \tau \sigma_{r_t}(X_{\Xi_t}) \leq \tau \sigma_{r_t}(Q) = \tau\sigma_{\min}(Q)\\
	\leq & \frac{\tau \sigma_{\min}(U_{t, \Omega}Q)}{\sigma_{\min}(U_{t,\Omega})} \leq \tau\bar{\lambda}_t \sigma_{\min}(X_{\Omega_t\times \Xi_t}), 
	\end{split}
	\end{equation}
	we can calculate that
	\begin{equation*}
	\begin{split}
	& \left\|A_{t1}\left(J_{t1}\right)^{-1}\right\| = \left\|\left(A_{t1}^{(X)}+A_{t1}^{(Z)}\right)\left(J_{t1}^{(X)} + J_{t1}^{(Z)}\right)^{-1}\right\| \\
	\leq & \left\|A_{t1}^{(X)}(J_{t1}^{(X)})^{-1}\right\| +  \left\|A_{t1}^{(X)}\left(\left(J_{t1}^{(X)} + J_{t1}^{(Z)}\right)^{-1} - \left(J_{t1}^{(X)}\right)^{-1}\right)\right\|\\
	& +\left\|A_{t1}^{(Z)}\left(J_{t1}^{(X)} + J_{t1}^{(Z)}\right)^{-1}\right\|\\
	\overset{\text{Lemma 5}}{=} & \left\|A_{t1}^{(X)}(J_{t1}^{(X)})^{-1}\right\| + \left\|A_{t1}^{(X)}\left(J_{t1}^{(X)}\right)^{-1} J_{t1}^{(Z)} \left(J_{t1}^{(X)} + J_{t1}^{(Z)}\right)^{-1}\right\|\\
	& + \left\|A_{t1}^{(Z)}\right\|\sigma_{\min}^{-1}\left(J_{t1}^{(X)} + J_{t1}^{(Z)}\right)\\
	\overset{\eqref{ineq:A_t1^-1J_t1}}{=} & \bar\lambda_t + \bar\lambda_t \frac{\|J_{t1}^{(Z)}\|}{\sigma_{\min}(J_{t1}^{(X)}) - \|J_{t1}^{(Z)}\|} + \frac{\|A_{t1}^{(Z)}\|}{\sigma_{\min}(J_{t1}^{(X)}) - \|J_{t1}^{(Z)}\|}\\
	\overset{\eqref{ineq:J_t1^Z, A_t1^Z leq}\eqref{ineq:sigma_min_J_t1}}{\leq} & \bar\lambda_t + \frac{2\bar\lambda_t \tau}{(1-\tau_1^2)(1-\tau^4/(1-2\tau)^2) - \tau}.
	\end{split}
	\end{equation*}
	This has finished the proof for \eqref{ineq:A^X+A^Z_J^X+J^Z}. 
	
	Next, we move on to \eqref{ineq:A_t2_J^X_t2^-1}. Recall the definitions of $A_{\hat{r}_t}$ and $J_{\hat{r}_t}$, and the fact that $K_{t1}, K_{t2}$; $L_{t1}, L_{t2}$ are orthogonal, we have $K_{t1}^\top K_{t1} = L_{t1}^\top L_{t1} = I$, $K_{t2}^\top K_{t2} = L_{t2}^\top L_{t2} = I$, $K_{t1}^\top K_{t2} = L_{t1}^\top L_{t2} = 0$. Then,
	\begin{equation*}
	\begin{split}
	& A_{\hat{r}_t} J_{\hat{r}_t}^{-1}\\
	= & \left(A_{\hat{r}_t}L_{t1}L_{t1}^\top +A_{\hat{r}_t}L_{t2}L_{t2}^\top\right)\left(K_{t1}K_{t1}^\top J_{\hat{r}_t}L_{t1}L_{t1}^\top + K_{t1}K_{t1}^\top J_{\hat{r}_t}L_{t2}L_{t2}^\top \right)^{-1}\\
	= & \left(A_{t1} L_{t1}^\top + A_{t2}L_{t2}^\top\right)\left(K_{t1} J_{t1} L_{t1}^\top + K_{t2}J_{t2} L_{t2}\right)^{-1}\\
	= & \left(A_{t1} L_{t1}^\top + A_{t2}L_{t2}^\top\right)\left(L_{t1} J_{t1}^{-1} K_{t1}^\top + L_{t2}J_{t2}^{-1} K_{t2}\right)\\
	= & A_{t1}J_{t1}^{-1}K_{t1}^\top + A_{21}J_{t2}^{-1}K_{t2}^\top.
	\end{split}
	\end{equation*}
	On the other hand, $\left\| A_{\hat{r}_t}J_{\hat{r}_t}^{-1}\right\| \leq \lambda_t$. Thus, \eqref{ineq:A_t2_J^X_t2^-1} can shown by
	\begin{equation*}
	\begin{split}
	& \left\| A_{t2} \left(J_{t2}\right)^{-1}\right\| \leq \left\|A_{\hat{r}_t}J_{\hat{r}_t}^{-1}\right\| + \left\|A_{t1} J_{t1}^{-1} K_{t1}^\top\right\| \\
	\leq & \left\|A_{\hat{r}_t}J_{\hat{r}_t}^{-1}\right\| + \left\|A_{t1}J_{t1}^{-1}\right\| \leq \lambda_t + \bar{\lambda}_t + \frac{2\bar{\lambda}_t\tau}{(1-\tau_1^2)(1-\tau^4/(1-2\tau)^2) - \tau}.
	\end{split}
	\end{equation*}
	\item Upper bounds in Frobenius and operator norm for ``arm$\cdot$(joint)$^{-1}$$\cdot$body:" for the major part:
	\begin{equation}\label{ineq:arm-joint-body-1}
	\begin{split}
	& \left\|A_{t1}\left(J_{t1}\right)^{-1} \mathcal{M}_t(B_{111}^{(X)}) - A_{t1}^{(X)}\left(J_{t1}^{(X)}\right)^{-1} \mathcal{M}_t(B_{ 111}^{(X)})\right\|_F\\
	\leq & C\bar{\lambda}_t\xi_t\|Z_{\Omega_t\times \Xi_t}\|_F + C\xi_t\|Z_{\Xi_t}\|_F.
	\end{split}
	\end{equation}
	\begin{equation}\label{ineq:arm-joint-body-2}
	\begin{split}
	& \left\|A_{t1}\left(J_{t1}\right)^{-1} \mathcal{M}_t(B_{111}^{(X)}) - A_{t1}^{(X)}\left(J_{t1}^{(X)}\right)^{-1} \mathcal{M}_t(B_{ 111}^{(X)})\right\|\\
	\leq & C\bar{\lambda}_t\xi_t\|Z_{\Omega_t\times \Xi_t}\| + C\xi_t\|Z_{\Xi_t}\|.
	\end{split}
	\end{equation}
	In order to show these two results, we need to use Lemma 5 and \cite{zhang2016semi}, which is an expansion formula for inverse matrix. Actually,
	\begin{equation*}
	\begin{split}
	& \left\|A_{t1}\left(J_{t1}\right)^{-1} \mathcal{M}_t(B_{111}^{(X)}) - A_{t1}^{(X)}\left(J_{t1}^{(X)}\right)^{-1} \mathcal{M}_t(B_{ 111}^{(X)})\right\|_F \\
	= & \Big\|(A_{t1}^{(X)}+A_{t1}^{(Z)})\left(J_{t1}^{(X)} + J_{t1}^{(Z)}\right)^{-1} \mathcal{M}_t(B_{111}^{(X)})\\
	&\quad- A_{t1}^{(X)}\left(J_{t1}^{(X)}\right)^{-1} \mathcal{M}_t(B_{ 111}^{(X)})\Big\|_F\\
	\leq & \left\|A_{t1}^{(X)}\left(\left(J_{t1}^{(X)} + J_{t1}^{(Z)}\right)^{-1} - \left(J_{t1}^{(X)}\right)^{-1}\right) \mathcal{M}_t(B_{111}^{(X)})\right\|_F\\
	& + \left\|A_{t1}^{(Z)}\left(J_{t1}^{(X)} + J_{t1}^{(Z)}\right)^{-1} \mathcal{M}_t(B_{111}^{(X)})\right\|_F\\
	\overset{\text{Lemma 5}}{\leq} & \Big\|A_{t1}^{(X)}\left(J_{t1}^{(X)}\right)^{-1}  \left(J_{t1}^{(Z)} - J_{t1}^{(Z)}\left(J_{t1}^{(X)} + J_{t1}^{(Z)}\right)^{-1}J_{t1}^{(Z)}\right)\\
	& \quad\quad \cdot \left(J_{t1}^{(X)}\right)^{-1}\mathcal{M}_t(B_{111}^{(X)})\Big\|_F\\
	& + \left\|A_{t1}^{(Z)}\left(-\left(J_{t1}^{(X)} + J_{t1}^{(Z)}\right)^{-1}J_{t1}^{(Z)} + I\right)\left(J_{t1}^{(X)}\right)^{-1} \mathcal{M}_t(B_{111}^{(X)})\right\|_F\\
	\overset{\eqref{ineq:A_t1^-1J_t1}\eqref{ineq:J_t1^-1B_111}}{\leq} & C\bar{\lambda}_t\xi_t\left\|J_{t1}^{(Z)} - J_{t1}^{(Z)}\left(J_{t1}^{(X)} + J_{t1}^{(Z)}\right)^{-1}J_{t1}^{(Z)}\right\|_F\\
	& + C\xi_t\left\|A_{t1}^{(Z)}\right\|_F \cdot \left\|-\left(J_{t1}^{(X)} + J_{t1}^{(Z)}\right)^{-1}J_{t1}^{(Z)} + I\right\|\\
	\overset{\eqref{ineq:J_t1_Z}\eqref{ineq:sigma_min_J_t1}}{\leq} & C\bar{\lambda}_t\xi_t \|J_{t1}^{(Z)}\|_F + C\xi_t\|A_{t1}^{(Z)}\|_F \leq C\bar{\lambda}_t\xi_t \|Z_{\Omega_t\times\Xi_t}\|_F + C\xi_t\|Z_{\Xi_t}\|_F.
	\end{split}
	\end{equation*}
	which has proved \eqref{ineq:arm-joint-body-1}. The proof for \eqref{ineq:arm-joint-body-2} essentially follows from the proof for \eqref{ineq:arm-joint-body-1} when we replace the Frobenius norms with the spectral norm. 
	\item Upper bounds for minor ``body" part:
	\begin{equation}\label{ineq:B_s1s2s3}
	\begin{split}
	& \text{if } (s_1, s_2, s_3)\in \{1, 2\}^3 \text{ and } s_t = 2, \quad \text{then }\\ 
	& \left\{\begin{array}{ll}
	\left\|\B_{s_1s_2s_3}\right\|_{\rm HS} \leq 2\xi_t\left\|Z_{\Omega_t\times \Xi_t}\right\|_F + \left\|\Z_{[\Omega_1, \Omega_2, \Omega_3]}\right\|_{\rm HS},\\
	\left\|\B_{s_1s_2s_3}\right\|_{\rm op} \leq 2\xi_t\left\|Z_{\Omega_t\times \Xi_t}\right\| + \left\|\Z_{[\Omega_1, \Omega_2, \Omega_3]}\right\|_{\rm op}.
	\end{array}\right.
	\end{split}
	\end{equation}
	Instead of considering the ``body" part above directly, we detour and discuss the ``joint" part first. It is noteworthy that $J_{t2}$ is the $r_t+1, \ldots, \hat{r}_t$-th principle components for $J_{\hat{r}_t} = J_{\hat{r}_t}^{(X)} + J_{\hat{r}_t}^{(Z)}$. As $\rank(J_{\hat{r}_t}^{(X)}) = r_t$, by Lemma 1 in \cite{cai2016structured} (which provides inequalities of singular values in the low-rank perturbed matrix), 
	\begin{equation*}
	\begin{split}
	& \sigma_{i}(J_{t2}) = \sigma_{r_t+i}(J_{\hat{r}_t}) \leq \sigma_i(J_{\hat{r}_t}^{(Z)}), \quad \forall 1\leq i \leq \hat{r}_t - r_t \\ 
	\Rightarrow\quad & \|J_{t2}\|_F = \sqrt{\sum_{i=1}^{\hat{r}_t - r_t} \sigma_i^2(J_{t2})} \leq \sqrt{\sum_{i=1}^{\hat{r}_t - r_t} \sigma_i^2(J_{\hat{r}_t}^{(Z)})}\\
	&\quad\quad\quad = \|J_{\hat{r}_t}^{(Z)}\|_F \leq \|Z_{\Omega_t\times \Xi_t}\|_F,\\
	\Rightarrow \quad & \|J_{t2}^{(X)}\|_F \leq \|J_{t2}\|_F + \|J_{t2}^{(Z)}\|_F \leq 2\|Z_{\Omega_t\times \Xi_t}\|_F,\\
	\overset{\eqref{ineq:J_t1^-1B_111}}{\Rightarrow} \quad & \left\|\mathcal{M}_t(\B^{(X)}_{s_1s_2s_3})\right\|_F \leq \left\|J_{t2}^{(X)}\right\|_F \left\|\left(J_{t2}^{(X)}\right)^{-1}\mathcal{M}_t(\B^{(X)}_{s_1s_2s_3})\right\|\\
	& \quad\quad\quad\quad \quad\quad\quad \leq 2\xi_t \|Z_{\Omega_t\times \Xi_t}\|_F\\
	{\Rightarrow} \quad & \|\B_{s_1s_2s_3}\|_{\rm HS} \leq \left\|\B_{s_1s_2s_3}^{(X)}\right\|_{\rm HS} + \left\|\B_{s_1s_2s_3}^{(Z)}\right\|_{\rm HS}\\ 
	& \quad\quad\quad\quad\quad \leq \|\mathcal{M}_t(\B_{s_1s_2s_3})\|_F + \|\Z_{[\Omega_1, \Omega_2, \Omega_3]}\|_{\rm HS}\\
	& \quad\quad\quad\quad\quad \leq 2\xi_t\|Z_{\Omega_t\times \Xi_t}\|_F + \|\Z_{[\Omega_1, \Omega_2, \Omega_3]}\|_{\rm HS}.
	\end{split}
	\end{equation*}
	We can similarly derive that $\|\B_{s_1s_2s_3}\|_{\rm op} \leq 2\xi_t\|Z_{\Omega_t\times \Xi_t}\| + \|\Z_{[\Omega_1, \Omega_2, \Omega_3]}\|_{\rm op}$, which has proved \eqref{ineq:B_s1s2s3}.
	
	\item Equality for the original tensor $\X$:
	\begin{equation}\label{eq:X=A_11^X...}
	\X = \B_{111}^{(X)} \times_1 A_{11}^{(X)}\left(J_{11}^{(X)}\right)^{-1} \times_2 A_{21}^{(X)}\left(J_{21}^{(X)}\right)^{-1} \times_3 A_{31}^{(X)}\left(J_{31}^{(X)}\right)^{-1}.
	\end{equation}
	In fact, according to the definitions \eqref{eq:A_ts^X^Z} - \eqref{eq:B_s1s2s3}, 
	\begin{equation*}
	\begin{split}
	& \B_{111}^{(X)}\times_1 A_{11}^{(X)} \times_2 A_{21}^{(X)} \times_3 A_{31}^{(X)} \\
	= & \X_{[\Omega_1, \Omega_2, \Omega_3]}  \times_{t=1}^3 \Bigg\{ X_{\Xi_t} \left(V_{t, [:, 1:\hat{r}_t]}^{(A)} L_{t1} \right)\\  
	& \quad\quad \quad\cdot \left(\left(U^{(B)}_{t,[:, 1:\hat{r}_t]}K_{t1}\right)^\top X_{\Omega_t\times \Xi_t} \left(V_{t, [:, 1:\hat{r}_t]}^{(A)} L_{t1} \right)\right)^{-1}\left(U^{(B)}_{1,[:, 1:\hat{r}_1]}K_{11}\right)^\top \Bigg\}\\
	\end{split}
	\end{equation*}
	Based on \eqref{ineq:sigma_min_J_t1}, $J_{t1}^{(X)} = K_{t1}^\top (U^{(B)}_{1, [:, 1:\hat{r}_t]})^\top X_{\Omega_t\times \Xi_t} V_{t, [:, 1:\hat{r}_t]}^{(A)} L_{t1}$ is non-singular, then by Theorem \ref{th:noiseless}, we can show the term above equals $\X$,  which has proved \eqref{eq:X=A_11^X...}.
\end{itemize}

\ \par

\noindent (\emph{Step 5.})  Now we are ready to analyze the estimation error of $\hat{\X}$ based on all the preparations in the previous steps. Based on the decompositions of $\hat{\X}$ \eqref{eq:decomposition_hat_X} and $\X$ \eqref{eq:X=A_11^X...}, one has
\begin{equation}
\begin{split}
& \left\|\hat{\X}- \X\right\|_{\rm HS}\\
\leq & \Big\| \B_{111} \times_1 A_{11}J_{11}^{-1}\times_2 A_{21}J_{21}^{-1} \times_3 A_{31}J_{31}^{-1}\\
& -  \B_{111}^{(X)} \times_1 A_{11}^{(X)}\left(J_{11}^{(X)}\right)^{-1} \times_2 A_{21}^{(X)}\left(J_{21}^{(X)}\right)^{-1} \times_3 A_{31}^{(X)}\left(J_{31}^{(X)}\right)^{-1} \Big\|_{\rm HS}\\
& + \sum_{\substack{s_1, s_2, s_3 = 1\\(s_1, s_2, s_3) \neq (1, 1, 1)}}^{2} \left\|\B_{s_1s_2s_3}\times_1 A_{1s_1}J_{1s_1}^{-1}\times_2 A_{2s_2}J_{2s_2}^{-1} \times_3 A_{3s_3} J_{3s_3}^{-1}\right\|_{\rm HS}\\
\leq & \left\|\left(\B_{111} - \B_{111}^{(X)}\right)\times_1 A_{11}J_{11}^{-1}\times_2 A_{21}J_{21}^{-1} \times_3 A_{31}J_{31}^{-1}\right\|_{\rm HS}\\
& + \left\|\B_{111}^{(X)}\times_1 A_{11}J_{11}^{-1}\times_2 A_{21}J_{21}^{-1} \times_3 \left(A_{31}J_{31}^{-1} - A_{31}^{(X)} (J_{31}^{(X)})^{-1}\right) \right\|_{\rm HS}\\
& + \left\|\B_{111}^{(X)} \times_1 A_{11}J_{11}^{-1}\times_2 \left(A_{21}J_{21}^{-1} - A_{21}^{(X)}J_{21}^{(X)-1}\right) \times_3 A_{31}^{(X)}J_{31}^{(X)-1}\right\|_{\rm HS}\\
& + \left\|\B_{111}^{(X)}\times_1 \left(A_{11}J_{11}^{-1} - A_{11}^{(X)}J_{11}^{(X)-1}\right)\times_2 A_{21}^{(X)}J_{21}^{(X)-1} \times_3 A_{31}^{(X)}J_{31}^{(X)-1} \right\|_{\rm HS}\\
& + \sum_{\substack{s_1, s_2, s_3 = 1\\(s_1, s_2, s_3) \neq (1, 1, 1)}}^{2} \left\|\B_{s_1s_2s_3}\times_1 A_{1s_1}J_{1s_1}^{-1}\times_2 A_{2s_2}J_{2s_2}^{-1} \times_3 A_{3s_3} J_{3s_3}^{-1}\right\|_{\rm HS}.
\end{split}
\end{equation}
For each term separately above, we have
\begin{equation*}
\begin{split}
&\quad\quad\ \left\|\left(\B_{111} - \B_{111}^{(X)}\right)\times_1 A_{11}J_{11}^{-1}\times_2 A_{21}J_{21}^{-1} \times_3 A_{31}J_{31}^{-1} \right\|_{\rm HS}\\
&~~\quad\leq \left\|A_{11}J_{11}^{-1}\right\|\cdot \left\|A_{21}J_{21}^{-1}\right\|\cdot \left\|A_{31}J_{31}^{-1}\right\| \left\| \B_{111}^{(Z)}\right\|_{\rm HS}\\
&\overset{\eqref{ineq:B^Z}\eqref{ineq:A_t1^-1J_t1}}{\leq}  C\lambda_1\lambda_2\lambda_3 \|\Z_{[\Omega_1, \Omega_2, \Omega_3]}\|_{\rm HS}.
\end{split}
\end{equation*}
\begin{equation*}
\begin{split}
& \left\|\B_{111}^{(X)}\times_1 A_{11}J_{11}^{-1} \times_2 A_{21}J_{21}^{-1} \times_3 \left(A_{31}J_{31}^{-1} - A_{31}^{(X)}(J_{31}^{(X)})^{-1}\right)\right\|_{\rm HS} \\
& \leq \left\|A_{11}J_{11}^{-1} \right\| \cdot \left\|A_{21}J_{21}^{-1}\right\| \cdot \left\|\left(A_{31}J_{31}^{-1} - A_{31}^{(X)}(J_{31}^{(X)})^{-1}\right)\mathcal{M}_3(\B^{(X)}_{111})\right\|_F\\
& \overset{\eqref{ineq:arm-joint-body-1}\eqref{ineq:A_t1^-1J_t1}}{\leq} C\lambda_1\lambda_2\lambda_3\xi_3 \|Z_{\Omega_3\times\Xi_3}\|_F + C\lambda_1\lambda_2\xi_3 \|Z_{\Xi_3}\|_F,
\end{split}
\end{equation*}
\begin{equation*}
\begin{split}
& \left\|\B_{111}^{(X)} \times_1 A_{11}J_{11}^{-1}\times_2 \left(A_{21}J_{21}^{-1} - A_{21}^{(X)}J_{21}^{(X)-1}\right) \times_3 A_{31}^{(X)}J_{31}^{(X)-1}\right\|_{\rm HS} \\
&\leq \left\|A_{11}J_{11}^{-1}\right\| \cdot \left\|A_{31}^{(X)}(J_{31}^{(X)})^{-1}\right\|\cdot \left\|\left(A_{21}J_{21}^{-1} - A_{21}^{(X)}(J_{21}^{(X)})^{-1}\right)\mathcal{M}_2(\B_{111}^{(X)})\right\|_F\\
& \overset{\eqref{ineq:A^X+A^Z_J^X+J^Z}\eqref{ineq:arm-joint-body-1}\eqref{ineq:A_t1^-1J_t1}}{\leq} C\lambda_1\lambda_2\lambda_3 \xi_2\|Z_{\Omega_2\times \Xi_2} \|_F + C\lambda_1\lambda_3 \xi_2 \|Z_{\Xi_2}\|_F,
\end{split}
\end{equation*}
\begin{equation*}
\begin{split}
& \left\|\B_{111}^{(X)}\times_1 \left(A_{11}J_{11}^{-1} - A_{11}^{(X)}J_{11}^{(X)-1}\right)\times_2 A_{21}^{(X)}J_{21}^{(X)-1} \times_3 A_{31}^{(X)}J_{31}^{(X)-1} \right\|_{\rm HS}\\
\leq & \left\|A_{21}^{(X)}(J_{21}^{(X)})^{-1}\right\| \cdot \left\|A_{31}^{(X)}(J_{31}^{(X)})^{-1}\right\|\\
& \cdot \left\|\left(A_{11}J_{11}^{-1} - A_{11}^{(X)}(J_{11}^{(X)})^{-1}\right)\mathcal{M}_1(\B_{111}^{(X)})\right\|_F\\
& \overset{\eqref{ineq:A^X+A^Z_J^X+J^Z}\eqref{ineq:arm-joint-body-1}\eqref{ineq:A_t1^-1J_t1}}{\leq} C\lambda_1\lambda_2\lambda_3 \xi_1\|Z_{\Omega_1\times \Xi_1} \|_F + C\lambda_2\lambda_3 \xi_1 \|Z_{\Xi_1}\|_F.
\end{split}
\end{equation*}
Last but not least, for any $s_1, s_2, s_3\in \{1, 2\}^3$ such that $(s_1, s_2, s_3)\neq (1, 1, 1,)$, let us specify that $s_t = 2$ for some $1\leq t\leq 1$. Then
\begin{equation*}
\begin{split}
& \left\|\B_{s_1s_2s_3} \times_1 A_{1s_1}J_{1s_1}^{-1}\times_2 A_{2s_2}J_{2s_2}^{-1} \times_3 A_{3s_3} J_{3s_3}^{-1}\right\|_{\rm HS}\\
\leq & \left\|A_{1s_1}J_{1s_1}^{-1}\right\| \cdot \left\|A_{2s_2}J_{2s_2}^{-1}\right\| \cdot \left\|A_{2s_2}J_{1s_1}^{-1}\right\|  \cdot \left\|\B_{s_1s_2s_3}\right\|_{\rm HS}\\
\overset{\eqref{ineq:A_t1^-1J_t1}\eqref{ineq:A_t2_J^X_t2^-1}\eqref{ineq:B_s1s2s3}}{\leq} & C\lambda_1\lambda_2\lambda_3 \left(\xi_t\|Z_{\Omega_t\times \Xi_t}\|_F + \|\Z_{[\Omega_1, \Omega_2, \Omega_3]}\|_{\rm HS}\right).
\end{split}
\end{equation*}

Combing all terms above, we have proved the targeted upper bound for $\|\hat{\X}-\X\|_{\rm HS}$.
By similar argument, we can show the upper bound for $\|\hat{\X}-\X\|_{\rm op}$. Therefore, we have finished the proof for Theorem \ref{th:noisy}. \quad $\square$

\subsection{Proof of Proposition \ref{pr:degree_of_freedom}} 
\begin{figure}
	\begin{center}
		\includegraphics[height = 1.4in, width=1.2in]{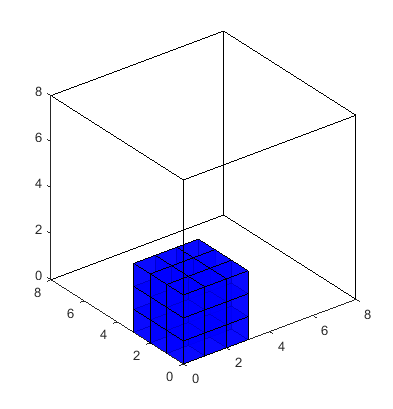} 
		\includegraphics[height = 1.4in, width=1.2in]{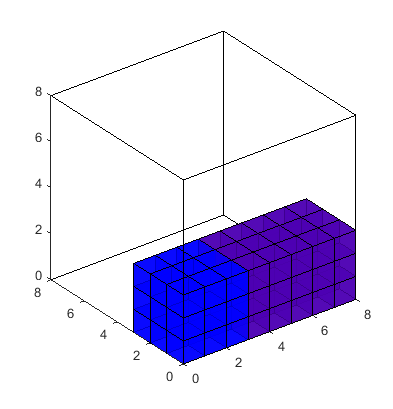}
		\includegraphics[height = 1.4in, width=1.2in]{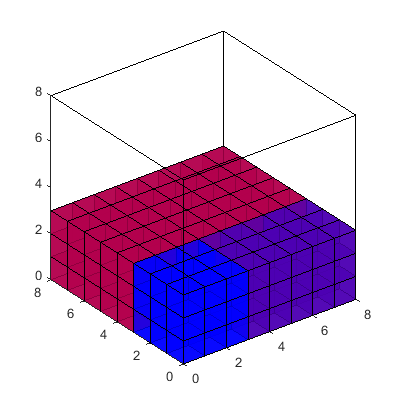}
		\includegraphics[height = 1.4in, width=1.2in]{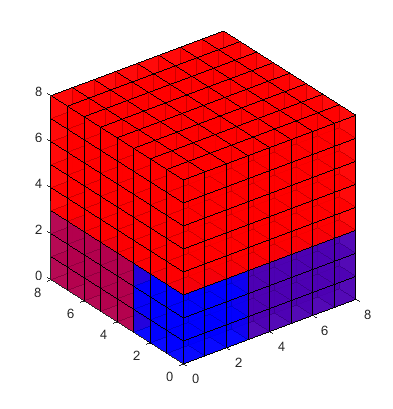}
		\caption{Illustrative example with $p_1 =p_2 = p_3 = 8, r_1 = r_2 = r_3 = 3$ for four steps in Proposition \ref{pr:degree_of_freedom}. }\label{fig:degree_of_freedom}\vspace{-.2in}
	\end{center}
\end{figure}

In order to calculate the degrees of freedom for rank-$(r_1, r_2, r_3)$ tensors in $\mathbb{R}^{p_1\times p_2\times p_3}$, we consider the following process to generate such tensors. A pictorial illustration of the whole process is provided in Figure \ref{fig:degree_of_freedom}.
\begin{enumerate}[label=(\alph*)]
	\item First, the top corner $\X_{[1:r_1, 1:r_2, 1:r_3]}$ is free to choose all values, which includes $r_1r_2r_3$ degrees of freedom (Panel (a)). 
	\item After $\X_{[1:r_1, 1:r_2, 1:r_3]}$ is set up, the following $p_1 - r_1$ slices $\X_{[r_1+1, 1:r_2, 1:r_3]},\ldots, \X_{[p_1, 1:r_2, 1:r_3]}$ are the linear combinations of $r_1$ slices -- $\X_{[1, 1:r_2, 1:r_3]}, \ldots, X_{[r_1, 1:r_2, 1:r_3]}$, which contributes $r_1(p_1 - r_1)$ degrees of freedom. (Panel (b))
	\item Next, the $p_2 - r_2$ slices $\X_{[1:p_1, r_2+1, 1:r_3]},\ldots, \X_{[1:p_1, p_2, 1:r_3]}$ are the linear combination of $r_2$ slices -- $\X_{[1, 1:r_2, 1:r_3]}, \ldots, X_{[r_1, 1:r_2, 1:r_3]}$, which means there are $r_2(p_2 - r_2)$ degrees of freedom. (Panel (c))
	\item Finally, the rest of the undetermined block can be divided into $p_3 - r_3$ slices:
	$$\X_{[1:p_1, 1:p_2, r_3+1]}, \ldots, \X_{[1:p_1, 1:p_2, p_3]}.$$ 
	According to the low-rank assumption, these slices are linear combinations of 
	$$\X_{[1:p_1, 1:p_2, r_3+1]}, \ldots, \X_{[1:p_1, 1:p_2, p_3]}.$$ 
	Then there are $r_3(p_3-r_3)$ degrees of freedom in the selection. (Panel (d))
\end{enumerate}
To sum up, the total number of degrees of freedom of rank-$(r_1, r_2, r_3)$ tensors in $\mathbb{R}^{p_1\times p_2\times p_3}$ is $r_1r_2r_3 + (p_1-r_1)r_1 + (p_2-r_2)r_2 + (p_3 -r_3)r_3$. \quad $\square$

\subsection{Proof of Theorem \ref{th:lower_bound}} 
The idea for proving this theorem is to construct two pairs of tensors $(\X_1, \Z_1)$ and $(\X_2, \Z_2)$ such that they both satisfy the criterion and share the same values in the observed indices, while retaining different values in the others. This is characterized by the following lemma. 
\begin{Lemma}\label{lm:lower_bound_difference}
	Suppose $\mathcal{G}$ is any collection of tuples $(\X, \Z, \Omega_t, \Xi_t)$.
	If there exists $\Omega_t, \Xi_t$ and two pairs of tensors $(\X_1, \Z_1)$ and $(\X_2, \Z_2)$ such that 
	$$(\X_1, \Z_1, \Omega_t, \Xi_t), (\X_2, \Z_2, \Omega_t, \Xi_t) \in \mathcal{G},$$
	$$\left(\X_1 + \Z_1\right)_{\bOmega} = \left(\X_2 + \Z_2\right)_{\bOmega},\quad \text{where $\bOmega$ represents the observable entries;}$$ 
	then for any tensor norms $\|\cdot\|$,
	\begin{equation*}
	\inf_{\hat{\X}}\sup_{(\X, \Z, \Omega_t, \Xi_t)\in\mathcal{G}}\left\|\hat{\X} - \X \right\| \geq \frac{1}{2}\|\X_1 - \X_2\|.
	\end{equation*}
\end{Lemma}
The proof for Lemma \ref{lm:lower_bound_difference} is provided later. For the proof of Theorem \ref{th:lower_bound}, let $r = \min\{r_1, r_2, r_3\}$, we focus on the scenario that $r$ is a even number first. The case for odd number is slightly more complicated but essentially follows if we replace $r_t / 2$ by $\lfloor r_t/2\rfloor$. For convenience, we also treat the indices of $t$ as mod-3, e.g., $\Omega_4 = \Omega_1, \Xi_5 = \Xi_2$, etc.
\begin{enumerate}[leftmargin=*]
	\item We first set $\Omega_t = [1:r]$, $\Xi_t = \left[1: (r/2) \right]\times \left[(r/2 + 1): r\right] \subseteq \Omega_{t+1}\times \Omega_{t+2}$. It is easy to see that the following sets have no overlap: 
	\begin{equation*}
	\begin{split}
	& \Omega_1 \times \Xi_1 = \{(i, j, k): i\in \Omega_1, (j, k)\in \Xi_1\},\\
	& \Omega_2 \times \Xi_2 = \{(i, j, k): j\in \Omega_2, (k, i)\in \Xi_2\},\\
	& \Omega_3 \times \Xi_3 = \{(i, j, k): k\in \Omega_3, (i, j)\in \Xi_3\},\\
	& \left[1:\frac{r}{2}, 1:\frac{r}{2}, 1:\frac{r}{2}\right], \quad \left[\left(\frac{r}{2}+1\right) : r, \left(\frac{r}{2}+1\right):r, \left(\frac{r}{2}+1\right): r\right].
	\end{split}
	\end{equation*}	
	
	\item In this step, we construct a full rank core tensor $\S \in \mathbb{R}^{r\times r\times r}$ with the following procedure.	We first construct $\A, \B\in \mathbb{R}^{(r/2)\times (r/2)\times (r/2)}$,
	$$\A_{ijk} = \left\{\begin{array}{ll}
	1, & i=j=k,\\
	0, & \text{otherwise};
	\end{array}\right., \quad \B_{ijk} = \left\{\begin{array}{ll}
	1, & i = (j-1) = (k - 2) \text{ Mod } (r/2),\\
	0, & \text{otherwise}.
	\end{array}\right. $$ 
	Then we partition $\S$ into eight parts:
	\begin{equation}\label{eq:S_construction1}
	\S_{s_1s_2s_3} = \S_{\left[\frac{r(s_1-1)}{2} + 1:\frac{rs_1}{2}, \frac{r(s_2-1)}{2} + 1:\frac{rs_2}{2}, \frac{r(s_3-1)}{2} + 1:\frac{rs_3}{2}\right]}, \quad s_1, s_2, s_3\in\{1, 2\},
	\end{equation}
	and assign 
	\begin{equation}\label{eq:S_construction2}
	\begin{split}
	& \S_{111} = \A, \quad \S_{112} = \frac{3}{\xi_1}\B, \quad \S_{121} = \frac{3}{\xi_3}\B, \quad \S_{122} = \frac{3}{\xi_3}\A,\\
	& \S_{211} = \frac{3}{\xi_2}\B, \quad \S_{212} = \frac{3}{\xi_1}\A, \quad \S_{221} = \frac{3}{\xi_2}\A,\quad \S_{222} = \B.
	\end{split}
	\end{equation}
	Now we denote $S_t = \mathcal{M}_t(\S), S_{\Xi_t} = (\mathcal{M}_t(\S))_{[:, \Xi_t]}$. It is not hard to see from the definition of $\S$ that
	$$S_{\Xi_t} = \begin{bmatrix}
	\frac{3}{\xi_t} \mathcal{M}_t(\A)\\
	\frac{3}{\xi_t} \mathcal{M}_t(\B)
	\end{bmatrix} \quad \text{(equality holds may up to row/column permutation),}$$ 
	Given the definition of $\A$ and $\B$, we can see
	\begin{equation}\label{ineq:lower_bound_positive_probablity}
	\sigma_{\min}(S_{\Xi_t}) = \sigma_{\max}(S_{\Xi_t}) = \frac{3}{\xi_t}
	\end{equation}
	holds for $t = 1, 2, 3$. So there exists a construction of $\S$ in the way of \eqref{eq:S_construction1} and \eqref{eq:S_construction2}, which also satisfies \eqref{ineq:lower_bound_positive_probablity}.
	\item In this step, we prove that based on the construction of $\S$ in the last step, $\| S_tS_{\Xi_t}^\dagger\| < \xi_t$. Here,
	$S_t = \mathcal{M}_t(\S)$ is the matricization of $\S$. First, by \eqref{eq:S_construction1}, \eqref{eq:S_construction2}, $S_1$ can be written as (up to row/column permutation)
	\begin{equation*}
	S_1 = \begin{bmatrix}
	\mathcal{M}_1(\A) & \frac{3}{\xi_1}\mathcal{M}_1(\B) & \frac{3}{\xi_3}\mathcal{M}_1(\B) & \frac{3}{\xi_3}\mathcal{M}_1(\A)\\
	\frac{3}{\xi_2}\mathcal{M}_1(\B) & \frac{3}{\xi_1}\mathcal{M}_1(\A) & \frac{3}{\xi_2}\mathcal{M}_1(\A) & \mathcal{M}_1(\B).
	\end{bmatrix}
	\end{equation*}
	Then given $\xi_t \geq 3$, we have
	\begin{equation*}
	\begin{split}
	& \sigma_{\max}\left(S_1\right)\\
	\leq & \left(\left\|\begin{bmatrix}
	\mathcal{M}_1(\A)\\
	\frac{3}{\xi_2}\mathcal{M}_1(\B)
	\end{bmatrix}\right\|^2
	+ \left\|\begin{bmatrix}
	\frac{3}{\xi_1}\mathcal{M}_1(\B)\\
	\frac{3}{\xi_1}\mathcal{M}_1(\A)
	\end{bmatrix}\right\|^2
	+ \left\|\begin{bmatrix}
	\frac{3}{\xi_3}\mathcal{M}_1(\B)\\
	\frac{3}{\xi_2}\mathcal{M}_1(\A)
	\end{bmatrix}\right\|^2
	+ \left\|\begin{bmatrix}
	\frac{3}{\xi_3}\mathcal{M}_1(\A)\\
	\mathcal{M}_1(\B)
	\end{bmatrix}\right\|^2\right)^{1/2}\\
	\leq & 2\left\|\begin{bmatrix}
	\mathcal{M}_1(\A)\\
	\mathcal{M}_1(\B)
	\end{bmatrix}\right\| = 2\cdot \frac{\xi_1}{3} \left\|S_{\Xi_1}\right\| \overset{\eqref{ineq:lower_bound_positive_probablity}}{\leq} 2.
	\end{split}
	\end{equation*}
	Similar results also hold for $S_2, S_3$. As a consequence, 
	\begin{equation*}
	\begin{split}
	\left\|S_{\Xi_t}^\dagger S_t\right\| \leq & \sigma_{\min}^{-1}(S_{\Xi_t})\sigma_{\max}(S_t) \leq \frac{ 2}{\frac{3}{\xi_t}} < \xi_t.
	\end{split}
	\end{equation*}	
	\item In this step we construct $\X_1$ and $\Z_1$. Let
	\begin{equation*}
	E_t = \begin{blockarray}{cc}
	& r\\
	\begin{block}{c[c]}
	r &  I_{r}\\
	r  & \sqrt{\lambda_t^2/2 - 1}\cdot I_{r}\\
	p_t - 2r & 0\\
	\end{block}
	\end{blockarray}, \quad t = 1, 2, 3,
	\end{equation*}
	and
	\begin{equation*}
	\X_1 = T\cdot \S \times_1 E_1 \times_2 E_2 \times_3 E_3 \in \mathbb{R}^{p_1\times p_2\times p_3}, \quad \Z_1 = 0 \in \mathbb{R}^{p_1\times p_2\times p_3}.
	\end{equation*}
	Here $T>0$ is a large constant to be determined later. Then,
	\begin{equation}
	\sigma_{\min}\left((X_1)_{\Omega_t\times \Xi_t}^\dagger (X_1)_{t, \Omega}\right) = \sigma_{\min}\left(S_{\Xi_t}^\dagger S_{t}\right) < \xi_t;
	\end{equation}
	\begin{equation}
	\sigma_{\max}\left((X_1)_{\Xi_t}(X_1)_{\Omega_t\times \Xi_t}^\dagger\right) = \sigma_{\max}\left(\begin{bmatrix}
	S_{\Xi_t}\\
	\sqrt{\lambda_t^2/2-1}\cdot S_{\Xi_t}
	\end{bmatrix} \cdot S_{\Xi_t}^\dagger\right) = \lambda_t/\sqrt{2} < \lambda_t.
	\end{equation}
	Similarly, $\|\X_{[\Omega_1,\Omega_2, \Omega_3]}\times_t (X_{\Omega_t\times \Xi_t}^\dagger)^\top \|_{\rm op} <\xi_t$.
	In contrast, $\Z_1 = 0$. Therefore, $(\X_1, \Z_1, \Omega_t, \Xi_t)$ satisfies all criterion in \eqref{eq:F_{lambda,Xi}}, \eqref{ineq:Z_upper_bound} and \eqref{ineq:Z_upper_bound_op}.
	
	In order to show \eqref{ineq:lower_bound}, we can separately prove the following inequalities:
	\begin{equation}\label{ineq:lower_bound_C^B}
	\begin{split}
	& \inf_{\hat{\X}}\sup_{\substack{(\X, \Z, \Omega_t, \Xi_t) \in \mathcal{F}\\ \Z \text{ satisfies \eqref{ineq:Z_upper_bound}}}}\left\|\hat{\X} - \X \right\|_{\rm HS} \geq c\lambda_1\lambda_2\lambda_3C^{(B)},\\
	& \inf_{\hat{\X}}\sup_{\substack{(\X, \Z, \Omega_t, \Xi_t) \in \mathcal{F}\\ \Z \text{ satisfies \eqref{ineq:Z_upper_bound_op}}}}\left\|\hat{\X} - \X \right\|_{\rm op} \geq c\lambda_1\lambda_2\lambda_3C^{(B)};
	\end{split}
	\end{equation}
	\begin{equation}\label{ineq:lower_bound_C^A}
	\begin{split}
	& \inf_{\hat{\X}}\sup_{\substack{(\X, \Z, \Omega_t, \Xi_t) \in \mathcal{F}\\ \Z \text{ satisfies \eqref{ineq:Z_upper_bound}}}}\left\|\hat{\X} - \X \right\|_{\rm HS} \geq  c\lambda_1\lambda_2\lambda_3\frac{\xi_t}{\lambda_t}C_t^{(A)},\\
	& \inf_{\hat{\X}}\sup_{\substack{(\X, \Z, \Omega_t, \Xi_t) \in \mathcal{F}\\ \Z \text{ satisfies \eqref{ineq:Z_upper_bound_op}}}}\left\|\hat{\X} - \X \right\|_{\rm op} \geq  c\lambda_1\lambda_2\lambda_3\frac{\xi_t}{\lambda_t}C_t^{(A)};
	\end{split}
	\end{equation}
	\begin{equation}\label{ineq:lower_bound_C^J}
	\begin{split}
	& \inf_{\hat{\X}}\sup_{\substack{(\X, \Z, \Omega_t, \Xi_t) \in \mathcal{F}\\ \Z \text{ satisfies \eqref{ineq:Z_upper_bound}}}}\left\|\hat{\X} - \X \right\|_{\rm HS} \geq  c\lambda_1\lambda_2\lambda_3\xi_tC^{(J)}_t,\\
	& \inf_{\hat{\X}}\sup_{\substack{(\X, \Z, \Omega_t, \Xi_t) \in \mathcal{F}\\ \Z \text{ satisfies \eqref{ineq:Z_upper_bound_op}}}}\left\|\hat{\X} - \X \right\|_{\rm op} \geq  c\lambda_1\lambda_2\lambda_3\xi_tC^{(J)}_t.
	\end{split}
	\end{equation}
	In the next three steps, we prove \eqref{ineq:lower_bound_C^B}, \eqref{ineq:lower_bound_C^A} and \eqref{ineq:lower_bound_C^J}, respectively. It is noteworthy that the construction of $(\X_2, \Z_2)$ is different for each of the three scenarios.
	\item We prove \eqref{ineq:lower_bound_C^B} in this step, while we focus on the first part as the second part essentially follows. Let
	\begin{equation*}
	\S' \in \mathbb{R}^{r_1\times r_2\times r_3}, \quad S_{ijk}' = \left\{\begin{array}{ll}
	(1+\varepsilon) \S_{ijk}, & i \in [1: r_1/2], j\in [1: r_2/2], k\in [1: r_3/2];\\
	\S_{ijk}, & \text{otherwise}
	\end{array}\right.
	\end{equation*}
	where $\varepsilon = C^{(B)}/(T\|S_{[1:r_1/2, 1:r_2/2, 1:r_3/2]}\|_{\rm HS})$. In other words, $\S'$ and $\S$ are only slightly different in the top  $[1:r_1/2, 1:r_2/2, 1:r_3/2]$ block. Next, let
	\begin{equation*}
	\X_2 = T\cdot \S' \times_1 E_1 \times_2 E_2 \times_3 E_3, \quad \Z_2 = T\cdot (\S - \S')\times_1E_1\times_2 E_2 \times_3 E_3.
	\end{equation*}
	Based on our construction, 
	\begin{equation*}
	(\Z_2)_{\Xi_t} = 0, \quad (\Z_2)_{\Omega_t\times \Xi_t} = 0,
	\end{equation*} 
	\begin{equation*}
	\left\|(\Z_2)_{[\Omega_1, \Omega_2, \Omega_3]}\right\|_{\rm HS} = \left\|\varepsilon \cdot T\cdot \S_{[1:r_1/2, 1:r_2/2, 1:r_3/2]}\right\|_{\rm HS} = C^{(B)}.
	\end{equation*}
	\begin{equation*}
	\begin{split}
	& \left\|\X_1 - \X_2\right\|_{\rm HS} = \left\|(\S - \S')\times_1 E_1 \times_2 E_2 \times_3 E_3\right\|_{\rm HS}\\
	\overset{\text{Lemma \ref{lm:property}}}{\geq} & \varepsilon\left\|T\cdot\S_{[1:r_1/2, 1:r_2/2, 1:r_3/2]}\right\|_{\rm HS} \lambda_1\lambda_2\lambda_3/(2\sqrt{2}) = \lambda_1 \lambda_2 \lambda_3 C^{(B)}/(2\sqrt{2}).
	\end{split}
	\end{equation*}
	When $T\to \infty$, $\varepsilon \to 0$, we have 
	$$(X_2)^\dagger_{\Omega_t\times \Xi_t}(X_2)_{t, \Omega} \to (X_1)^\dagger_{\Omega_t\times \Xi_t}(X_1)_{t, \Omega},  \quad (X_2)_{\Xi_t} (X_2)_{\Omega_t\times \Xi_t}^\dagger \to (X_1)_{\Xi_t} (X_1)_{\Omega_t\times \Xi_t}^\dagger.$$ 
	Thus, by selecting $T>0$ large enough, we are able to ensure the following inequalities hold for $t = 1, 2, 3$,
	\begin{equation*}
	\left\|\left((X_2)^\dagger_{\Omega_t\times \Xi_t}(X_2)_{t, \Omega}\right)\right\| \leq \xi_t,\quad \left\|\left((X_2)_{\Xi_t} (X_2)_{\Omega_t\times \Xi_t}^\dagger\right)\right\| \leq \lambda_t,
	\end{equation*}
	\begin{equation*}
	\left\|(Z_2)_{\Xi_t}\right\| \leq \frac{1}{5} \sigma_{r}\left((X_2)_{\Xi_t}\right), \quad \left\|(Z_2)_{\Omega_t\times \Xi_t}\right\| \leq \frac{1}{5}\sigma_{r}\left((X_2)_{\Omega_t\times \Xi_t}\right),
	\end{equation*}
	\begin{equation*}
	\left\|(Z_2)_{t, \Omega}\right\| \leq \frac{1}{5}\sigma_{r}\left((X_2)_{t, \Omega}\right).
	\end{equation*}
	In this case, $(\X_2, \Z_2, \Omega_t, \Xi_t)\in \mathcal{F}$ and $\Z_1, \Z_2$ both satisfy \eqref{ineq:Z_upper_bound}. Since $\X_1 + \Z_1 = \X_2 + \Z_2,$ by Lemma \ref{lm:lower_bound_difference}, we have proved the first part of \eqref{ineq:lower_bound_C^B}. The second part of \eqref{ineq:lower_bound_C^B} essentially follows as we only need to replace the value of $\varepsilon$ by $C^{(B)}/(T\|\S_{[1:r_1/2, 1:r_2/2, 1:r_3/2]}\|_{\rm op})$.
	
	\item In this step, we prove \eqref{ineq:lower_bound_C^A}. By symmetry, we only need to prove the inequality for $t = 1$. To be specific, let
	\begin{equation*}
	\tilde{E}_1 = \begin{blockarray}{cc}
	& r\\
	\begin{block}{c[c]}
	r_1 &  I_{r_1}\\
	r_1  & \left(\sqrt{\lambda_1^2/2 - 1}+\varepsilon\right) I_{r_1}\\
	p_1 - 2r_1 & 0\\
	\end{block}
	\end{blockarray}, \quad E_t = \begin{blockarray}{cc}
	& r\\
	\begin{block}{c[c]}
	r &  I_{r}\\
	r  & \sqrt{\lambda_t^2/2 - 1} I_{r}\\
	p_t - 2r & 0\\
	\end{block}
	\end{blockarray}, \quad t = 2, 3,
	\end{equation*}
	where $\varepsilon = C_1^{(A)}/\left(T\|S_{\Xi_1}\|_{F}\right)$ is a small constant. Now we define
	\begin{equation*}
	\X_2 = T\cdot \S \times_1\tilde{E}_1\times_2 E_2 \times_3 E_3,\quad 
	\Z_2 = T\cdot \S \times_1 \left(E_1 - \tilde{E}_1\right)\times_2 E_2 \times_3 E_3.
	\end{equation*}
	Based on the construction,
	\begin{equation*}
	\X_1 + \Z_1 = \X_2 + \Z_2;
	\end{equation*}
	\begin{equation*}
	\begin{split}
	& \left\|Z_{\Xi_1}\right\|_F = \varepsilon T\cdot \left\|S_{\Xi_1}\right\|_F \leq C_1^{(A)},\quad Z_{\Xi_2} = 0, \quad Z_{\Xi_3}=0;
	\end{split}
	\end{equation*}
	\begin{equation*}
	\Z_{[\Omega_1, \Omega_2, \Omega_3]}=0; \quad Z_{\Omega_t\times \Xi_t} = 0, \quad t= 1, 2, 3;
	\end{equation*}
	\begin{equation*}
	\begin{split}
	& \left\|\X_1 - \X_2\right\|_{\rm HS}\\
	= & \left\|T\cdot \S \times_1 \left(E_1 - \tilde{E}_1\right)\times_2 E_2 \times_3 E_3\right\|_{\rm HS} \\
	\geq & \left\|T\cdot \S\right\|_{\rm HS} \sigma_{\min}(E_1 - \tilde{E}_1) \sigma_{\min}(E_2)\sigma_{\min}(E_3)\\
	= & T\|S_1\|_F \varepsilon\cdot \lambda_2\lambda_3/2 = C^{(A)} \|S_1\|_F/\|S_{\Xi_1}\|_F \cdot  \lambda_2\lambda_3/2 \\
	\geq & C_1^{(A)}\lambda_2\lambda_3/2\cdot \sqrt{\frac{\|\S_{111}\|^2_{\rm HS} + \|\S_{222}\|^2_{\rm HS}}{\|\S_{112}\|_{\rm HS}^2 + \|\S_{212}\|_{\rm HS}^2}} = C^{(A)}\lambda_2\lambda_3\xi_1/28. \\
	& \quad \text{(by \eqref{eq:S_construction2} and definition of $\Xi_1$)}
	\end{split}
	\end{equation*}	
	Similarly to the previous step, select large enough $T$ to ensure 
	$$(\X_2, \Z_2, \Omega_t, \Xi_t) \in \mathcal{F}, \quad\text{while}\quad \Z_1 \Z_2 \text{ both satisfy \eqref{ineq:Z_upper_bound}}.$$
	By Lemma \ref{lm:lower_bound_difference}, we have proved the first part of \eqref{ineq:lower_bound_C^A}. The second part of \eqref{ineq:lower_bound_C^A} can be shown similarly, where we only need to specify $\varepsilon$ as $C_1^{(A)}/(T\|S_{\Xi_t}\|)$.
	\item We finally prove \eqref{ineq:lower_bound_C^J} in this step, where we still focus on the case of $t = 1$. Particularly, let
	\begin{equation*}
	\begin{split}
	\S'' \in \mathbb{R}^{r_1\times r_2\times r_3}, \quad S_{ijk}'' = \left\{\begin{array}{ll}
	(1+\varepsilon)S_{ijk}, \quad (i, j, k)\in \Omega_1\times \Xi_1;\\
	S_{ijk}, \quad \text{otherwise},
	\end{array}\right.
	\end{split}
	\end{equation*}
	\begin{equation*}
	\bar{E}_1 = \begin{blockarray}{cc}
	& r\\
	\begin{block}{c[c]}
	r_1 &  I_{r_1}\\
	r_1  & \frac{\sqrt{\lambda_1^2/2 - 1}}{1+\varepsilon} I_{r_1}\\
	p_1 - 2r_1 & 0\\
	\end{block}
	\end{blockarray}, \quad E_t = \begin{blockarray}{cc}
	& r\\
	\begin{block}{c[c]}
	r &  I_{r}\\
	r  & \sqrt{\lambda_t^2/2 - 1} I_{r}\\
	p_t - 2r & 0\\
	\end{block}
	\end{blockarray}, \quad t = 2, 3,
	\end{equation*}
	$$\X_2 = T\cdot \S'' \times_1 \bar{E}_1 \times_2 E_2 \times_3 E_3,$$
	$$\Z_2 = \X_1 + \Z_1 - \X_2 = T\cdot (\S - \S'')\times_1 \bar{E}_1 \times_2 E_2 \times_3 E_3 + T\cdot \S \times_1 \left(E_1 - \bar{E}_1\right) \times_2 E_2 \times_3 E_3. $$
	Here $\varepsilon = C^{(J)}_1/(T\|S_{\Xi_1}\|_F)$. We also assume $T$ is large enough such that $\varepsilon\leq 1$, $(\X_2, \Z_2, \Omega_t, \Xi_t)\in \mathcal{F}$.
	Based on our construction, $\S''$ and $\S$ are different only in $\Omega_1\times \Xi_1$ blocks, but not in the other parts of $[\Omega_1, \Omega_2, \Omega_3]$. Thus,
	\begin{equation*}
	\left\|Z_{\Omega_1, \Xi_1}\right\|_F = C^{(J)}_1, \quad \left\|Z_{\Omega_2, \Xi_2}\right\|_F = \left\|Z_{\Omega_3, \Xi_3}\right\|_F = 0.
	\end{equation*}
	According to the assumption that $C^{(J)}_t \leq \min\{C^{(B)}, C_t^{(A)}\}$, we have
	\begin{equation*}
	\left\|(\Z_2)_{[\Omega_1, \Omega_2, \Omega_3]}\right\|_{\rm HS} = \left\|T\S - T\S''\right\|_{\rm HS} = \varepsilon T\left\|S_{\Xi_1}\right\|_F = C^{(J)}_1 \leq C^{(B)},
	\end{equation*}
	\begin{equation*}
	\begin{split}
	& \left\|(Z_2)_{\Xi_1}\right\|_{F} = \left\|\begin{bmatrix}
	(Z_2)_{\Omega_1\times \Xi_1}\\
	(Z_2)_{\Omega_1^c\times \Xi_1}
	\end{bmatrix}\right\|_F = \left\|\begin{bmatrix}
	\varepsilon TS_{\Xi_1}\\
	(1+\varepsilon)TS_1 \cdot \frac{\sqrt{\lambda_1^2/2-1}}{1+\varepsilon}I_{r_1} - TS_1 \sqrt{\lambda_1^2/2-1}
	\end{bmatrix}\right\|_F\\
	= & \varepsilon T\|S_{\Xi_1}\|_F = C_1^{(J)} \leq C_1^{(A)}.
	\end{split}
	\end{equation*}
	Thus, $\Z_1, \Z_2$ both satisfy \eqref{ineq:Z_upper_bound}.
	Besides, we shall note that $S_{\Xi_1}'' = (1+\varepsilon)S_{\Xi_1}$, $S_{\Xi_1^c}''=S_{\Xi_c^c}$. Therefore,
	\begin{equation*}
	\begin{split}
	& \left\|\X_1 - \X_2\right\|_{\rm HS} = \|\Z_2\|_{\rm HS} \geq \left\|T\S''\times_1 \bar{E}_1 - T\S \times E_1\right\|_{\rm HS} \cdot \sigma_{\min}(E_2)\cdot \sigma_{\min}(E_3)\\
	\overset{\text{Lemma \ref{lm:property}}}{=} & \left\|\mathcal{M}_1(\S'') \bar{E}_1^\top - \mathcal{M}_1(\S) E_1^\top\right\|_F\cdot T\lambda_2\lambda_3/2\\
	= & \left\|\begin{bmatrix}
	S''_{\Xi_1} - S_{\Xi_1} & S''_{\Xi_1^c} - S_{\Xi_1^c}\\
	\frac{\sqrt{\lambda_1^2/2-1}}{1+\varepsilon}S''_{\Xi_1} - \sqrt{\lambda_1^2/2-1}S_{\Xi_1} & \frac{\sqrt{\lambda_1^2/2-1}}{1+\varepsilon}S''_{\Xi_1^c} - \sqrt{\lambda_1^2/2-1}S_{\Xi_1^c}
	\end{bmatrix}\right\|_F \cdot T\lambda_2\lambda_3/2\\
	= & T\lambda_2\lambda_3/2\sqrt{\varepsilon^2\|S_{\Xi_1}\|_F^2 + \frac{\left(\lambda_1^2/2-1\right)\varepsilon^2}{(1+\varepsilon)^2}\|S_{\Xi_1^c}\|_F^2}\\
	\geq & T\lambda_2\lambda_3/2\cdot  \sqrt{\lambda_1^2/2-1}\cdot \frac{\varepsilon}{1+\varepsilon} \cdot \|S_{\Xi_1^c}\|_F\\
	\geq & T\lambda_2\lambda_3/2 \cdot (\lambda_1/2)\cdot (\varepsilon/2)\cdot \sqrt{\|\S_{111}\|_{\rm HS}^2 + \|\S_{222}\|_{\rm HS}^2} \quad \text{(by $\lambda_1\geq 2, \varepsilon<1$ and \eqref{eq:S_construction2})}\\
	\geq & C_1^{(J)}\lambda_1\lambda_2\lambda_3/8 \cdot \sqrt{\frac{\|\S_{111}\|^2_{\rm HS} + \|\S_{222}\|^2_{\rm HS}}{\|\S_{112}\|_{\rm HS}^2 + \|\S_{212}\|_{\rm HS}^2}} \quad \text{(by \eqref{eq:S_construction2} and definition of $\Xi_1$)}\\
	\geq & C_1^{(J)}\lambda_1\lambda_2\lambda_3\xi_1/112.
	\end{split}
	\end{equation*}
	By Lemma \ref{lm:lower_bound_difference} again, we obtain the first part of \eqref{ineq:lower_bound_C^J}. For the second part of \eqref{ineq:lower_bound_C^J}, it can be shown based on the similar construction of $\X_2, \Z_2$ if we choose $\varepsilon = C_1^{(J)}/(T\|S_{\Xi_1}\|)$.
\end{enumerate}
To sum up, we have finished the proof for this theorem. \quad $\square$

\subsection{Proof of Theorem \ref{th:random}.}

In order to prove this theorem, we first introduce the following lemma. The lemma contains two parts that treat the least singular values sub-matrix and sub-tensor, respectively. The proof is postponed to Section \ref{sec.proof_technical_lemmas}.
\begin{Lemma}\label{lm:sub-matrix-tensor}
	The following results hold regarding the sub-matrix and sub-tensors.
	\begin{itemize}[leftmargin=*]
		\item Suppose $D \in \mathbb{R}^{p\times d}$ satisfies $\rank(D) = r$ and the incoherence condition with constant $\rho$, i.e. $\frac{r}{p}\|\mathbb{P}_{D}e_i\| \leq \rho$. Suppose $\Omega$ contains uniformly randomly selected $m$ numbers from $[1:p]$ with or without replacement and $D_{\Omega} = D_{[:, \Omega]}$ is the collection of columns of $D$. Then for any $m > r$, $0 < \varepsilon < 1$,
		\begin{equation}\label{ineq:lm_submatrix_bound}
		\begin{split}
		& P\left(\rank(D_{\Omega}) = r, \quad \text{and}\quad \frac{p}{(1+\varepsilon)m}\leq \sigma_{r}^2(DD_{\Omega}^\dagger) \leq \sigma_{\max}^2(DD_{\Omega}^\dagger) \leq \frac{p}{(1-\varepsilon)m} \right)\\
		\geq & 1 - 2r\exp\left(-m(1-\varepsilon)^2/(4\rho r)\right).
		\end{split}
		\end{equation}
		\item Suppose $\X \in \mathbb{R}^{p_1 \times p_2\times p_3}$ is a tensor with Tucker decomposition $\X = \S \times_1 U_1\times_2 U_2 \times_3 U_3$, where $\S\in \mathbb{R}^{r_1\times r_2\times r_3}$, $U_t\in \mathbb{O}_{p_t, r_t}$. Assume that $U_1, U_2, U_3, \mathcal{M}_1(\S\times_2 U_2 \times_3 U_3), \mathcal{M}_2(\S\times_3 U_3 \times_1 U_1), \mathcal{M}_3(\S\times_1 U_1 \times_2 U_2)$ are all full rank and satisfy the incoherence condition, i.e.
		\begin{equation*}
		\begin{split}
		& \frac{p_t}{r_t}\left\|\mathbb{P}_{U_t} e_j\right\|_2^2 \leq \rho, 
		\quad \forall e_j\in \mathbb{R}^{p_t} \\ 
		& \frac{p_{t+1}p_{t+2}}{r_t}\|\mathbb{P}_{\mathcal{M}_t(\S\times_{t+1} U_{t+1} \times_{t+2} U_{t+2})} e_j\|_2^2 \leq \rho,\quad \forall e_j \in \mathbb{R}^{p_{t+1}p_{t+2}}.
		\end{split}
		\end{equation*} 
		Suppose $\Omega_1, \Omega_2, \Omega_3$ are uniform randomly selected $m_1$, $m_2$ and $m_3$ numbers from $[1:p_1]$, $[1:p_2]$ and $[1:p_3]$, respectively. Conditioning on the selection of $\Omega_1, \Omega_2$, $\Omega_3$, let $\Xi_t$ be uniform randomly selected $g_t$ values from $\Omega_{t+1}\times \Omega_{t+2}$. Denote the body, joint and arm matricizations as
		$$X_{t, \Omega} = \mathcal{M}_t(\X_{[\Omega_1, \Omega_2, \Omega_3]}), \quad X_{\Omega_t\times \Xi_t} = (\mathcal{M}_t(\X))_{[\Omega_t, \Xi_t]}, \quad X_{\Xi_t} = (\mathcal{M}_t(\X))_{[1:p_t, \Xi_t]}.$$ 
		Then with probability at least 
		$$1 - 2\sum_{t=1}^3 r_t\exp(-m_t(1-\varepsilon)^2/(4\rho r_t)) - 2\sum_{t=1}^3r_t\exp\left(-g_t(1-\varepsilon)^4/(4\rho r_t)\right),$$ 
		the following hold: for $t =1, 2, 3$,
		\begin{equation}\label{ineq:lm_subtensor_bound}
		\begin{split}
		& \rank(X_{\Omega_t\times \Xi_t}) = r_t, \\
		& \sigma_{r_t}\left(X_{\Xi_t}X_{\Omega_t\times \Xi_t}^\dagger\right) = \sigma_{\max}^{-1}(U_{t, [\Omega_t, :]}),\quad \sigma_{\max}\left(X_{\Xi_t}X_{\Omega_t\times \Xi_t}^\dagger\right) = \sigma_{\min}^{-1}(U_{t, [\Omega_t, :]}),\\
		& \frac{p_t}{(1+\varepsilon)m_t} \leq \sigma_{r_t}^2\left(X_{\Xi_t}X_{\Omega_t\times \Xi_t}^\dagger \right) \leq \sigma_{\max}^2\left(X_{\Xi_t}X_{\Omega_t\times \Xi_t}^\dagger \right) \leq \frac{p_t}{(1-\varepsilon)m_t}, \\
		& \frac{m_{t+1}m_{t+2}}{(1+\varepsilon)g_t} \leq \sigma_{r_t}^2\left(X^\dagger_{\Omega_t\times \Xi_t}X_{t, \Omega}\right) \leq \sigma_{\max}^2\left(X^\dagger_{\Omega_t\times\Xi_t}X_{t, \Omega}\right) \leq \frac{m_{t+1}m_{t+2}}{(1-\varepsilon)g_t}.
		\end{split}
		\end{equation}
	\end{itemize}
\end{Lemma}
Now let us move back to the proof for Theorem \ref{th:random}. By the second part of Lemma \ref{lm:sub-matrix-tensor}, setting $\varepsilon = 1/2$, we know
\begin{equation}
\begin{split}
\xi_t = \|X_{\Omega_t\times \Xi_t}^\dagger X_{t, \Omega}\| \leq \sqrt{\frac{2m_2m_3}{g_1}}.
\end{split}
\end{equation}
\begin{equation}\label{ineq:lambda_t geq U}
\begin{split}
& \sqrt{\frac{2p_t}{3m_t}} \leq \sigma_{r_t}\left(X_{\Xi_t} X_{\Omega_t\times \Xi_t}^\dagger\right) \leq \left\|X_{\Xi_t} X_{\Omega_t\times \Xi_t}^\dagger\right\| \leq  \sqrt{\frac{2p_t}{m_t}}.
\end{split}
\end{equation}
with probability at least $1 - 2\sum_{t=1}^3 r_t\left\{\exp(-m_t/(16r_t\rho )) + \exp(-g_t/(64 r_t\rho))\right\}$. Under such situation, we also have
\begin{equation*}
\begin{split}
& 2\left\|X_{\Xi_t} X_{\Omega_t\times \Xi_t}^\dagger\right\| \leq \lambda_t = 3\sqrt{p_t/m_t},\\
& \sqrt{\frac{m_t}{2p_t}} \leq \sigma_{\min}\left(U_{t, [\Omega_t, :]}\right) \leq \sigma_{\max}\left(U_{t, [\Omega_t, :]}\right) \leq \sqrt{\frac{3m_t}{2p_t}},
\end{split}
\end{equation*}
which means the arm-joint ratio is bounded by $\lambda_t = 3\sqrt{p_t/m_t}$. Since $\S \times_1 U_{1, [\Omega_1, :]} \times_2 U_{2, [\Omega_2, :]} \times_3 U_{3, [\Omega_3, :]} = \X_{[\Omega_1,\Omega_2, \Omega_3]}$, by Lemma \ref{lm:property}, we obtain
\begin{equation*}
X_{1, \Omega} = U_{1, [\Omega_1, :]} \cdot  \mathcal{M}_1(\S) \cdot \left(U_{2, [\Omega_2, :]}\otimes U_{3, [\Omega_3, :]}\right)^\top,
\end{equation*}
where $U_{1, [\Omega_1, :]} \in \mathbb{R}^{m_1\times r_1}$, $\mathcal{M}_1(\S)\in \mathbb{R}^{r_1\times (r_2r_3)}, \left(U_{2, [\Omega_2, :]}\otimes U_{3, [\Omega_3, :]}\right)^\top \in \mathbb{R}^{(r_2r_3)\times (m_2m_3)}$. Similar equalities can be obtained for $X_{2, \Omega}$ and $X_{3, \Omega}$. Thus,
\begin{equation*}
\begin{split}
& \sigma_{r_t}\left(X_{t, \Omega}\right) \overset{\text{Lemma \ref{lm:property}}}{\geq} \sigma_{r_t}\left(U_{t, [\Omega, :]}\right)\cdot \sigma_{r_t}(\mathcal{M}_t(\S)) \cdot \sigma_{\min}\left(U_{t+1, [\Omega_{t+1}, :]}\right) \sigma_{\min}\left(U_{t+2, [\Omega_{t+2}, :]}\right) \\
\overset{\eqref{ineq:th_min_S_condition}\eqref{ineq:lambda_t geq U}}{\geq} & 13\sqrt{\frac{p_1p_2p_3}{m_1m_2m_3}}\|Z_{1, \Omega}\| \cdot \sqrt{\frac{2m_1}{3p_1}}\cdot \sqrt{\frac{2m_2}{3p_2}} \cdot \sqrt{\frac{2m_3}{3p_3}} = 13\sqrt{8/27}\|Z_{1, \Omega}\| \geq 5\|Z_{1, \Omega}\|.
\end{split}
\end{equation*}
Next, since
$X_{t, \Omega} = \mathbb{P}_{X_{\Omega_t\times \Xi_t}}X_{t,\Omega} = X_{\Omega_t\times \Xi_t} X_{\Omega_t\times \Xi_t}^\dagger X_{t, \Omega}$,
\begin{equation*}
\sigma_{r_t}(X_{t, \Omega}) \overset{\text{Lemma \ref{lm:property}}}{\leq} \sigma_{r_t}\left(X_{\Omega_t\times \Xi_t}\right) \|X_{\Omega_t\times \Xi_t}^\dagger X_{t, \Omega}\| \leq \sqrt{\frac{2m_{t+1}m_{t+2}}{g_t}} \sigma_{r_t}\left(X_{\Omega_t\times \Xi_t}\right).
\end{equation*}
Thus,
\begin{equation}\label{ineq:signal_dominate_noise_joint}
\begin{split}
& \sigma_{r_t}\left(X_{\Omega_t\times \Xi_t}\right) \geq \frac{\sigma_{r_t}(X_{t, \Omega})}{\sqrt{2m_{t+1}m_{t+2}/g_t}}\\ 
\geq & \frac{\sigma_{\min}\left(U_{1, [\Omega_1, :]}\right) \sigma_{\min}\left(U_{2, [\Omega_2, :]}\right) \sigma_{\min}\left(U_{3, [\Omega_3, :]}\right)\cdot \sigma_{r_t}(\mathcal{M}_t(\S))}{\sqrt{2m_{t+1}m_{t+2}/g_t}}\\
\geq & \frac{\sqrt{8m_1m_2m_3/(27p_1p_2p_3)} \cdot 10\sqrt{p_1p_2p_3/m_tg_t}\|Z_{\Omega_t\times \Xi_t}\|}{\sqrt{2m_{t+1}m_{t+2}/g_t}}\\
\geq & 10\sqrt{8/27}\|Z_{\Omega_t\times \Xi_t}\| \geq 5\|Z_{\Omega_t\times \Xi_t}\|.
\end{split}
\end{equation}
Then, we shall note that 
$$X_{\Xi_1} = \left(\mathcal{M}_1\left(\X\right)\right)_{[1:p_1]\times \Xi_1} = \left(U_1\cdot \mathcal{M}_1(\S)\cdot (U_2\otimes U_3)^\top \right)_{[:, \Xi_1]},$$
$$X_{\Omega_1\times \Xi_1} = \left(\mathcal{M}_1\left(\X_{[\Omega_1, :, :]}\right)\right)_{[1:p_1]\times\Xi_1} = \left(U_{1, [\Omega_1, :]}\cdot \mathcal{M}_1(\S)\cdot (U_2\otimes U_3)^\top \right)_{[:, \Xi_1]},$$
which means $U_{1, [\Omega_1, :]} U_1^\top X_{\Xi_1} = X_{\Omega_1\times \Xi_1}$. Similarly equalities also hold for $X_{\Omega_2\times \Xi_2}$ and $X_{\Omega_3\times \Xi_3}$. Then
\begin{equation*}
\begin{split}
\sigma_{r_t}(X_{\Xi_t}) \overset{\text{Lemma \ref{lm:property}}}{\geq} & \frac{\sigma_{r_t}(X_{\Omega_t\times \Xi_t})}{\sigma_{\max}(U_{t, [\Omega_1, :]}U_t^\top)}\\
\overset{\eqref{ineq:signal_dominate_noise_joint}}{\geq}& \frac{\sigma_{\min}\left(U_{1, [\Omega, :]}\right) \sigma_{\min}\left(U_{2, [\Omega, :]}\right) \sigma_{\min}\left(U_{2, [\Omega, :]}\right)\cdot \sigma_{r_t}(\mathcal{M}_t(\S))}{\sqrt{2m_{t+1}m_{t+2}/g_t}\sigma_{\max}(U_{t, [\Omega_t, :]})}\\
\geq & \frac{\sqrt{8m_1m_2m_3/(27p_1p_2p_3)}}{\sqrt{2m_{t+1}m_{t+2}/g_t} \cdot\sqrt{2m_t/p_t}}\cdot 19\sqrt{\frac{p_{t+1}p_{t+2}}{g_t}}\|Z_{\Xi_t}\|\\
\geq & 5\|Z_{\Xi_t}\|.
\end{split}
\end{equation*}
Thus, the singular value gap condition in Theorem \ref{th:noisy} holds. Finally, by Theorem \ref{th:noisy}, we have obtained the targeted result:
\begin{equation*}
\begin{split}
\left\|\hat{\X} - \X\right\|_{\rm HS} \leq & C\lambda_1\lambda_2\lambda_3\| \Z_{[\Omega_1, \Omega_2, \Omega_3]}\|_{\rm HS} + C\lambda_1\lambda_2\lambda_3 \sum_{t=1}^3 \left(\xi_t\|Z_{\Omega_t\times \Xi_t}\|_F + \frac{\xi_t}{\lambda_t}\|Z_{\Xi_t}\|_F\right)\\
\leq & C\sqrt{\frac{p_1p_2p_3}{m_1m_2m_3}}\| \Z_{[\Omega_1, \Omega_2, \Omega_3]}\|_{\rm HS} + C\sqrt{p_1p_2p_3} \sum_{t=1}^3 \left(\frac{\|Z_{\Omega_t\times \Xi_t}\|_F}{\sqrt{g_tm_t}} + \frac{\|Z_{\Xi_t}\|_F}{\sqrt{g_t p_t}} \right);
\end{split}
\end{equation*}
\begin{equation*}
\begin{split}
\left\|\hat{\X} - \X \right\|_{\rm op} \leq & C\lambda_1\lambda_2\lambda_3\| \Z_{[\Omega_1, \Omega_2, \Omega_3]}\|_{\rm op} + C\lambda_1\lambda_2\lambda_3 \sum_{t=1}^3 \left(\xi_t\|Z_{\Omega_t\times \Xi_t}\| + \frac{\xi_t^{(op)}}{\lambda_t}\|Z_{\Xi_t}\|\right)\\ 
\leq & C\sqrt{\frac{p_1p_2p_3}{m_1m_2m_3}} \left\|\Z_{[\Omega_1, \Omega_2, \Omega_3]}\right\|_{\rm op} + C\sqrt{p_1p_2p_3} \sum_{t=1}^3 \left(\frac{\|Z_{\Omega_t\times \Xi_t}\|}{\sqrt{g_tm_t}} + \frac{\|Z_{\Xi_t}\|}{\sqrt{g_t p_t}} \right).
\end{split}
\end{equation*}
\quad $\square$

\section{Technical Lemmas}

We collect some important technical lemmas in this section. The first lemma is about some basic properties of order-3 tensors. It is also noteworthy that the result can be further extended to order-4 or higher tensors.
\begin{Lemma}[Some properties of tensors]\label{lm:property} The following properties hold for order-3 tensors.
	\begin{enumerate}
		\item (Tensor operator norm and matricization spectral norm) For any tensor $\X$,
		$$\left\|\X\right\|_{\rm op} \leq \min\{\|\mathcal{M}_1(\X)\|, \|\mathcal{M}_2(\X)\|, \|\mathcal{M}_3(\X)\|\}. $$
		\item (Matricization of tensor mode product) For two matrices $A\in \mathbb{R}^{p_1\times p_2}, B\in \mathbb{R}^{q_1\times q_2}$, $A\otimes B\in  \mathbb{R}^{(p_1q_1) \times (p_2q_2)}$ is defined as
		\begin{equation*}
		\begin{split}
		& (A\otimes B)_{i_1 +j_1 (p_1 - 1), i_2 + j_2(p_2 - 1)} = A_{i_1, i_2} B_{j_1, j_2}, 
		\end{split}
		\end{equation*} 
		for $1\leq i_1 \leq p_1, 1\leq i_2\leq p_2, 1\leq j_1 \leq q_1, 1\leq j_2 \leq q_2$. Now suppose $\S\in \mathbb{R}^{r_1\times r_2\times r_3}$, $E_t\in \mathbb{R}^{p_t\times r_t}$ for $t = 1, 2, 3$, then
		$$\mathcal{M}_1\left(\S \times_1 E_1\times_2 E_2 \times_3 E_3\right) = E_1\cdot \mathcal{M}_t(\X) \cdot (E_2\otimes E_3)^\top,$$
		where $``\cdot"$ is the usual matrix product, $``\otimes"$ is the outer product.
		Similar results hold for $\mathcal{M}_2\left(\X \times_1 E_1\times_2 E_2 \times_3 E_3\right)$ and $\mathcal{M}_3\left(\X \times_1 E_1\times_2 E_2 \times_3 E_3\right)$.

		\item (Tensor operator and Hilbert-Schmidt norm for mode product) Suppose $\S\in \mathbb{R}^{r_1\times r_2 \times r_3}$, $E_\in \mathbb{R}^{p_t\times r_t}$, then
		$$\|\S\|_{\rm HS}\prod_{t=1}^3 \sigma_{r_t}(E_t) \leq \left\|\S \times_1 E_1 \times_2 E_2 \times_3 E_3\right\|_{\rm HS} \leq \|\S\|_{\rm HS} \prod_{t=1}^3 \left\|E_t\right\|. $$
		$$\|\S\|_{\rm op}\prod_{t=1}^3 \sigma_{r_t}(E_t) \leq \left\|\S \times_1 E_1 \times_2 E_2 \times_3 E_3\right\|_{\rm op} \leq \|\S\|_{\rm op} \prod_{t=1}^3 \left\|E_t\right\|.$$
	\end{enumerate}
\end{Lemma}

Lemma \ref{lm:sigma_r_matrix} discusses the relationship between the singular values among $A, B$ and $A\cdot B$.
\begin{Lemma}\label{lm:sigma_r_matrix}
	For any two matrices $A\in\mathbb{R}^{p\times r}$, $B\in \mathbb{R}^{r\times q}$ (without specifying the order of $p, r, q$), we have
	\begin{equation}\label{ineq:sigma_r(AB)}
	\sigma_r(AB) \geq \sigma_r(A)\sigma_r(B), 
	\end{equation}
	\begin{equation}\label{ineq:sigma_p(AB)}
	\sigma_p(AB) \geq \sigma_p(A)\sigma_r(B), \quad \sigma_q(AB) \geq \sigma_r(A)\sigma_q(B).
	\end{equation}
\end{Lemma}

The next lemma focus on expansion of the inverse matrix $(G+H)^{-1}$.
\begin{Lemma}\label{lm:matrix_inverse}
	Suppose $G, H$ are two squared matrices such that $G, G+H$ are both invertible, then for any $k \geq 1$ and $0\leq i \leq k$,
	\begin{equation}\label{eq:lm_matrix_inverse_target}
	(G+H)^{-1} = \sum_{i=0}^{k-1} (-1)^{i} G^{-1} \left(HG^{-1}\right)^{i} + (-1)^{k}\cdot (G^{-1}H)^{i} (G+H)^{-1}\left(HG^{-1}\right)^{k-i}.
	\end{equation}
	In particular,
	\begin{equation*}
	\begin{split}
	(G+H)^{-1} = & G^{-1} - (G+H)^{-1}HG^{-1} = G^{-1} - G^{-1}H(G+H)^{-1}\\
	= & G - G^{-1}HG^{-1} + G^{-1}H(G+H)^{-1}HG^{-1}. 	
	\end{split}
	\end{equation*}
\end{Lemma}

\subsection{Proofs of Technical Lemmas}\label{sec.proof_technical_lemmas}

In this section, we provide the proofs for the technical lemmas used in the main content.

{\noindent\bf Proof of Lemma \ref{lm:property}.} We prove the statements of Lemma \ref{lm:property} one by one.
\begin{enumerate}
	\item By definition, 
	\begin{equation*}
	\begin{split}
	\left\|\X\right\|_{op} \leq & \max_{u, v, w} \frac{\X \times_1 u \times_2 v \times_3 w}{\|u\|_2\|v\|_2\|w\|_2} = \max_{u, v, w} \frac{u^\top \mathcal{M}_1(\X) (v\otimes w)}{\|u\|_2 \|v\otimes w\|_2} \\
	\leq & \max_{u, x} \frac{u^\top \mathcal{M}_1(\X)x}{\|u\|_2\|x\|_2} = \|\mathcal{M}_1(\X)\|.
	\end{split}
	\end{equation*}
	We can similarly prove that $\left\|\X\right\|_{op} \leq \|\mathcal{M}_2(\X)\|$ and $\left\|\X\right\|_{op} \leq \|\mathcal{M}_3(\X)\|$.
	\item For any $(i, j, k)\in [1:p_1, 1:p_2, 1:p_3]$, by definition
	\begin{equation*}
	\begin{split}
	& \left\{\mathcal{M}_1(\S\times_1 E_1\times_2 E_2 \times_3 E_3)\right\}_{i, j + p_2(k-1)}= \left\{\S \times_1 \E_1 \times_2 E_2 \times_3 E_3 \right\}_{i, j, k}\\
	= & \sum_{i'=1}^{r_1}\sum_{j'=1}^{r_2} \sum_{k'=1}^{r_3} \X_{i'j'k'} (E_{1})_{i, i'}(E_{2})_{j, j'}(E_{3})_{k, k'}\\
	= & \sum_{i'=1}^{r_1}\sum_{j'=1}^{r_2} \sum_{k'=1}^{r_3} (E_{1})_{i, i'} \left(E_2\otimes E_3\right)_{j+p_2(k-1), j' + p_2(k'-1)}\\
	= & \left\{E_1\cdot \mathcal{M}_1(\S) \cdot (E_2\otimes E_3)^\top\right\}_{i, j+p_2(k-1)},
	\end{split}
	\end{equation*}
	which means $\mathcal{M}_1(\S \times_1 E_1 \times_2 E_2 \times_3 E_3) = E_1\cdot \mathcal{M}_1(\X) \cdot (E_2\otimes E_3)^\top$.
	
	\item Based on the result in Part 2, 
	\begin{equation*}
	\begin{split}
	& \left\|\S \times_1 E_1 \times_2 E_2 \times_3 E_3\right\|_{\rm HS} \\
	= & \left\|\mathcal{M}_1(\S \times_1 E_1 \times_2 E_2 \times_3 E_3)\right\|_F =  \left\|E_1 \mathcal{M}_1(\S) (E_2 \otimes E_3)^\top \right\|_{F}\\
	\leq & \left\|\mathcal{M}_1(\S) \right\|_F \|E_1\| \cdot\|E_2\otimes E_3\| = \left\|\S \right\|_{\rm HS} \|E_1\|\|E_2\| \| E_3\|.
	\end{split}
	\end{equation*}
	In addition,
	\begin{equation*}
	\begin{split}
	& \left\|\S \times_1 E_1 \times_2 E_2 \times_3 E_3\right\|_{\rm HS} \\
	= & \left\|\mathcal{M}_1(\S \times_1 E_1 \times_2 E_2 \times_3 E_3)\right\|_F =  \left\|E_1 \mathcal{M}_1(\S) (E_2 \otimes E_3)^\top \right\|_{F}\\
	\geq & \left\|\mathcal{M}_1(\S) \right\|_F \sigma_{r_1}(E_1) \cdot \sigma_{r_2r_3}(E_2\otimes E_3) \geq \left\|\S \right\|_{\rm HS} \sigma_{r_1}(E_1) \sigma_{r_2}(E_2) \sigma_{r_3}(E_3).
	\end{split}
	\end{equation*}	
	
	For the operator norm,
	\begin{equation*}
	\begin{split}
	& \left\|\S \times_1 E_1 \times_2 E_2 \times_3 E_3\right\|_{\rm op} = \max_{u, v, w} \frac{\S \times_1 (u^\top E_1) \times_2 (v^\top E_2) \times_3 (w^\top E_3)}{\|u\|_2\|v\|_2\|w\|_2}\\
	\leq & \frac{\|\S\|_{\rm op} \|u^\top E_1\|_2 \|v^\top E_2\|_2 \|w^\top E_3\|_2}{\|u\|_2\|v\|_2\|w\|_2} \leq \|\S\|_{\rm op} \|E_1\| \|E_2\| \|E_3\|.
	\end{split}
	\end{equation*}
	Finally it remains to prove 
	$$\|\S\|_{\rm op}\prod_{t=1}^3 \sigma_{r_t}(E_t) \leq \left\|\S \times_1 E_1 \times_2 E_2 \times_3 E_3\right\|_{\rm op}.$$
	This clearly holds if $\sigma_{r_1}(E_1) = 0$, $\sigma_{r_2}(E_2) = 0$, or $\sigma_{r_3}(E_3) = 0$. When $\sigma_{r_1}(E_1) > 0$, $\sigma_{r_2}(E_2) > 0$, or $\sigma_{r_3}(E_3) > 0$, suppose $x \in \mathbb{R}^{r_1}, y \in \mathbb{R}^{r_2}, z \in \mathbb{R}^{r_3}$ satisfy 
	$$\frac{\S \times_1 x \times_2 y \times_3 z}{\|x\|_2\|y\|_2\|z\|_2} = \|\S\|_{\rm op}.$$ Since $E_t\in \mathbb{R}^{p_t\times r_t}$, there exists $u_\ast\in \mathbb{R}^{p_1}, v_\ast \in \mathbb{R}^{p_2}, w_\ast \in \mathbb{R}^{p_3}$ such that $x = E_1^\top u_\ast$, $y = E_2^\top v_\ast$, $z = E_3^\top w_\ast$. In this case,
	\begin{equation*}
	\|u_\ast\|_2 \leq \frac{\|x\|_2}{\sigma_{r_1}(E_1)}, \quad \|v_\ast\|_2 \leq \frac{\|y\|_2}{\sigma_{r_2}(E_2)}, \quad \|w_\ast\|_2 \leq \frac{\|z\|_2}{\sigma_{r_3}(E_3)}.
	\end{equation*}
	Thus,
	\begin{equation*}
	\begin{split}
	& \left\|\S \times_1 E_1 \times_2 E_2 \times_3 E_3\right\|_{\rm op} = \max_{u, v, w} \frac{\S \times_1 (u^\top E_1) \times_2 (v^\top E_2) \times_3 (w^\top E_3)}{\|u\|_2\|v\|_2\|w\|_2}\\
	\geq & \frac{\S \times_1 x \times_2 y \times_3 z}{\|u_\ast\|_2\|v_\ast\|_2\|w_\ast\|_2} \geq \|\S\|_{\rm op} \sigma_{r_1}(E_1) \sigma_{r_2}(E_2) \sigma_{r_3}(E_3),
	\end{split}
	\end{equation*}
	which has finished the proof of this lemma. \quad $\square$
\end{enumerate}

{\noindent\bf Proof of Lemma \ref{lm:sigma_r_matrix}.} 
When $\sigma_r(A), \sigma_r(B) >0$, we have $\sigma_{r}(BB^\top) = \sigma_{r}^2(B)$, then $BB^\top  \succeq \sigma_{r}^2(B) I_r$. Additionally, $ABB^\top A^\top \succeq \sigma_r^2(B) AA^\top$. Therefore,
\begin{equation*}
\begin{split}
\sigma_r^2(AB) = \lambda_r(ABB^\top A^\top) \geq \lambda_r(\sigma_r^2(B)AA^\top) = \sigma_r^2(B) \sigma_r^2(A),
\end{split}
\end{equation*}
\begin{equation*}
\begin{split}
\sigma_p^2(AB) = \lambda_p(ABB^\top A^\top) \geq \lambda_p(\sigma_r^2(B)AA^\top) = \sigma_r^2(B) \sigma_p^2(A),
\end{split}
\end{equation*}	
which implies \eqref{ineq:sigma_r(AB)} and the first part of \eqref{ineq:sigma_r(AB)}. The second part of \eqref{ineq:sigma_r(AB)} can be shown by symmetry as the first part of \eqref{ineq:sigma_r(AB)}. \quad $\square$

{\bfseries\noindent Proof of Lemma \ref{lm:matrix_inverse}.} This lemma essentially follows from Lemma 6.2 in \cite{zhang2016semi}. For completeness, we still provide the complete proof here. First, we note that
\begin{equation*}
\begin{split}
& (G+H)(G^{-1}H)^i = (HG^{-1})^i(G+H)\\ 
\overset{\text{Multiply both sides by } (G+H)^{-1}}{\Rightarrow} \quad & (G+H)^{-1}(HG^{-1})^i = (G^{-1}H)^i(G+H)^{-1},\\
\end{split}
\end{equation*}
thus
\begin{equation*}
\begin{split}
& \left(G+H\right) \left\{\sum_{i=0}^{k-1} (-1)^{i} G^{-1} \left(HG^{-1}\right)^{i} + (-1)^{k}\cdot (G^{-1}H)^{i} (G+H)^{-1}\left(HG^{-1}\right)^{k-i}\right\} \\
= & \sum_{i=0}^{k-1} (-1)^i \left((HG^{-1})^i + (HG^{-1})^{i+1}\right) + (-1)^k(G+H) (G+H)^{-1} \left(HG^{-1}\right)^{i}\left(HG^{-1}\right)^{k-i}\\
= & \sum_{i=0}^{k-1} (-1)^i (HG^{-1})^i - \sum_{i=1}^k(-1)^i (HG^{-1})^i + (-1)^k (HG^{-1})^k = I.
\end{split}
\end{equation*}
This has proved \eqref{eq:lm_matrix_inverse_target}. \quad $\square$

{\bf\noindent Proof of Lemma \ref{lm:lower_bound_difference}.} According to the assumptions of Lemma \ref{lm:lower_bound_difference},
\begin{equation*}
\begin{split}
&\inf_{\hat{\X}} \sup_{(\X, \Z, \Omega_t, \Xi_t) \in \mathcal{G}} \left\|\hat{\X} - \X\right\| \geq \max\left\{ \left\|\hat{\X} - \X_1\right\|, \left\|\hat{\X} - \X_2\right\|\right\}\\
\geq & \frac{1}{2} \left(\left\|\hat{\X} - \X_1\right\|, \left\|\hat{\X} - \X_2\right\|\right) \geq \frac{1}{2} \left(\left\|\X_1 - \X_2\right\|\right).
\end{split}
\end{equation*}
\quad $\square$

{\bf\noindent Proof of Lemma \ref{lm:sub-matrix-tensor}.}
\begin{itemize}
	\item We first prove \eqref{ineq:lm_submatrix_bound}. Suppose $D = U\Sigma V^\top$, where $U\in \mathbb{O}_{p, r}, \Sigma\in \mathbb{R}^{r\times r}, V\in \mathbb{O}_{d, r}$. Note $U_{\Omega} = U_{[\Omega, :]}$, $\Sigma V^\top$ has full row rank, so $\left(U_{\Omega}\Sigma V^\top\right)^\dagger = \left(\Sigma V^\top\right)^\dagger U_{\Omega}^\dagger$ and
	\begin{equation*}
	D D_{\Omega}^\dagger = U\Sigma V^\top \left(U_{\Omega} \Sigma V^\top\right)^\dagger = U U_{\Omega}^\dagger.
	\end{equation*}
	Since $U$ is orthogonal, the above equality implies 
	$$\sigma_{\max}^{-1}(U_{\Omega}) = \sigma_{\min}(DD_{\Omega}^\dagger) \leq \sigma_{\max}(DD_{\Omega}^\dagger) = \sigma_{\min}^{-1}(U_{\Omega}), \quad \text{ when } \rank(U_{\Omega}) = r.$$
	Thus, we only need to focus on the largest and least singular values of $U_{\Omega}$ if $\sigma_{\min}(U_{\Omega}) > 0$. Note that
	\begin{equation*}
	\begin{split}
	\mathbb{P}_{D}e_i = & D (D^\top D)^{-1} D^\top e_i = U\Sigma^{-1}V^\top \left(V\Sigma U^\top U\Sigma V^\top \right)^{-1} D^\top e_i\\
	= & UU^\top e_i, 
	\end{split}
	\end{equation*}
	thus, $\|\mathbb{P}_{D} e_i\|_2^2 = \|UU^\top e_i\|_2^2 = \|U_{[i, :]}\|_2^2$. By the assumption of the incoherence condition, we have 
	$$\max_{1\leq i\leq p} \frac{p}{r}\|U_{[i, :]}\|_2^2 \leq \rho.$$
	In this case, for any $x>0$,
	\begin{equation*}
	\begin{split}
	& P\left(\frac{p}{(1+\varepsilon)m} \leq \sigma_{\min}^2(DD_{\Omega}^\dagger) \leq \sigma_{\max}^2(DD_{\Omega}^\dagger) \leq \frac{p}{(1-\varepsilon)m}\right)\\
	= & P\left(\frac{(1-\varepsilon)m}{p} \leq \sigma_{\min}^2(U_{\Omega}) \leq \sigma_{\max}^2(U_{\Omega}) \leq \frac{(1+\varepsilon)m}{p}\right)\\
	= & P\left(\frac{(1-\varepsilon)m}{p} \leq \sigma_{\max}(U_{\Omega}^\top U_{\Omega}) \leq \frac{(1+\varepsilon)m}{p}\right)=  P\left(\left\|U_{\Omega}^\top U_{\Omega} - \frac{m}{p} I_{r}\right\| \leq \frac{\varepsilon m}{p}\right)\\
	= & P\left(\left\|\sum_{k=1}^m \left(U_{[\Omega(k), :]}^\top U_{[\Omega(k), :]} - \frac{1}{p}I_r\right)\right\| \leq \frac{\varepsilon m}{p}\right).
	\end{split}
	\end{equation*}
	To bound the tail probability of the random matrix above, we first calculate that
	$$\E \left(U_{[\Omega(k), :]}^\top U_{[\Omega(k), :]} - \frac{1}{p}I_r\right) = 0,$$
	\begin{equation*}
	\left\| U_{[\Omega(k), :]}^\top U_{[\Omega(k), :]} - \frac{1}{p}I_r\right\| \leq \max\left\{\max_{1\leq k \leq p}\|U_{[k, \cdot]}\|_2^2, \left\|\frac{1}{p}I_r\right\|\right\} \leq \frac{\rho r}{p}.
	\end{equation*}
	\begin{equation*}
	\begin{split}
	& \left\|\E \left(U_{[\Omega(k), :]}^\top U_{[\Omega(i), :]} - \frac{1}{p}I_r\right)^2\right\| =  \left\|\frac{1}{p}\sum_{i=1}^p \left(U_{[i, :]}^\top U_{[i, :]}U_{[i, :]}^\top U_{[i, :]}\right) - \frac{1}{p^2}I_r\right\|\\
	& \quad \left(\text{since $\left(U_{[\Omega(k), :]}^\top U_{[\Omega(i), :]} - \frac{1}{p}I_r\right)^2$ is non-negative definite}\right)\\
	\leq & \left\|\frac{1}{p}\sum_{i=1}^p \left(U_{[i, :]}^\top \|U_{[i, :]}\|_2^2 U_{[i, :]}\right)\right\| - \frac{1}{p^2} \\
	\leq & \frac{\rho r}{p^2} \left\|\sum_{i=1}^p U_{[i, :]}^\top U_{[i, :]}\right\| - \frac{1}{p^2} = \frac{\rho r-1}{p^2}.
	\end{split}
	\end{equation*}
	Then by matrix Bernstein's inequality (Theorem 1 in \cite{gross2010note}),
	\begin{equation*}
	\begin{split}
	& P\left(\left\|\sum_{k=1}^m \left(U_{[\Omega(k), :]}^\top U_{[\Omega(k), :]} - \frac{1}{p}I_r\right)\right\| \leq \frac{\varepsilon m}{p}\right)\\
	\geq & 1 - 2r\exp\left(-\min\left(\frac{(1-\varepsilon)m/p)^2}{4m(\rho r-1)/p^2}, \frac{(1-\varepsilon)m/p}{2\rho r/p}\right)\right) \geq 1 - 2r\exp\left(-\frac{m(1-\varepsilon)^2}{4\rho r}\right).
	\end{split}
	\end{equation*}
	
	\item Next, we consider the second part of the lemma. Since $U_1, U_2, U_3$ satisfies the incoherence condition with constant $\rho$, by the first part of this lemma, we have
	\begin{equation}\label{ineq:lm_subtensor_event}
	\begin{split}
	& P\left(\rank(U_{t, [\Omega_t, :]}) = r_t,  \frac{(1-\varepsilon)m_t}{p_t} \leq \sigma_{\min}^2(U_{t, [\Omega_t, :]}) \leq \sigma_{\max}^2(U_{2, [\Omega_2, :]}) \leq \frac{(1+\varepsilon)m_t}{p_t}\right)\\ 
	\geq & 1 - 2r_t\exp\left(-\frac{(1-\varepsilon)^2m_t}{p_t}\right).
	\end{split}
	\end{equation}
	Suppose the inequality above holds. By Lemma \ref{lm:property}, we also have,
	\begin{equation*}
	\begin{split}
	\frac{p_{t+1}p_{t+2}}{(1+\varepsilon)^2m_{t+1}m_{t+2}} \leq & \sigma_{r_t}^2\left(\left(U_{t+1} U_{t+1, [\Omega_{t+1}, :]}^\dagger\right)^\top \otimes \left(U_{t+2} U_{t+2, [\Omega_{t+2}, :]}^\dagger\right)^\top\right)\\
	\leq & \sigma_{\max}^2\left(\left(U_{t+1} U_{t+1, [\Omega_{t+1}, :]}^\dagger\right)^\top \otimes \left(U_{t+2} U_{t+2, [\Omega_{t+2}, :]}^\dagger\right)^\top\right) \\
	\leq &  \frac{p_{t+1}p_{t+2}}{(1-\varepsilon)^2m_{t+1}m_{t+2}},
	\end{split}
	\end{equation*}
	for $t =1, 2, 3$. By the same procedure as Theorem \ref{th:noiseless}, we can see $\rank(X_{1, \Omega_1}) = \rank(\mathcal{M}_1(\X_{[\Omega_1, \Omega_2, \Omega_3]})) = r_t$ and
	\begin{equation*}
	\X_{[\Omega_1, :, :]} = \X_{[\Omega_1, \Omega_2, \Omega_3]} \times_2 \left(U_2 U_{2, [\Omega_2, :]}^\dagger \right) \times_3 \left(U_3 U_{3, [\Omega_3, :]}^\dagger \right).
	\end{equation*}
	Take mode-1 matricization to the equality above and denote $X_{1, [\Omega_1, :]} = \mathcal{M}_1(\X_{[\Omega_1, :, :]})$, $X_{2, [\Omega_2, :]} = \mathcal{M}_2(\X_{[:, \Omega_2, :]})$ and $X_{3, [\Omega_3, :]} = \mathcal{M}_1(\X_{[:, :, \Omega_1]})$, we have
	\begin{equation*}
	X_{1, [\Omega_1, :]} = X_{1, \Omega} \cdot \left\{\left(U_2 U_{2, [\Omega_2, :]}^\dagger \right)^\top \otimes \left(U_3 U_{3, [\Omega_3, :]}^\dagger \right)^\top\right\}.
	\end{equation*}
	Since $\rank(X_{1, \Omega}\in \mathbb{R}^{r}) = \rank(X_{1, [\Omega_1, :]}) = r_1$, it implies
	\begin{equation*}
	X_{1, \Omega}^\dagger = \left\{\left(U_2 U_{2, [\Omega_2, :]}^\dagger \right)^\top \otimes \left(U_3 U_{3, [\Omega_3, :]}^\dagger \right)^\top\right\} X_{1, [\Omega_1, :]}^\dagger.
	\end{equation*}
	Since $X_{1, \Omega}\in \mathbb{R}^{m_1 \times(m_2m_3)}$ is a subset of columns of $X_{1, [\Omega_1, :]}\in \mathbb{R}^{m_1\times (p_2p_3)}$, for any $j\in [1:m_2m_3]$, there exists $k(j)\in [1:p_2p_3]$ such that the $j$-th column of $X_{1, \Omega}$ is equal to $X_{1, [\Omega_1, :]}$, namely 
	$$X_{1, \Omega}e_j = X_{1, [\Omega_1, :]} \tilde{e}_{k(j)}, $$
	where $e_j\in \mathbb{R}^{m_2m_3}$ and $\tilde{e}_k(j)\in \mathbb{R}^{p_2p_3}$ are both canonical vectors of different dimensions. In this case,
	\begin{equation*}
	\begin{split}
	& \mathbb{P}_{X_{1, \Omega}} e_j = X_{1, \Omega}^\dagger X_{1, \Omega}e_j = \left\{\left(U_2 U_{2, [\Omega_2, :]}^\dagger \right)^\top \otimes \left(U_3 U_{3, [\Omega_3, :]}^\dagger \right)^\top\right\} X_{1, [\Omega_1, :]}^\dagger X_{1, [\Omega_1, :]} \tilde{e}_{k(j)}\\
	= & \left\{\left(U_2 U_{2, [\Omega_2, :]}^\dagger \right)^\top \otimes \left(U_3 U_{3, [\Omega_3, :]}^\dagger \right)^\top\right\} \mathbb{P}_{X_{1, [\Omega_1, :]}}\tilde{e}_{k(j)},
	\end{split}
	\end{equation*}
	thus,
	\begin{equation*}
	\begin{split}
	& \left\|\mathbb{P}_{X_{1, \Omega}}\right\|_2^2 \leq \left\|\left\{\left(U_2 U_{2, [\Omega_2, :]}^\dagger \right)^\top \otimes \left(U_3 U_{3, [\Omega_3, :]}^\dagger \right)^\top\right\}\right\|^2 \|\mathbb{P}_{X_{1, [\Omega_1, :]}}\tilde{e}_{k(j)}\|_2^2\\
	\leq & \frac{p_2p_3}{(1-\varepsilon)^2m_2m_3} \cdot \frac{\rho r_1}{p_2p_3} = \frac{\rho r_1}{(1-\varepsilon)^2m_2m_3}.
	\end{split}
	\end{equation*}
	In other words, under the event that \eqref{ineq:lm_subtensor_event} holds, $X_{1, \Omega}$ satisfies the incoherence condition with constant $\rho/(1-\varepsilon)^2$. Since $\Xi_1$ is uniformly randomly selected $g_1$ values from $\Omega_2\times \Omega_3$, apply the first part of Lemma \ref{lm:sub-matrix-tensor} again, we have, with probability at least $1 - 2r_1\exp\left(-g_1(1-\varepsilon)^4/(4\rho r_1)\right)$,
	\begin{equation*}
	\rank(X_{\Omega_1\times \Xi_1}) = r_t,\quad \text{and}\quad \frac{p_1}{(1+\varepsilon)g_1} \leq \sigma_{\max}\left(X_{\Omega_1\times \Xi_1}^\dagger X_{t, \Omega} \right) \leq \frac{p_1}{(1+\varepsilon)g_1}.
	\end{equation*}
	Clearly, similar results hold for $t= 2, 3$. In conclusion, with probability at least $1 - 2\sum_{t=1}^3 \exp(-m_t(1-\varepsilon)^2/(4\rho r_t)) - 2\sum_{t=1}^3\exp\left(-g_t(1-\varepsilon)^4/(4\rho r_t)\right)$, \eqref{ineq:lm_subtensor_bound} holds. \quad $\square$
	
\end{itemize}

\begin{table}[h]
	\begin{footnotesize}
		\begin{center}
			\begin{tabular}{lll}
				Symbol & Definition & Note  \\\hline
				$\X$, $\S$, $U_t$ & \eqref{eq:Tucker-decomposition} & Tucker decomposition of low-rank $\X$\\
				$\bOmega$, $\Omega_t$, $\Xi_t$ & \eqref{eq:bOmega}\eqref{eq:index_Omega_Xi} & Index set of all, body, and arm measurements \\
				$Y_{\Xi_t}$, $X_{\Xi_t}$, $Z_{\Xi_t}$ & \eqref{eq:arm_measurements} & Arm matricizations for $\Y$, $\X$ and $\Z$\\
				$Y_{t, \Omega}$, $X_{t, \Omega}$, $Z_{t, \Omega}$ & \eqref{eq:body_matricization} & Body matricizations for $\Y$, $\X$ and $\Z$\\
				$Y_{\Omega_t\times \Xi_t}$, $X_{\Omega_t\times \Xi_t}$, $Z_{\Omega_t\times \Xi_t}$ & \eqref{eq:joint_matricization} & Joint matricizations for $\Y$, $\X$ and $\Z$\\
				$\hat{\X}$ & \eqref{eq:hat_X} & Final estimator for $\X$\\
				$\bar{R}_t$ & \eqref{eq:hat_R_t} & Expanding matrix \\
				$\xi_t$ & \eqref{eq:assumption_body_ratio} & body-arm ratio for pure signal $\X$\\
				$\lambda_t$ & \eqref{ineq:assumption_arm_body} & Tunning parameter to control arm-joint ratio \\
				$\bar{\lambda}_t$ & \eqref{ineq:bar_lambda_t} & Arm-joint ratio for pure signal $\X$\\
				$V^{(A)}$, $U^{(B)}$ & \eqref{eq:V^A-U^B} & Right singular vectors of $Y_{\Xi_t}$ and left singular vectors of $Y_{t, \Omega}$\\
				$J_t \in \mathbb{R}^{m_t\times g_t}$, $A_t \in \mathbb{R}^{p_t\times g_t}$ & \eqref{eq:A_t-J_t} & Joint and arm after rotations\\
				$r_t$, $\hat{r}_t$ & & True rank  of $\X$ and estimated rank by Algorithm \ref{al:noisy}\\
				$M_t, N_t \in \mathbb{O}_{p_t, r_t}$ & \eqref{eq:N_t,M_t} & First $r_t$ left, right singular vectors of $X_{t, \Omega}, X_{\Xi_t}$, respectively\\
				$\hat{M}_t, \hat{N}_t \in \mathbb{O}_{p_t, r_t}$ & \eqref{eq:N_t,M_t}\eqref{eq:hat_N_t_hat_M_t_U_V_related} & First $r_t$ left, right singular vectors of $Y_{t, \Omega}, Y_{\Xi_t}$, respectively\\
				$\tau$ & & = 1/5, convenient notation measuring signal noise gap (see \eqref{ineq:assumption_gap})\\
				$(N_t)_{\perp}, (M_t)_{\perp}$ & & Orthogonal complement matrices of $N_t$, $M_t$\\
				$Q$ & \eqref{eq:Q-U-X_Xi-association} & Connection between $X_{\Omega_t\times \Xi_t}$ and $X_{\Xi_t}$\\
				$U_{t, \Omega}$ & & Equivalent to $U_{t, [\Omega_t, :]}$, i.e., collection of rows of $U_t$\\
				$J_{\hat{r}_t}$, $A_{\hat{r}_t}$ &\eqref{eq:J_hat_r_t}\eqref{eq:A_hat_r_t} & Submatrices of $J_t, A_t$, a.k.a., $J_t$, $A_t$ after trimming\\
				$J_{\hat{r}_t}^{(X)}$, $J_{\hat{r}_t}^{(Z)}$ & \eqref{eq:J_hat_r_t^X^Z} & Signal and noise parts of $J_{\hat{r}_t}$  \\
				$A_{\hat{r}_t}^{(X)}$, $A_{\hat{r}_t}^{(Z)}$ & \eqref{eq:A_hat_r_t^X^Z} & Signal and noise parts of $A_{\hat{r}_t}$\\
				$K_{t1}, K_{t2}, L_{t1}, L_{t2} $ & \eqref{eq:K_t1_Lt1} & SVDs of $J_{\hat{r}_t}$ \\
				$A_{ts}$ & \eqref{eq:A_ts} & The first $r_t$ principle components of $A_{\hat{r}_t}$\\
				$A_{ts}^{(X)}, A_{ts}^{(Z)}$ & \eqref{eq:A_ts^X^Z} & Signal part and noise part of $A_{\hat{r}_t}$\\
				$J_{ts}$ & \eqref{eq:J_ts} & The first $r_t$ principle components of $J_{\hat{r}_t}$\\
				$J_{ts}^{(X)}, J_{ts}^{(Z)}$ & \eqref{eq:J_ts^X^Z} & Signal part and noise part of $J_{\hat{r}_t}$\\
				$\B_{s_1s_2s_3}$ & \eqref{eq:B_s1s2s3} & Tensor $\Y_{[\Omega_1, \Omega_2, \Omega]}$ after rotation\\
				$\B_{s_1s_2s_3}^{(X)}$, $\B_{s_1s_2s_3}^{(Z)}$ & \eqref{eq:B_s1s2s3^X},\eqref{eq:B_s1s2s3^Z} & Signal and noise part of $\B_{s_1s_2s_3}$\\
				$\tau_1$ & \eqref{ineq:sigma_min_J_t1} & Another convenient notation measuring signal noise gap\\
				$\bar{K}_t, \bar{L}_t$ & \eqref{eq:def_K_t_L_t} & Left and right singular vectors of $J_{\hat{r}_t}^{(X)}$\\\hline\hline
			\end{tabular}
		\end{center}
		\caption{List of symbols used in the proof for Theorem \ref{th:noisy}.}
		\label{tb:sympbols}
	\end{footnotesize}
\end{table}

 \end{document}